%% file: ms.tex
\documentclass[apj,tighten]{emulateapj}
\usepackage{apjfonts}
\usepackage{verbatim}
\newcommand{\zsun}{$Z_\odot$}
\newcommand{\msun}{$M_\odot$}

\newcommand{\hi}{H\,{\sc i}\rm}
\newcommand{\hii}{H\,{\sc ii}\rm}

\newcommand{\heii}{He\,{\sc ii}\rm}

\newcommand{\nii}{[N\,{\sc ii}]}

\newcommand{\oiii}{[O\,{\sc iii}]}
\newcommand{\oii}{[O\,{\sc ii}]}
\newcommand{\oi}{[O\,{\sc i}]}

\newcommand{\sii}{[S\,{\sc ii}]}

\newcommand{\expone}{$^{-1}$}

\newcommand{\eq}{\,=\,}

\newcommand{\te}{$T_e$}

\newcommand{\hbeta}{H$\beta$}
\newcommand{\halpha}{H$\alpha$}
\newcommand{\lin}{$\,\lambda$}
\newcommand{\llin}{$\,\lambda\lambda$}

\newcommand{\rtf}{$R_{25}$}

\newcommand{\oh}{12\,+\,log(O/H)}

\newcommand{\rtwothree}{R$_{23}$}
\newcommand{\vs}{vs.}

\slugcomment{}

\shorttitle{Metallicities in NGC~1512 and NGC~3621}
\shortauthors{Bresolin et al.}

\begin{document}

\title{Gas metallicities in the extended disks of NGC~1512 and NGC~3621. Chemical signatures of metal mixing or enriched gas accretion?\footnotemark[1]} 

\footnotetext[1]{Based on observations collected at the European Southern Observatory, Chile, under program 084.B-0107.}

\author{Fabio Bresolin} \affil{Institute for Astronomy, 2680 Woodlawn
Drive, Honolulu, HI 96822 \\bresolin@ifa.hawaii.edu}

\author{Robert C. Kennicutt} \affil{Institute of Astronomy, University of Cambridge, Madingley Road, Cambridge CB3 0HA, UK \\robk@ast.cam.ac.uk}

\and

\author{Emma Ryan-Weber} \affil{Centre for Astrophysics \& Supercomputing, Swinburne University of Technology, Mail H39, PO Box 218, Hawthorn, 3122 VIC, Australia \\eryanweber@swin.edu.au}

\begin{abstract}
We have obtained spectra of 135  \hii\ regions located in the inner and extended disks of the spiral galaxies NGC~1512 and NGC~3621, spanning the range of galactocentric distances  0.2\,--\,2\,$\times$\,\rtf\,(from $\sim$2-3~kpc to $\sim$18-25~kpc). We find that the excitation properties of nebulae in the outer ($R$\,$>$\,\rtf) disks are similar to those of the inner disks, but on average younger \hii\ regions tend to be selected in the bright inner disks. Reddening by dust is not negligible in the outer disks, and subject to significant large-scale spatial variations. 
For both galaxies the radial abundance gradient  flattens to a constant value 
outside of the isophotal radius. The outer disk O/H abundance ratio is highly homogeneous, with a scatter of only $\sim$0.06 dex. In the case of the interacting galaxy NGC~1512 we find a number of \hii\ regions with peculiar metallicity for their radius, a result which can be interpreted by gas flows activated by the gravitational encounter with NGC~1510.
Based on the excitation and chemical (N/O ratio) analysis we find no compelling evidence for variations in the upper initial mass function of ionizing clusters of extended disks.
The O/H abundance in the outer disks of the target galaxies corresponds to $\sim$35\% of the solar value (or higher, depending on the metallicity diagnostic). This agrees with our earlier measurements in M83 and NGC~4625, and conflicts with the notion that metallicities in extended disks of spiral galaxies are low and on the order of $\sim$0.1$\times$\zsun.
We show that, in general, the observed metal enrichment cannot be produced with the current level of star formation, even if the latter extends over a Hubble time. 
We discuss the possibility that metal transport mechanisms from the inner disks lead to metal pollution of the outer disks. Gas accretion from the intergalactic medium, enriched by outflows, offers an alternative solution, justified within the framework of hydrodynamic simulations of galaxy evolution. Specific model predictions of the 
chemical enrichment and the flat gradients in extended disks of nearby galaxies will be valuable to discriminate between these different scenarios.

\end{abstract}

 
\section{Introduction}

Optical and far-UV observations reveal that a large fraction of star forming  galaxies in the local Universe harbor significant levels of recent star formation in their extended disks,  in the form of young star clusters (ages $<$ 1 Gyr: \citealt{Dong:2008, Herbert-Fort:2009}) distributed well beyond the optical edges of the disks (\citealt{Gil-de-Paz:2005, Zaritsky:2007, Thilker:2005, Thilker:2007, Lemonias:2011}). 
This discovery has prompted several recent studies (e.g.~\citealt{Boissier:2007, Bigiel:2010, Goddard:2010, Alberts:2011}) concerning the conversion of gas into stars in a poorly explored galactic environment, characterized by low gas densities, long dynamical timescales, and presumably a lower  chemical enrichment compared to the inner disks of galaxies. Remarkably, also a number of early-type galaxies, which are 
generally quiescent in terms of star formation, have been found to host UV-emitting star clusters  in their outskirts (\citealt{Salim:2010, Thilker:2010, Moffett:2011}), and thus appear to be experiencing a `rejuvenation' in their extended disks.

A variety of mechanisms can be invoked for the assembly of  the large \hi\ envelopes observed around outer disk host galaxies, providing the raw material for the formation of new stars, and restructure the gas distribution into the observed extended, star forming disks. Among these are
tidal interactions, mergers,  bar-induced gas flows and gas accretion from the intergalactic medium. In particular,  3/4 of the Type~1 UV disks defined by \citet{Thilker:2007} display signatures of recent or ongoing gravitational encounters. However, the processes that trigger and sustain the star forming activity in outer disks are not always obvious, and still poorly understood in the case of isolated galaxies.
Once the gas reservoir is in  place, spiral density waves are thought to be able to propagate from the inner into the outer disks (\citealt{Bertin:2010})
and trigger star formation in overdensities of the gas distribution  (\citealt{Bush:2008, Bush:2010}), even if the spatially-averaged gas density remains below the critical value for gravitational instability (\citealt{Dong:2008}). However, this is unlikely to be the only process involved in the generation of extended UV disks, even for isolated galaxies, because the phenomenon is also observed around dwarf galaxies (\citealt{Hunter:2011}), where spiral arms are generally absent.

The star forming activity in extended disks proceeds with a very small efficiency, with values of the star formation rate surface density
that are one to two orders of magnitude smaller than those observed in the inner disks, with depletion timescales that are on the order of 10$^{11}$ years  (\citealt{Bigiel:2010}). Under these conditions a very slow buildup of metals can be reasonably anticipated, but the present-day oxygen abundances measured in outer disks seem to contrast this expectation. In the few detailed studies of the chemical composition of the interstellar medium (ISM) of extended disks carried out so far from \hii\ region spectroscopy, typical O/H values corresponding to at least 0.3\,--\,0.4 times the solar value\footnote{We adopt the solar oxygen abundance \oh$_\odot$ \,=\,8.69 by \citet{Asplund:2009}.} are found (\citealt{Bresolin:2009, Goddard:2011, Werk:2011}), which are difficult to reconcile with the significantly lower metallicities predicted by simple models of galactic evolution for systems that, like extended disks, have extremely high gas fractions ($>$\,90\%).
In addition, these pioneering studies have also found that beyond the isophotal radii (\rtf)
the metallicity radial distribution flattens to a virtually constant value, irrespective of the galactocentric distance,  in contrast with the negative exponential radial abundance gradients present at smaller distances. This behavior has also been seen in the  old stellar populations (\citealt{Vlajic:2009}). 

Various processes of angular momentum redistribution and gas mixing in galaxy disks have been invoked to explain these observations, including resonance scattering with transient spirals (\citealt{Roskar:2008}) and spiral-bar resonance overlap (\citealt{Minchev:2010}). Alternatively, \citet{Salim:2010} and \citet{Lemonias:2011}, among others, have argued that the gas that provides the raw material for the ongoing star formation in extended disks results from cold-mode accretion, as predicted by cosmological simulations (\citealt{Keres:2005, Dekel:2006, Ocvirk:2008}). This idea, however, contrasts with the low metal content expected for the infalling gas. In order to explain the relatively  high metallicity found for the ISM of the outer disks we would also need to consider a mechanism similar to the enriched infall scenario predicted by simulations involving wind recycling (\citealt{Oppenheimer:2008, Dave:2011}). We can of course envision a  combination of {\it internally-} and {\it externally-}driven processes  to enrich and homogenize the ISM at large galactocentric distances.

Measuring chemical abundances is of primary importance in the study of outer disks, in order to provide observational tests for the mechanisms mentioned above, and to understand whether extended disks truly represent young, chemically unevolved portions of galaxies, as implied by the inside-out disk growth paradigm  (\citealt{Matteucci:1989, Boissier:1999, Naab:2006}), and whether secular processes, too, affect the radial distribution of metals in the external regions of galaxies.

While only the youngest stellar clusters, still containing short-lived massive ionizing stars, produce a measurable emission line spectrum, the optical spectroscopy of the associated ionized gas provides a unique probe into the present-day metallicity of outer galactic disks.
In this paper we continue our exploration of the chemical abundances of the faint \hii\ regions located in the extended disks of nearby spiral galaxies. In \citet{Bresolin:2009} and \citet{Goddard:2011} we presented a chemical abundance analysis  of the prototypical outer disk galaxies M83 and NGC~4625, respectively. Here we apply the same multi-object spectroscopic technique to obtain the chemical abundances of a large number of \hii\ regions in the outer disks of the southern hemisphere galaxies NGC~1512 and NGC~3621.  Our radial coverage extends from the inner disk ($\sim$0.2\,$\times$\,\rtf)
 out to $\sim$2\,$\times$\,\rtf\, (19\,--\,25~kpc) for both galaxies.
 The comparison of their chemical properties  is of particular interest, because while NGC~3621 is a bulgeless, isolated spiral, NGC~1512 represents, together with its companion NGC~1510, an interacting system. Therefore, while both galaxies display a multitude of UV-bright clusters and \hii\ regions in their extended disks, their origin and chemical properties can in principle differ, since different  mixing processes could be at work. Further details about the two galaxies are introduced in Section~2, together with a description of our observations and the data reduction. 
In Section~3 we provide an account of the excitation properties of our \hii\ region sample. The chemical abundances and the radial abundance gradients are derived in Section~4.
Our interpretation of the results obtained in this work, and more generally in our project  on extended disks, in terms of metal mixing and enriched accretion, follows in Section~5. We summarize our results and conclusions in Section~6.

\section{Observations and data reduction}
In this section we first provide a brief description of the properties of the two galaxies investigated in this paper that are relevant for our analysis. The observations and the data reduction are presented next.

\begin{figure*}
\medskip
\center \includegraphics[width=0.75\textwidth]{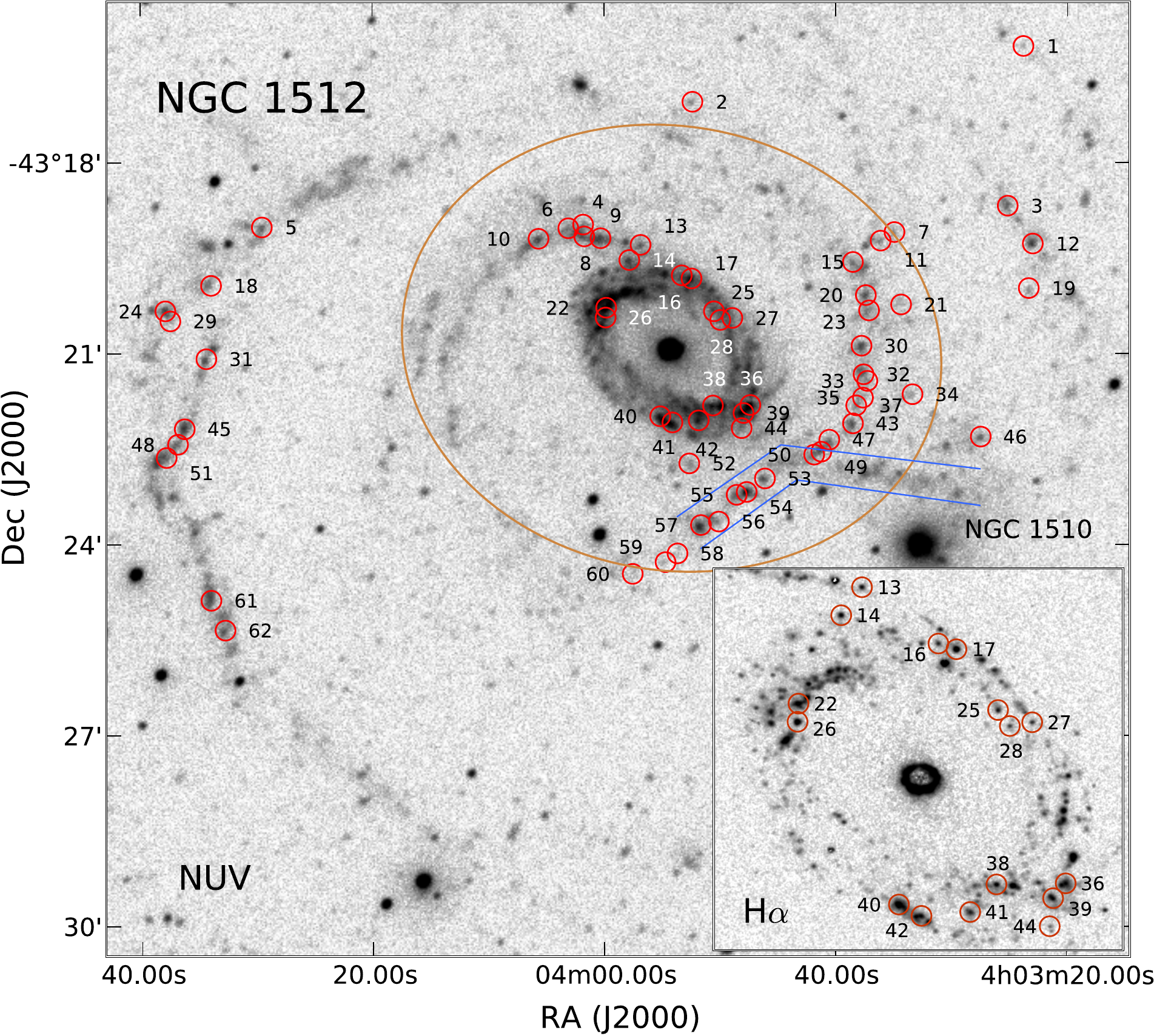}\medskip
\caption{Near-UV GALEX image of NGC~1512, identifying the target \hii\/ regions. The ellipse represents the projected circle with radius equal to the isophotal radius \rtf.
The companion galaxy NGC~1510 is identified in the south-west direction. The wedge-shaped area delimited by the blue lines marks the `tidal bridge' between NGC~1510 and NGC~1512. Regions numbered from 53 to 57 are located along this feature. The inset shows the inner 3\farcm2 of the galaxy from the continuum-subtracted SINGG (\citealt{Meurer:2006}) \halpha\ image. Region numbers correspond to those given in Table~\ref{fluxes1512}.
\label{image1}}
\end{figure*}

\subsection{The target galaxies}

\noindent
{\it NGC~1512 --} This galaxy and its companion, the blue compact dwarf NGC~1510 ($5'$ or $\sim$13.8~kpc to the SW), are a well-known interacting pair. The tidal interaction is likely responsible for the nuclear hotspot ring, the bar structure, the larger inner ring knotted with \hii\ regions (hence the SB(r) morphological classification), and the star forming activity seen in the asymmetric, filamentary outer spiral arm structure, composed of  a large number of UV-bright clusters and \hii\ regions (\citealt{Kinman:1978, Hawarden:1979, Thilker:2007}; see Fig.~\ref{image1}). The ongoing star formation in the compact NGC~1510 has also likely been triggered by the interaction. Evidence for tidally stripped material can be found in the zone between the two galaxies and in the wispy filaments to the NW of the pair. \citet{Koribalski:2009} made a detailed investigation of the extended  \hi\ envelope discovered by \citet{Hawarden:1979}, and discovered two tidal dwarf galaxy candidates at galactocentric distances between 64~kpc and 80~kpc. The spatial extent of the \hi\ distribution ($\sim$55~kpc in radius) is comparable to what is seen in other extended disk galaxies, such as M83.
The hydrodynamical simulations of the galaxy pair by \citet{Chakrabarti:2011} indicate a mass ratio of 1:50. A dynamical mass for NGC~1512 of $3\times10^{11}$\,\msun\, was measured by \citet{Koribalski:2009}. These authors have also found a good spatial correlation between the \hi\ emission and the distribution of UV clusters in the {\it Galaxy Evolution Explorer (GALEX)} images, with the clusters located at the peaks of the gas column density. Star formation occurs near the local critical density (\citealt{Kennicutt:1989}). \halpha\ images of NGC~1512 (and also NGC~3621) show that the spatial extent of the \hii\ region distribution is similar to that of the extended UV disk. \\

\begin{figure}
\medskip
\center \includegraphics[width=0.46\textwidth]{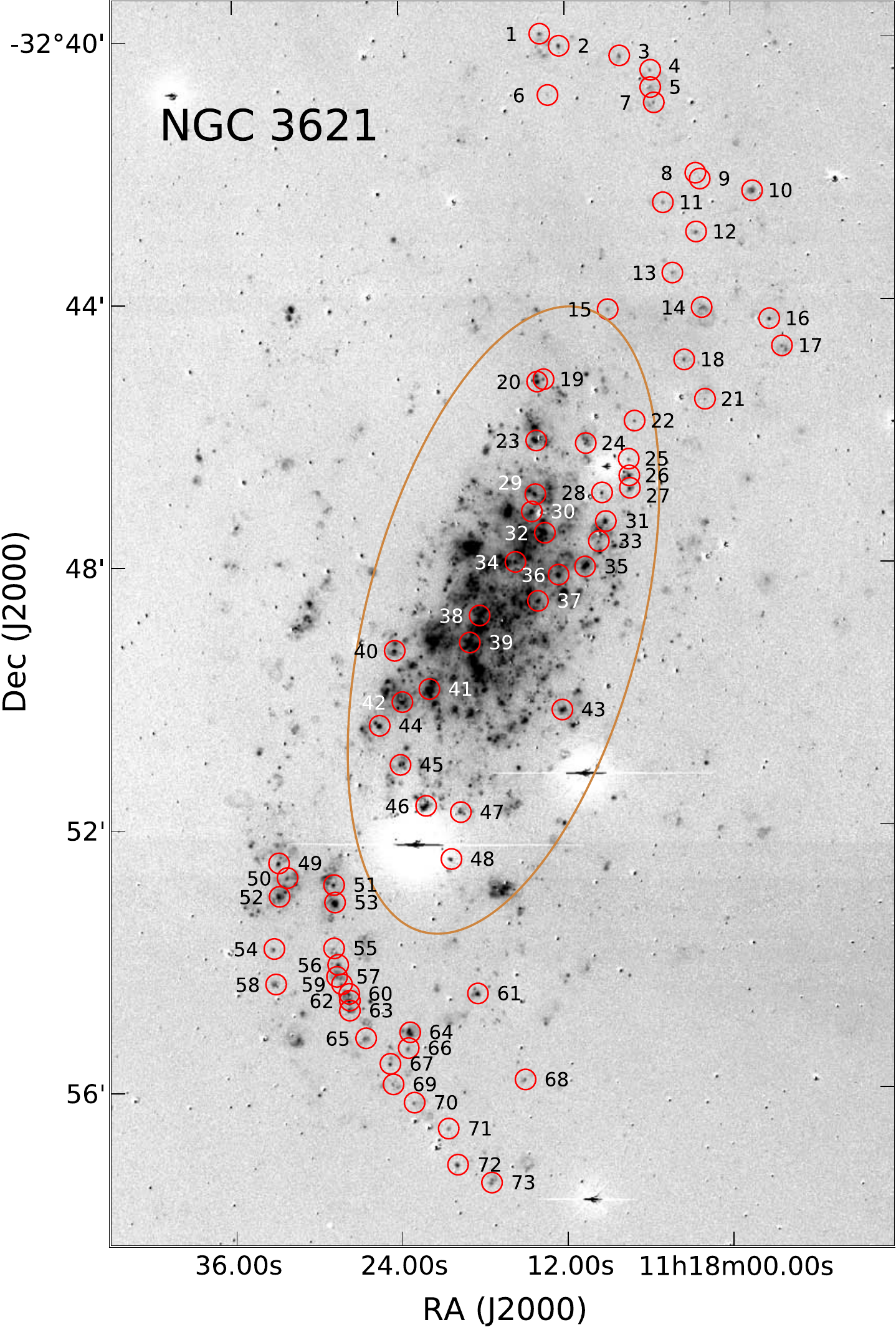}\medskip
\caption{Continuum-subtracted \halpha\ image of NGC~3621 (SINGS, \citealt{Kennicutt:2003a}), identifying the target \hii\/ regions. The ellipse shows the projected circle with radius equal to the isophotal radius \rtf. Region numbers correspond to those given in Table~\ref{fluxes3621}.
\label{image2}}
\end{figure}

\noindent
{\it NGC~3621 --} The GALEX and \halpha\ images (the latter is shown in Fig.~\ref{image2}) show a rather regular spiral structure extending out to a large galactocentric distance, and contained within the $34\times7$~kpc \hi\ envelope (\citealt{Koribalski:2004}). This galaxy is less massive than NGC~1512, with a dynamical mass of $2\times10^{10}$\,\msun\, (\citealt{de-Blok:2008}). NGC~3621 appears to be isolated (\citealt{Kraan-Korteweg:1979}). While the \hi\ velocity maps suggest  past interactions (\citealt{Thilker:2007}), rotation in a warped disk offers an alternative explanation (\citealt{de-Blok:2008}).
An analysis of GALEX and Spitzer images of the outer disk of this galaxy is presented by \citet{Alberts:2011}, who derived cluster ages between 1~Myr and 1~Gyr.

\input{tab1}

\subsection{Observations and data reduction} \label{observations}
Table~\ref{parameters} summarizes the main properties of the target galaxies used for this study. The center coordinates, as well as the positions of the individual \hii\ regions,
 were measured from astrometrically calibrated \halpha\ images obtained by the SINGG (\citealt{Meurer:2006}) and SINGS (\citealt{Kennicutt:2003a}) projects for NGC~1512 and NGC~3621, respectively (the preimaging data described below were used to measure the coordinates of a few \hii\ regions not included in the SINGG \halpha\ image of NGC~1512). For NGC~3621 we adopted the Cepheid distance given by \citet{Freedman:2001}, while for NGC~1512 the distance has been derived, following \citet{Koribalski:2004}, from the recessional velocity and a value of H$_0=75$~km\,s$^{-1}$\,Mpc$^{-1}$. The position angle and inclination to the line of sight of NGC~1512 correspond to the \hi\ kinematics values determined by \citet[see also \citealt{Hawarden:1979}]{Koribalski:2009}, which agree with the optical measurements by \citet{Buta:1988} and \citet{Dicaire:2008}.
For NGC~3621 we adopted the average values from the \hi\ study by \citet{de-Blok:2008}.
The remaining parameters included in Table~\ref{parameters} have been collected from \citet{de-Vaucouleurs:1991} and the HyperLeda database (\citealt{Paturel:2003}).

We obtained optical spectroscopy of \hii\ region candidates with the Focal Optical Reducer and Spectrograph (FORS2) attached to the 8\,m-aperture Antu unit of the European Southern Observatory (ESO) Very Large Telescope on Cerro Paranal. The observations were carried out in service mode between November 2009 and  March 2010. The  targets for the multi-object spectroscopy were selected from narrow-band \halpha\  pre-imaging data taken with the same instrument. The instrument setup was the same one we used in our study of the outer disk \hii\ regions of M83 (\citealt{Bresolin:2009}). Masks were cut with 1-arcsec wide slits.  The 600B grism with a central wavelength of 4650~\AA\ provided coverage of the blue spectral range ($\sim$3600-5200~\AA, at 4.5~\AA\ FWHM spectral resolution), while the 1200R grism and a central wavelength of 6500~\AA\ allowed us to cover the region around \halpha, \nii\llin6548,\,6583 and the \sii\llin6717,\,6731 lines with a spectral resolution of 2.4~\AA. 

The targets are distributed in four 6\farcm8 $\times$ 6\farcm8 FORS2 fields in NGC~1512, and in three fields in NGC~3621. \hii\ regions in both the inner and outer disks were included in the slit masks, to ensure a wide radial coverage for the resulting nebular chemical abundance gradient. In the final selection of the target \hii\ regions we accounted for both the \halpha\ brightness and the spatial distribution, observing some of the brightest \hii\ regions in the outer disks, but including several fainter ones in order to fill the slit maks.
The exposure times adopted were 6$\times$940\,s and 6$\times$970\,s for the blue exposures in NGC~1512 and NGC~3621, respectively. In the red 3$\times$667\,s exposures were used. For the easternmost field of NGC~1512 no red exposure could be obtained during the time allocated for this program. Therefore, for 10 targets in NGC~1512 the \halpha, \nii\ and \sii\ flux information is unavailable. We also secured spectra of the spectrophotometric standard stars LTT~1788 and LTT~3864, in addition to those included in the Paranal Observatory calibration plan, in order to
determine the instrumental spectral response.

The raw data were reduced with the EsoRex pipeline provided by ESO. This first step of the data reduction yielded wavelength-calibrated, two-dimensional spectra for each individual target. The final extractions and the flux calibration were obtained with standard {\sc iraf}\footnote{{\sc iraf} is distributed by the National Optical Astronomy
Observatories, which are operated by the Association of Universities for Research in Astronomy, Inc., under cooperative agreement with the National Science Foundation.} routines. The emission line intensities were then measured with the non-interactive {\tt\small fitprofs} program, by fitting one-dimensional gaussian profiles to the most important spectral features. The line intensity errors were estimated by accounting for both the uncertainty in placing the continuum level, as provided by the rms in regions adjacent to the emission lines, and the photon statistics. 

A number of targets were removed because they were either too faint to provide useful line emission information, or they were found to be background galaxies. In addition, we excluded from our analysis four objects (two in each galaxy) that we classified as supernova remnants (SNRs) from the strength of their \oi\ and \sii\ lines relative to \halpha. In NGC~3621 we also identified three emission-line star candidates, whose spectra are characterized by broad \halpha\ lines,  similar to those identified in the first stellar spectroscopic study of this galaxy by \citet{Bresolin:2001}.
The coordinates of these objects and of the SNRs are reported in Table~\ref{snr}.

Our final sample comprises 62 and 73 \hii\ regions in NGC~1512 and NGC~3621, respectively. Their celestial coordinates are presented in Appendix~A (Tables~\ref{fluxes1512} and \ref{fluxes3621}), where the targets are listed in order of decreasing declination,
while their spatial distributions are shown in Fig.~\ref{image1} and Fig.~\ref{image2}.

\input{tab2}

\section{Extinction}

\begin{figure*}
\medskip
\center
\includegraphics[height=0.5\textwidth]{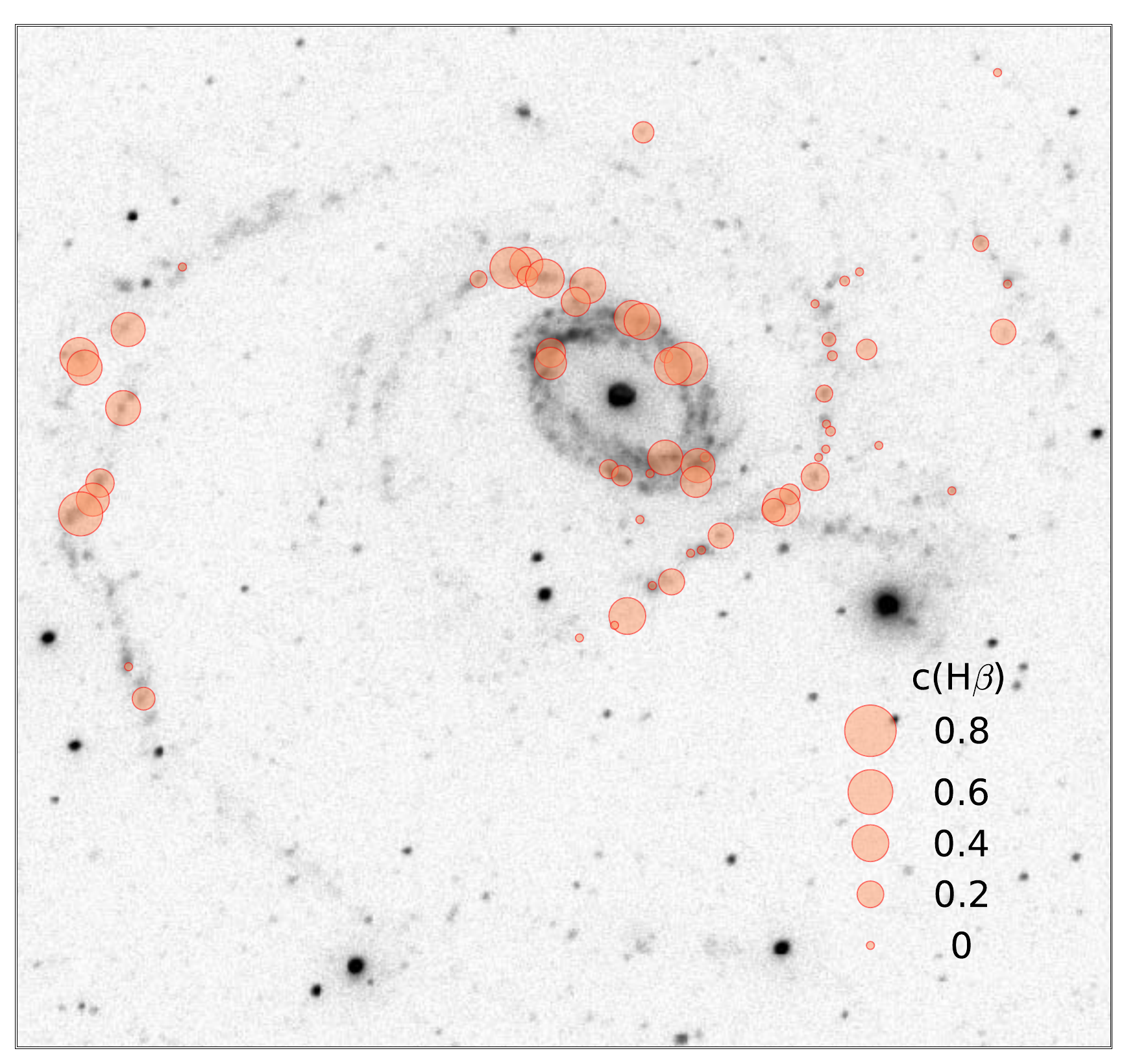}\medskip
\includegraphics[height=0.5\textwidth]{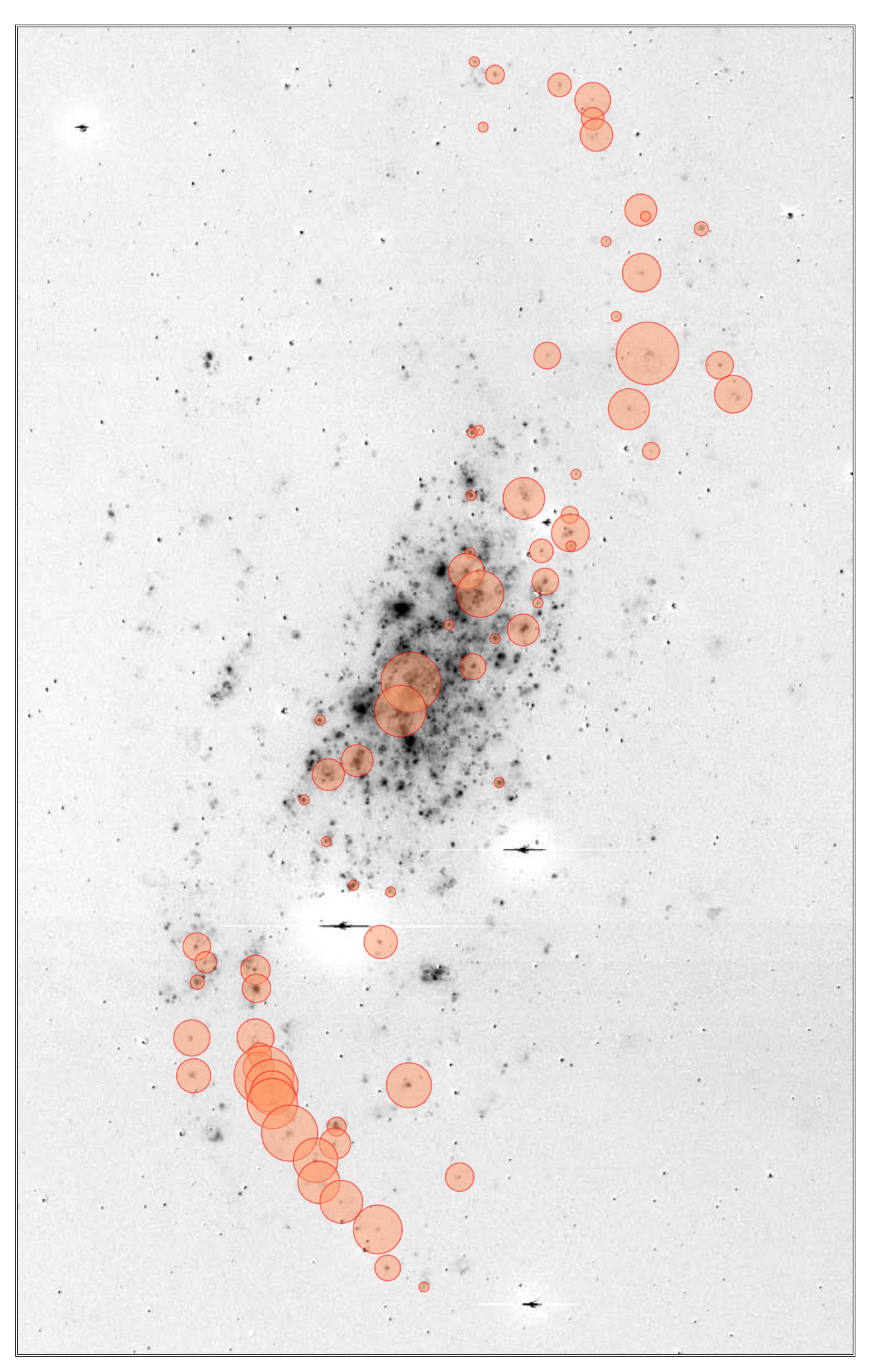}
\caption{The c(\hbeta) distribution across NGC~1512 (left) and NGC~3621 (right). Each \hii\ region in our sample is represented by a circle with a radius proportional to the measured extinction.
\label{extinction}}
\end{figure*}

The blue and red spectra have no overlap in wavelength. For this reason, the \halpha/\hbeta\  ratio could not be used to derive the value of the interstellar reddening from the Balmer decrement, and the strongest Balmer lines present in the blue spectral range (H$\beta$, H$\gamma$ and H$\delta$)
were used for this purpose, adopting the \citet{Seaton:1979} extinction curve. 
We accounted for the effects of underlying stellar absorption, by requiring that the H$\gamma$/H$\beta$ and H$\delta$/H$\beta$ ratios provided the same value for the extinction coefficient. In the few cases ($\sim 6$\% of the sample) where H$\delta$ was too faint to be measured reliably, we assumed an equivalent width of 2\,\AA\ for the absorption component.

The reddening-corrected  fluxes for the \oii\lin3727, \oiii\lin5007, \nii\lin6583 and \sii\llin6717,\,6731 emission lines are summarized in Tables~\ref{fluxes1512} and \ref{fluxes3621}. The associated errors include the line intensity uncertainties estimated above, and a term obtained by propagating the uncertainty in c(\hbeta), the logarithmic extinction at \hbeta\footnote{E(B$-$V) = 0.68\,c(\hbeta) (\citealt{Seaton:1979}).}. For the normalization of the \nii\ and \sii\ line fluxes to \hbeta\,=\,100 we assumed \halpha\,=\,286, which is valid for case~B emission at 10$^4$\,K.
The two tables also report the  c(\hbeta) values, and the {\it total}, extinction-corrected \halpha\ fluxes, obtained from the flux-calibrated narrow-band 
images used for the astrometry (a correction for the \nii\ contamination to the emission measured by the narrow-band filters was evaluated from our spectra).

We found that in the outer disks of the two galaxies studied here
the extinction is not negligible (we note that the foreground reddening values according to \citealt{Schlegel:1998} are 
quite small, E(B$-$V)\,=\,0.011 and 0.08 for NGC~1512 and NGC~3621, respectively).
Outer disks are therefore not dust-free, a result that is supported by the relatively high oxygen abundance we find (Section~4). We also find that the extinction is variable on large spatial scales. 
In particular, in the eastern outer arm of NGC~1512 the median extinction value is c(\hbeta)\,=\,0.34, while on the opposite side of the galaxy it is essentially zero. A similar asymmetry is present in NGC~3621, where the northern outer arm appears less dusty than the southern one. For the latter the section 
extending between regions 59 and 71 has a median c(\hbeta)\,=\,0.37, peaking at a value around 0.50 in the arm section between regions 59 and 65 (separated by 1.8~kpc in projection). Fig.~\ref{extinction} shows the spatial  distribution of the c(\hbeta) values across the disks of the two galaxies.
These asymmetrical distributions reflect those in the neutral gas density, as can be seen from the \hi\ maps  shown by \citet[NGC~1512]{Koribalski:2009} and
\citet[NGC~3621]{Walter:2008}.

\section{Excitation properties}
\begin{figure*}
\medskip
\includegraphics[width=0.51\textwidth]{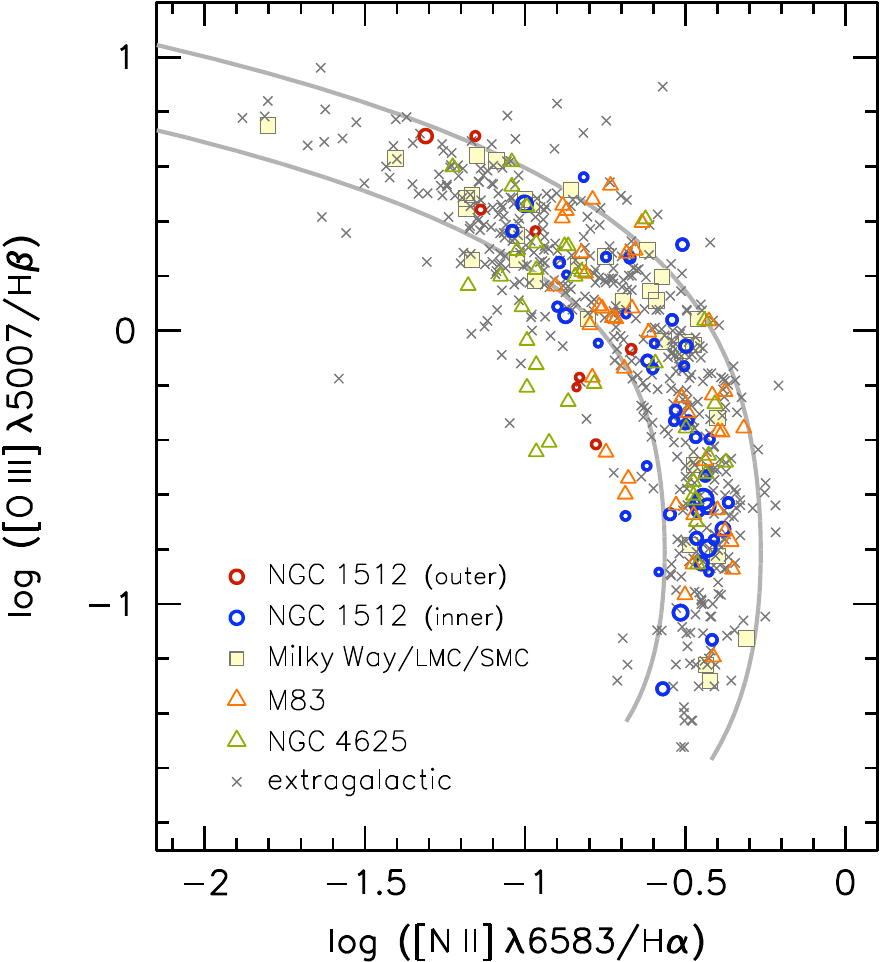}
\includegraphics[width=0.46\textwidth]{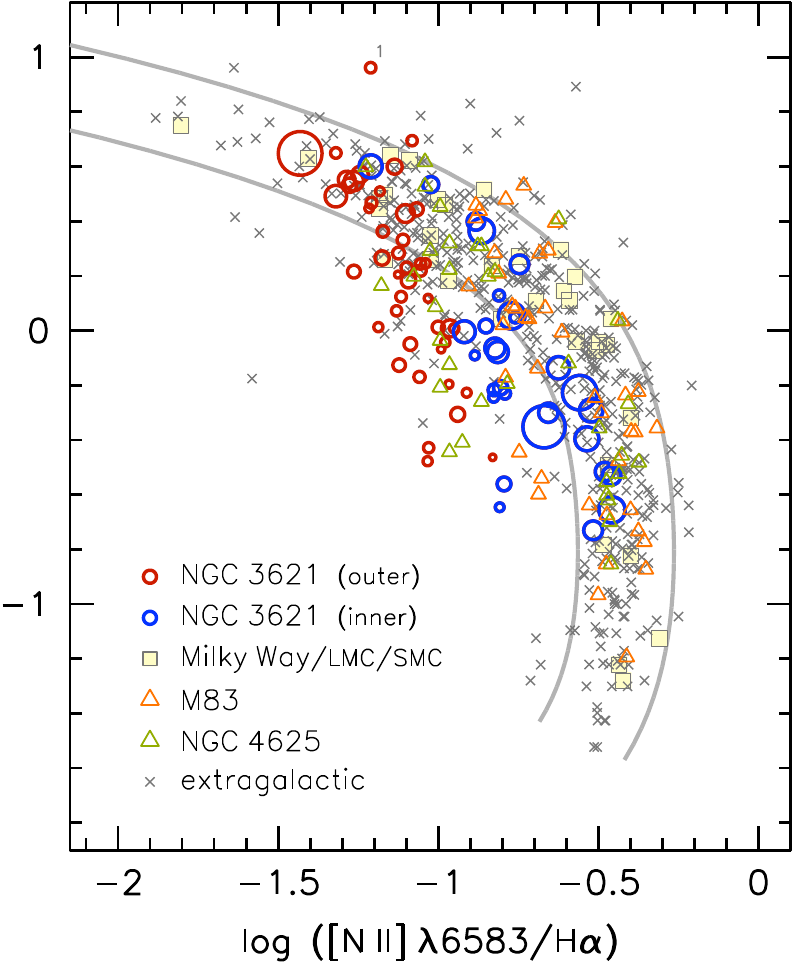}
\medskip
\caption{The \nii/\halpha\ \vs\ \oiii/\hbeta\ diagnostic diagram for the \hii\ regions observed in NGC~1512 (left) and NGC~3621 (right), divided into inner (blue circles) and outer (red circles) disk objects. The samples of ionized nebulae analyzed in the extended disk galaxies M83 (\citealt{Bresolin:2009}) and NGC~4625 (\citealt{Goddard:2011}) are included (triangles). Comparison samples from the Milky Way and the Magellanic Clouds (squares) and additional galaxies (crosses) are shown (see text for references). The two curves enclose \hii\ regions that deviate by $\pm$0.15 dex from the fit representing the empirical excitation sequence. The sizes of the symbols shown for NGC~1512 and NGC~3621 are proportional to the observed \halpha\ flux.
\label{bpt}}
\end{figure*}

We inspected the excitation properties of our \hii\ region sample  by plotting the \nii\lin6583/\halpha\ (=\,N2) line ratio against \oiii\lin5007/\hbeta\ (=\,O3), as shown in Fig.~\ref{bpt} (the left and right panels refer to NGC~1512 and NGC~3621, respectively). For comparison, we have included the samples of  inner and outer disk \hii\ regions studied in M83 by \citet{Bresolin:2009}
 and in NGC~4625 by \citet{Goddard:2011}, and shown with triangle symbols. Additional data points for nearly 500 extragalactic inner disk \hii\ regions in spiral and irregular galaxies (cross symbols) have been drawn from the following papers: \citet{McCall:1985}, \citet{Garnett:1997}, \citet{Bresolin:1999, Bresolin:2005, Bresolin:2009a}, \citet{van-Zee:1998a}, \citet{Kennicutt:2003}, and \citet{Esteban:2009}. Lastly, we also included the  39 \hii\ regions located in the Milky Way and the Magellanic Clouds studied by \citet[square symbols]{Kennicutt:2000}, because in many cases these nebulae only contain a single or very few ionizing stars, a situation resembling the one encountered in outer disks  (\citealt{Bresolin:2009}). For both NGC~1512 and NGC~3621 the area of the symbols is proportional to the observed \halpha\ flux, in order to provide information on the relative ionizing output of the embedded clusters.
   
As an aid in the interpretation of the excitation diagrams, we have fitted the sequence outlined by the data points with a four-degree polynomial, obtaining:

$$x = -0.585 - 0.576 y - 0.750 y^2 - 0.506 y^3 - 0.168 y^4$$

\noindent
where $x$\,=\,log(\nii\lin6583/\halpha),  $y$\,=\,log(\oiii\lin5007/\hbeta). In Fig.~\ref{bpt}  we show two curves, separated by $\pm0.15$ dex from this empirical sequence,  defining a region that encloses the bulk of the data points. Some important deviations from the main sequence occur both above and below this zone. The upper region of the diagnostic diagram hosts nebulae  with unusually hard ionizing fields. In fact, in virtually all cases,  \hii\ regions located in this area of the diagram display, in addition to large \oiii/\hbeta\ flux ratios, also nebular \heii\lin4686 emission, as summarized recently by \citet{Bresolin:2011a}. We note that for our target 1 in NGC~3621, identified in Fig.~\ref{bpt} (right) as the outlier located in the central top portion of the diagram, we also have detected the \heii\lin4686 emission line, and thus this nebula represents a new instance of this rare class of high-excitation extragalactic \hii\ regions, whose ionizing sources are still a matter of debate (\citealt{Kehrig:2011}). (The fact that we can spatially resolve this target in our \halpha\ images excludes the possibility that this high-excitation nebula is a planetary nebula).

\begin{figure*}
\medskip
\epsscale{0.8}
\plotone{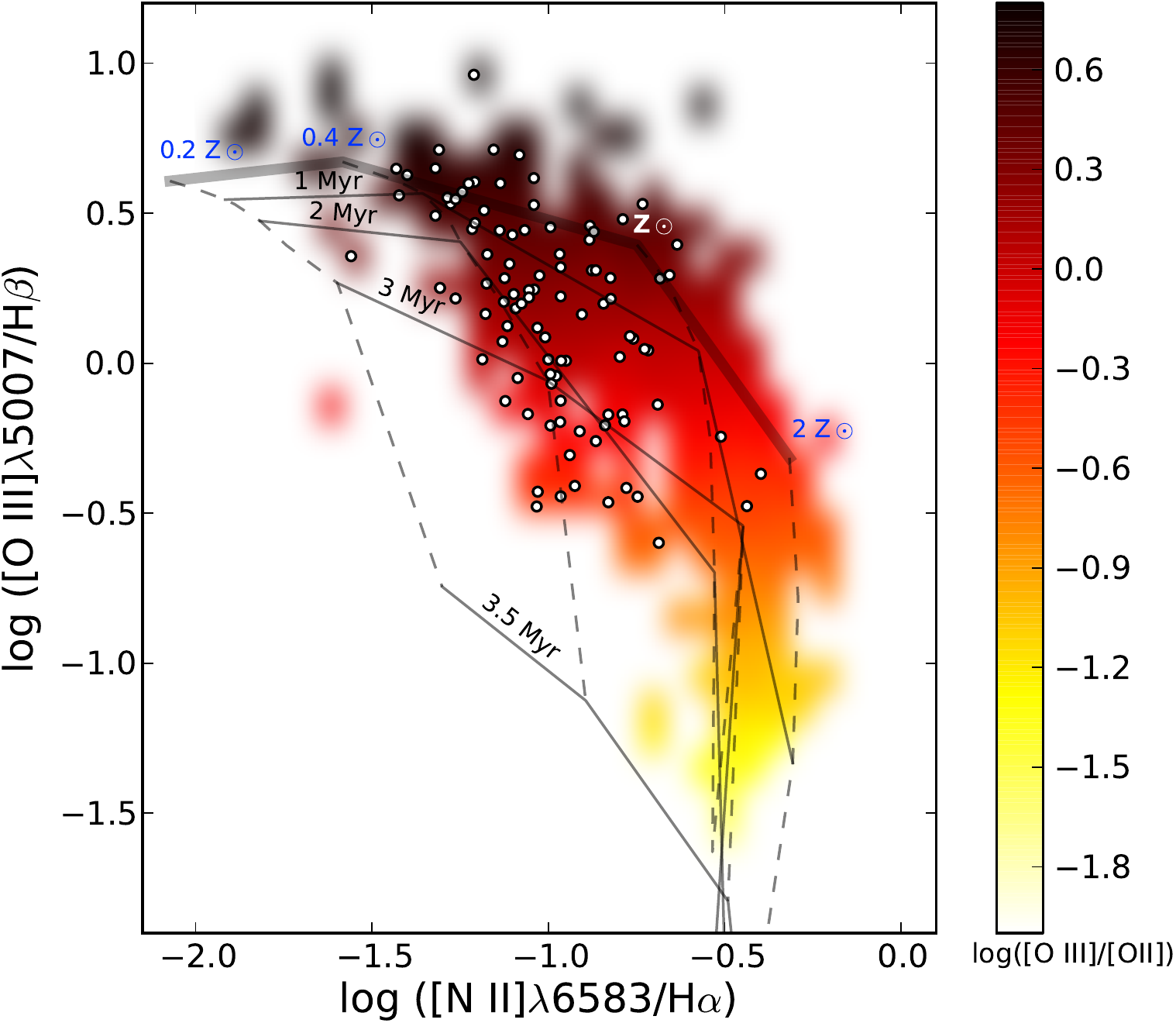}
\caption{The \nii/\halpha\ \vs\ \oiii/\hbeta\ diagnostic diagram for the objects shown in Fig.~\ref{bpt} (with additional \hii\ regions from \citealt{Ferguson:1998a}),
where the color code maps the 
\oiii\lin5007/\oii\lin3727 line ratio, a measure of the ionization parameter. The model grid by \citet{Dopita:2006a} for $\log (R)$\eq$-2$ is plotted for 
ages between  0.1~Myr (thick grey line) and 3.5~Myr, and metallicities Z\eq0.2~\zsun, 0.4~\zsun, \zsun\ and 2~\zsun. Dashed lines connect models of constant metallicity.
Outer disk \hii\ regions ($R$\,$>$\,\rtf) in different galaxies are represented by white dots.
\label{bpt-models}}
\end{figure*}

\citet{Goddard:2011} noticed that a significant fraction of outer disk \hii\ regions in NGC~4625 (green triangles in Fig.~\ref{bpt}) lies below the empirical  excitation sequence in the N2 \vs\ O3 diagram defined above. Fig.~\ref{bpt} shows that this region of the diagram, characterized by a lower excitation (\oiii/\hbeta\ ratio) compared to the main sequence, is populated also by inner disk \hii\ regions (cross symbols), even though these cases are apparently rare.
The few \hii\ regions in NGC~1512 that lie below the main excitation sequence belong to both the inner disk (defined as $R$/\rtf\,$<$\,1, and represented by blue symbols) and the outer disk ($R$/\rtf\,$>$\,1, red symbols) of the galaxy, and tend to have 
smaller ionizing fluxes compared to the majority of the objects located along the sequence, as indicated by the size of the symbols. This situation resembles the case of the \hii\ regions in the outer disk of M83 (orange triangles), where only very few points deviate from the main excitation sequence.

The case of  NGC~3621 is quite different. About half of the \hii\ regions in this galaxy, in both the inner and outer parts of the disk, deviate significantly from the excitation sequence.
\citet{Goddard:2011} attributed the low excitation of  outer disk \hii\ regions in  NGC~4625 to a combination of two effects. First,  the progressive depletion of massive stars belonging to the upper main sequence of the ionizing clusters as these evolve with time. Secondly,
bright and young \hii\ regions tend to be selected in spectroscopic studies of the inner portions of spiral galaxies, while in outer disks fainter, and possibly older, nebulae will be typically included, because of the relative scarcity of \hii\ regions. The study of NGC~4625 by \citet{Goddard:2011}, however, left open the possibility that outer disk \hii\ regions have {\it systematically} lower excitation levels relative to inner disk ones, which could imply systematic differences in the upper initial mass functions. In order to shed light on this issue, we have compared our excitation diagrams with the predictions from theoretical models of evolving \hii\ regions.\\

\subsection{Comparison with model predictions}
The region in the N2 \vs\ O3 diagnostic diagram  that lies below the main excitation sequence has seldom received attention in extragalactic \hii\ regions studies, possibly because of the relatively low number of objects found in it. Our investigations of outer spiral disks uncovered a significant population of \hii\ regions exhibiting excitation properties that deviate from the extragalactic \hii\ region excitation sequence.
 In order to clarify the nature of this population 
 we take a look at  theoretical predictions of the evolution of nebular emission line ratios with ionizing cluster age. We consider the models presented by \citet{Dopita:2006a}, in which the Starburst99 code (\citealt{Leitherer:1999}) was used to follow the evolution of ionizing clusters synthesized using the Geneva stellar tracks with high mass loss (\citealt{Meynet:1994}), and a stellar atmosphere prescription that includes the non-LTE O-star models of \citet{Pauldrach:2001} and the Wolf-Rayet star models of \citet{Hillier:1998}. The dynamical evolution of the surrounding \hii\ regions was followed by accounting for the mechanical energy input from stellar winds and supernova explosions, and was parameterized in terms of the ratio $R$\eq($M_{cl}/M_\odot$)\,/\,($P/k$) between the ionizing cluster mass and the pressure of the surrounding  interstellar medium (the latter term in units of 10$^4$\,cm$^{-3}$\,K). In these models the ionization parameter, instead of being treated as a free parameter, is  determined by the evolutionary status of the expanding \hii\ regions for  given values of $R$ and metallicity. The calculation of the model nebular spectra was carried out with the Mappings code (\citealt{Sutherland:1993}).
For a comparison with our empirical line flux ratios, we selected models with $\log (R)$\eq$-2$, which are appropriate for outer disk cluster masses of $\sim$10$^3$~\msun\ (\citealt{Goddard:2010, Goddard:2011}) and an interstellar medium pressure $\sim$10$^4$\,cm$^{-3}$\,K. We note, however, that our conclusions will be largely unaffected by the particular choice of $R$ (Fig.~3 of \citealt{Dopita:2006a} shows that variations of $R$ by orders of magnitude
affects mostly the ages derived from the model grids, but not their relative distribution).
We also remind the reader about the presence of systematic uncertainties in the predictions 
obtained from photoionization models of synthesized stellar population. The calculated \oiii/\hbeta\ ratio appears to be systematically underestimated, relative to
observations (\citealt{Dopita:2006a} attribute the discrepancy  to the inadequacy of the ionizing photon output predicted by the stellar atmosphere codes). However, while these uncertainties affect the quantitative predictions of the models to a certain degree (such as the \hii\ region ages and metallicities), the qualitative trends are regarded as being quite robust.

The comparison between the $\log (R)$\eq$-2$ model grid predicted by  \citet{Dopita:2006a} for the N2 \vs\ O3 diagnostic diagram
at metallicities Z\eq0.2~\zsun, 0.4~\zsun, \zsun\ and 2~\zsun\ and our extragalactic \hii\ region sample is shown in Fig.~\ref{bpt-models}. In addition to the observational data presented in Fig.~\ref{bpt}, we have included the \hii\ region sample of \citet{Ferguson:1998a}, which comprises a handful of nebulae in the outer disks of the galaxies NGC~628, NGC~1058 and NGC~6946. 
We mapped the variation of the \oiii\lin5007/\oii\lin3727 line ratio, which relates directly to the ionization parameter, using the color code found at the right hand of the figure. As expected, this ratio decreases from its maximum value at the top of the diagram, along the excitation sequence towards the bottom right. We show models for ionizing cluster ages of 0.1~Myr (thick grey line), 1~Myr, 2~Myr, 3~Myr and 3.5~Myr. Models of constant metallicity are connected by the dashed lines. The match between the model grid and the observational  data is satisfactory, although objects at the top of the diagram would be better described by models with higher $R$ values (see Fig.~3 in \citealt{Dopita:2006a}). This comparison indicates that \hii\ regions older than 3.5~Myr are absent  in the data samples we included in  this diagram. The decrease of the ionization parameter with increasing age is well illustrated by the decrease of the \oiii\lin5007/\oii\lin3727 line ratio along tracks of constant metallicity
(the plot also suggests a dependence of the  ionization parameter on metallicity).

The interesting point raised by Fig.~\ref{bpt-models} concerns the region of the diagram lying below the main excitation sequence, 
and the distribution of outer disk \hii\ regions in the diagram. The models illustrate how, as \hii\ regions of the upper branch of the excitation sequence evolve with time, they move to lower O3 values (at roughly constant N2 values), populating the region below the excitation sequence in about 3~Myr. At older ages the expanding \hii\ regions have reached such a low surface brightness  that their inclusion in spectroscopic samples becomes unlikely.
More importantly, \hii\ regions that are located at radii $R$\,$>$\,\rtf\ in their host galaxies, represented by white dots, span the full range of ages from 0 to 3-3.5~Myr, with a significant number of them located in the `zero-age', upper portion of the diagram. This indicates that outer disk \hii\ regions do not have systematically small \oiii/\hbeta\ line ratios compared to inner disk counterparts. The fact that many of them (including a few inner disk regions) are found in the lower part of the diagram appears to be consistent with the selection effect mentioned earlier, whereby inner disk \hii\ regions are preferentially observed when young and bright, while in the outer disks the observer's attention is also drawn by older nebulae.

\smallskip
In order to further test this interpretation we looked for trends in the \halpha-to-FUV flux ratio, $f_{H\alpha}/f_{FUV}$,  across the N2 \vs\ O3 diagram, since this ratio is expected to decrease with time, due to  the different typical timescales for the (nebular) \halpha\ ($<10^7$\,yr) and the (stellar) FUV ($\sim$10$^8$\,yr) emission of young star clusters. The comparison between \halpha\ and FUV emission is not straightforward on a cluster-by-cluster basis,
because of uncertainties in matching sources at the two different wavelengths (i.e.~due to different pixel scales between ground-based optical and spacecraft UV images, but also because the FUV emission is intrinsically more diffuse than the \halpha\ emission, so that objects that are clearly distinct in \halpha\ can often be seen as single clumps in FUV). Therefore, trends in the $f_{H\alpha}/f_{FUV}$ ratio
are expected to be rather crudely defined, and only meaningful in a statistical sense.

\begin{figure}
\medskip
\epsscale{1.1}
\plotone{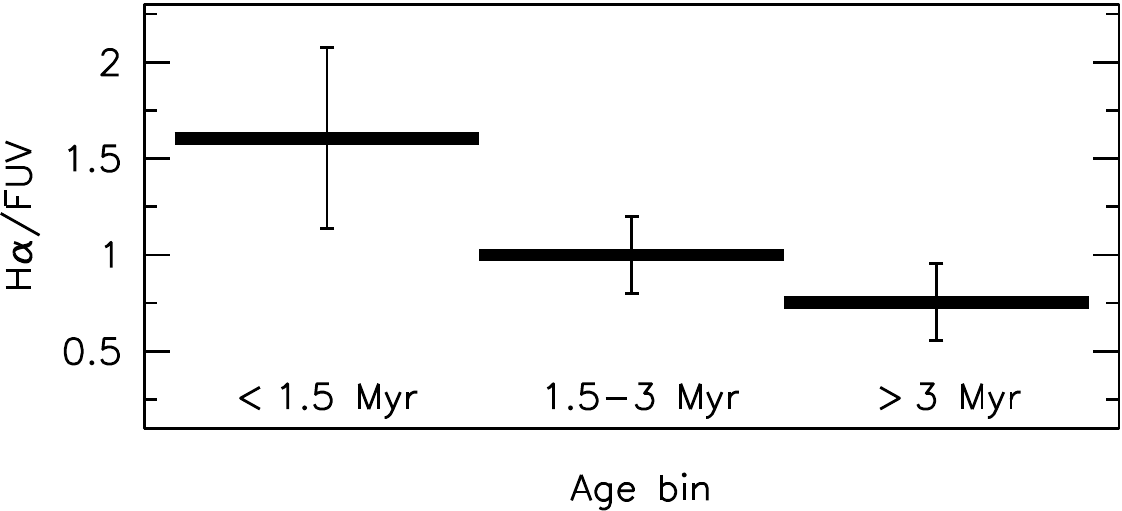}
\caption{The mean \halpha-to-FUV flux ratio for the \hii\ regions in NGC~3621 for three age bins, normalized to the value for the central bin.
\label{ages}}
\end{figure}

For our test we focused on NGC~3621, which hosts a large fraction of \hii\  regions falling outside the main sequence in the N2 \vs\ O3 diagram.
FUV fluxes were measured from the available GALEX images, using SExtractor (\citealt{Bertin:1996}). We then divided the \hii\  region sample in three broad categories, based on the age we derived for each object using  the model grid by \citet{Dopita:2006a} adopted earlier: $t$\,$<$\,1.5~Myr, $t$\,=\,1.5\,--\,3~Myr and $t$\,$>$\,3~Myr. 
Since we are looking for a statistical trend, and because of the uncertainties mentioned above, this approach is desirable over one with a finer subdivision in age groups. The mean $f_{H\alpha}/f_{FUV}$ values in these three age bins show a decreasing trend with age, as illustrated in Fig.~\ref{ages}, as expected. In this figure, we normalized the $f_{H\alpha}/f_{FUV}$ ratio to the central bin value, and included error bars to indicate the standard errors of the mean values. Although the significance of the result is only modest, given the size of the error bars, there is a significant difference in the $f_{H\alpha}/f_{FUV}$ ratio between the youngest and the oldest age bin. This ratio decreases by a factor of $\sim$2,  in good quantitative agreement with model predictions (see Fig.~7 of \citealt{Goddard:2011}). This result reinforces our conclusion that age plays an important  role in the distribution of \hii\ regions in the N2 \vs\ O3 diagram below the main sequence.

\smallskip
An apparent inconsistency in this picture concerns \hii\ regions hosting Wolf-Rayet (W-R) stars. Since these stars make their appearance in the ionizing clusters 2-3~Myr after the initial episode of star formation, nebulae containing such stars should be located below the main excitation sequence according to the theoretical predictions. However, they are not. As an example, the two \hii\ regions in NGC~3621 where we detected   the W-R \lin4650 feature (numbers 53 and 64 in Table~\ref{fluxes3621}) are both located on the excitation sequence. A clue to a possible explanation is provided by Fig.~\ref{bpt} (right), where object 53 is represented by the second largest open circle (i.e.~it possesses the second largest observed \halpha\ flux), in the upper part of the diagram. The presence of W-R stars appears to affect the location of the \hii\ regions  in this diagram by  boosting the ionizing flux output of the central clusters, and consequently providing a larger \oiii/\hbeta\ ratio. 
The problem, of course, is that this effect should be reproduced by the models, which include the W-R evolutionary phase. For a metallicity Z\eq0.4~\zsun\ this phase should last $\sim$2~Myr, which is long enough to be resolved by the model grid.

An important additional  point to make concerns the effects of metallicity on the evolution on the N2 \vs\ O3 diagram. At high metallicity ($\sim$\zsun\ and above) the evolution of the \hii\ regions occurs along the nearly-vertical, low-excitation arm of the main sequence, which is found at \nii\lin6583/\halpha\ $\simeq$ $-0.4$ (we remind the reader that the \nii/\halpha\ ratio is essentially a measure of metallicity). Therefore, if the metallicity of an aged outer disk \hii\ region approaches the solar value, its location in the diagnostic diagram will remain on, or near, the main excitation sequence. This appears to be the case for the galaxies M83 and NGC~1512, where very few objects lie outside the region of the diagram enclosed by the curves drawn in Fig.~\ref{bpt} (we must add that  in the case of NGC~1512 our interpretation is hampered by the small number of outer disk \hii\ regions with available \nii\ lines). For a galaxy of lower metallicity, such as NGC~3621, we see \hii\ regions straddling the space to the left of the near-vertical portion of the main excitation sequence, starting from \oiii\lin5007/\hbeta\,$\simeq$\,0.6 down to $\simeq$\,$-0.4$.
We conclude that the differences found in the  N2 \vs\ O3 diagnostic diagram  of outer disk \hii\ regions of different galaxies stem from differences in metallicity. We also infer from the inspection of  Fig.~\ref{bpt-models} that we do not need to invoke a different upper initial mass function for outer disk populations to explain their distribution in the excitation diagram, a possibility we could not rule out in our study of NGC~4625 (\citealt{Goddard:2011}), but that it is sufficient to account for the evolution of the emission line ratios as the ionizing clusters of the \hii\ regions become older.

\section{Chemical abundances}
\subsection{Abundance diagnostics}
For the determination of the oxygen abundances (`metallicities') 
of the target \hii\ regions we have used a variety of emission line diagnostics. In this way, we expect to be able to recover robust qualitative radial trends, regardless of the 
well-known, but unexplained, systematic uncertainties, which can amount to a factor of up to $\sim$0.7 dex, presently afflicting 
the determination of the chemical abundances of extragalactic \hii\ regions from nebular spectroscopy 
(see \citealt{Bresolin:2008, Bresolin:2009a, Kewley:2008} for recent discussions).
In fact, a well-established, but poorly understood, discrepancy exists between the nebular abundances obtained from the classical, {\em direct} \te-based method (\citealt{Menzel:1941}) and the theoretical predictions of photoionization model grids (e.g.~\citealt{McGaugh:1991}). 
Various strong-line abundance indicators have been introduced to obtain the oxygen abundances of ionized nebulae
from the strongest emission lines present in their spectra, e.g.~when the faint auroral lines 
required in the direct method remain undetected. Calibrations of these methods in terms of metallicity
have been 
proposed both from empirical abundances (based on auroral line detections)  and from the 
results of theoretical photoionization models. Consequently, the chemical abundances derived from strong-line methods  display large systematic differences when applied to the same observational data, with empirically calibrated  diagnostics providing measurements at the low end of the metallicity range.
Differential analysis methods are considered to be largely unaffected by these unexplained systematic discrepancies, although different calibrations of strong-line methods can yield small  differences in the slopes of galactic abundance gradients
(\citealt{Bresolin:2011}).

The faint \oiii\lin4363 auroral line, which is used, in combination with the nebular \oiii\llin4959,\,5007 lines, to obtain the electron temperature of the ionized gas in the direct method, was detected in six \hii\ regions in NGC~1512 and in 12 \hii\ regions in NGC~3621. These detections are important to verify the consistency of the results obtained from the remaining strong-line metallicity diagnostics, which do not provide direct measurements of the electron temperature.
Because the strength of \oiii\lin4363 relative to \oiii\lin5007 decreases with metallicity, most of the \hii\ regions in which the auroral line was measured are located in the outer disks, where metallicities are lower than in the inner disks. The direct oxygen abundances were derived from the observed line strengths with the {\tt nebular} package available in {\sc iraf}, adopting the updated atomic data used by \citet{Bresolin:2009a}.
We summarize in Table~\ref{te} the measured \oiii\lin4363 line fluxes and the derived electron temperatures, together with the O/H and N/O abundance ratios. The object identification numbers in Column~(1) refer to those found in the catalog of line fluxes presented in Tables~\ref{fluxes1512} and \ref{fluxes3621}.\smallskip

In addition, three strong-line nebular abundance diagnostics have been used:

\smallskip
\noindent
{\it (a)} the N2\eq\nii\lin6583/\halpha\ ratio, as calibrated as a function of metallicity by \citet{Pettini:2004} from a sample of \hii\ regions with \oiii\lin4363 detections. This diagnostic, therefore, provides oxygen abundances that are approximately on the same scale as the auroral line method. 

\begin{figure}
\epsscale{1.1}
\plotone{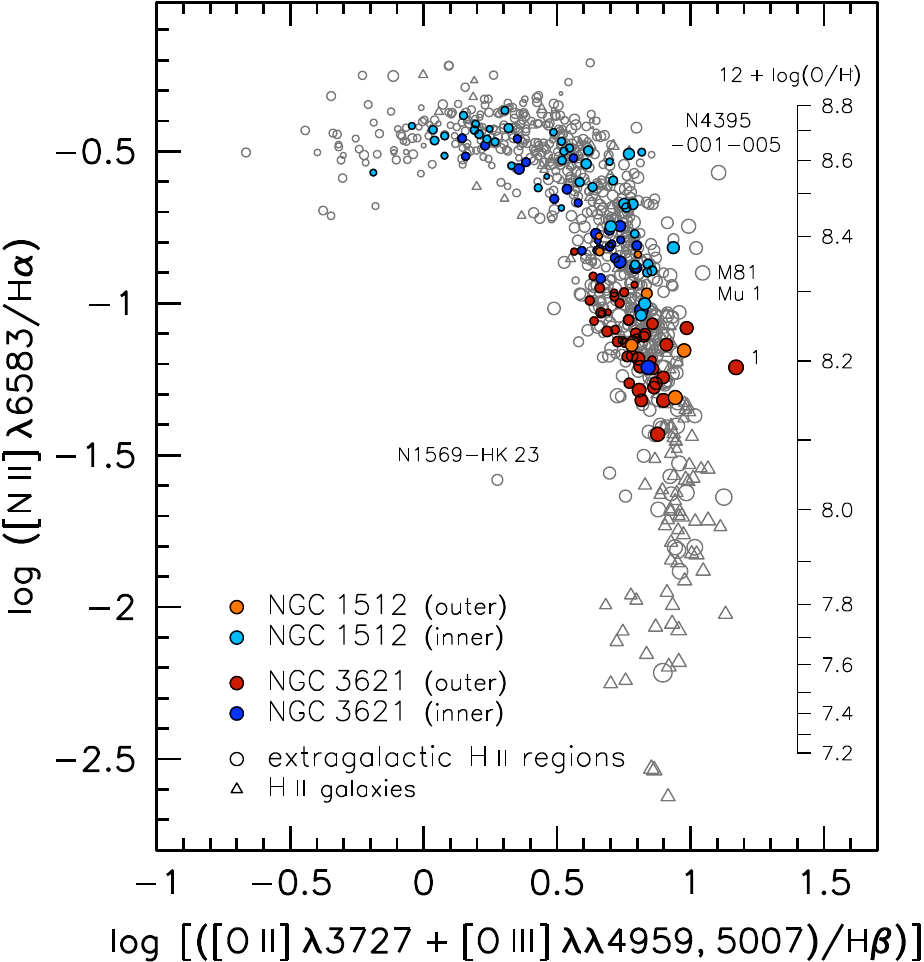}
\caption{The \rtwothree\eq(\oii\lin3727\,+\,\oiii\llin4959,\,5007)/\hbeta\ ratio as a function of N2\eq\nii\lin6583/\halpha\ for the sample of \hii\ regions shown in Fig.~\ref{bpt}. Colored symbols are used for \hii\ regions in the inner and outer disks of NGC~1512 and NGC~3621. 
\hii\ galaxies extending to low metallicities have been added (triangles). The metallicity scale at the far right refers to \oh\ values calculated from the N2 diagnostic. A few outlier \hii\ regions are identified.\label{bpt-r23}}
\end{figure}

\smallskip
\noindent
{\it (b)} the \rtwothree\eq(\oii\lin3727\,+\,\oiii\llin4959,\,5007)/\hbeta\ ratio, introduced by \citet{Pagel:1979}. Of the various  \rtwothree\ calibrations  as a function of metallicity available in the literature, we selected the one proposed by \citet{McGaugh:1991}, which is based on a grid of photoionization models, and which accounts for changes in the nebular ionization parameter through the value of the \oii/\oiii\ line ratio. It is well-known that this calibration, together with  other theoretical calibrations of \rtwothree, provides systematically higher oxygen abundances relative to empirical calibrations obtained from \oiii\lin4363 detections (e.g.~\citealt{Bresolin:2009a}), but  this offset is roughly independent of metallicity, so that relative metallicity trends remain well determined.
To address the difficulty represented by the fact that two different values of O/H can correspond to the same \rtwothree\ value, which gives rise to the notion of a `lower' and an `upper' branch of the calibration, we have plotted in Fig.~\ref{bpt-r23} the \rtwothree\ indicator as a function of N2, which is instead monotonic with O/H. In order to populate the diagram towards the low-metallicity end, we have considered, in addition to the \hii\ region samples already presented in the previous figures, the \hii\ galaxy samples of \citet{Terlevich:1991} and \citet{Guseva:2011}, shown by the triangle symbols.
It can be seen that for log(N2) values down to at least $-1.5$, corresponding to \oh\,$\simeq$\,8.1 in the empirical abundance scale (shown to the right of the \hii\ regions sequence), we can safely select the upper \rtwothree\ branch. All of the objects we observed in NGC~1512 and NGC~3621 fall in this range, and we have therefore applied the analytical expression for the upper branch calibration of \citet{McGaugh:1991}
provided  by \citet{Kuzio-de-Naray:2004}. 
The plot in Fig.~\ref{bpt-r23} identifies a few outliers relative to the \hii\ region sequence. To the right of the sequence, in particular, we find nebulae with particularly hard ionizing spectra, which can be related to the presence of \heii\lin4686 emission (e.g.~region 1 in NGC~3621; for other regions falling in this category and not shown here see \citealt{Bresolin:2011a}), or shocked gas signatures (e.g.~NGC~4395$-001-005$, see \citealt{McCall:1985}).

\smallskip
\noindent 
{\it (c)} the empirical method proposed by \citet{Pilyugin:2010}, which calibrates the strengths of the \oii\ and \oiii\ lines as a function of the O/H ratio, based on \oiii\lin4363-based abundances, but making use also of the \nii\llin6548,\,6583/\hbeta\ and \sii\llin6717,\,6731/\hbeta\ line ratios to empirically constrain the electron temperature. This method is being referred to as the ONS method. Once again, by construction this diagnostic should provide metallicities that are consistent with those derived from the \oiii\lin4363 line. Compared to previous empirical calibrations provided by the same authors, the ONS method has the advantage of being applicable to the full range of metallicities encountered in extragalactic \hii\ regions.

\subsection{Radial metallicity gradients}

The galactocentric radial gradients of the O/H abundance ratio are shown in Fig.~\ref{oh-1512} for NGC~1512 and Fig.~\ref{oh-3621} for NGC~3621.
The upper panels present the abundances obtained from the \oiii\lin4363 diagnostic (red-yellow circles) and N2 (open circles). The lower panels show the abundances obtained from \rtwothree\ (open triangles) and the ONS method (open circles). Linear regression fits to the inner disk data ($R$\,$<$\,\rtf) are represented by the sloping grey lines.  The different diagnostics show that the oxygen abundance distributions are flat beyond \rtf. Within the errors, 
gradient determinations in the outer disks are consistent with a flat radial trend. For simplicity, we have then represented the outer disk metallicities with 
the average O/H ratios, shown by the horizontal lines. The statistical test described below indicates that the change in slope takes place near the isophotal radius, in the radial range 0.8\,--\,1.2~\rtf. We did not attempt to pinpoint the exact radius at which this occurs, since a significantly higher number of targets in the transition zone would be required. The small discontinuities in O/H seen in Fig.~\ref{oh-1512} and \ref{oh-3621} reflect the difficulty to establish the radius at which the transition occurs.

The statistical significance of the flattening of the abundance gradient beyond the isophotal radius has been estimated with an F-test, following \citet{Zahid:2011a} in their study of NGC~3359. The test allows us to check whether the improved fit (i.e.~smaller residuals) obtained using a two-slope model for the radial abundance distribution (in which we set the outer gradient to be flat, for the reason given above) is statistically significant compared to a single-slope representation. 
In general we find that the flattening is significant at the $>$99.99\% level. The exception is represented by the gradient observed with the N2 diagnostic in NGC~1512, where a single slope (dashed line in Fig.~\ref{oh-1512}) reproduces the observed data as well as if we used two lines of different slopes. However, this is a result of the small number of \hii\  regions in the outer disk of this galaxy for which we can calculate N2-based abundances  (see below). The \rtwothree\ metallicity indicator, for which we have more data points in this galaxy, displays a statistically significant flattening.

In the following we discuss separately our results for NGC~1512 and NGC~3621.

\subsubsection*{NGC~1512}
\begin{figure}
\medskip
\epsscale{1.15}
\plotone{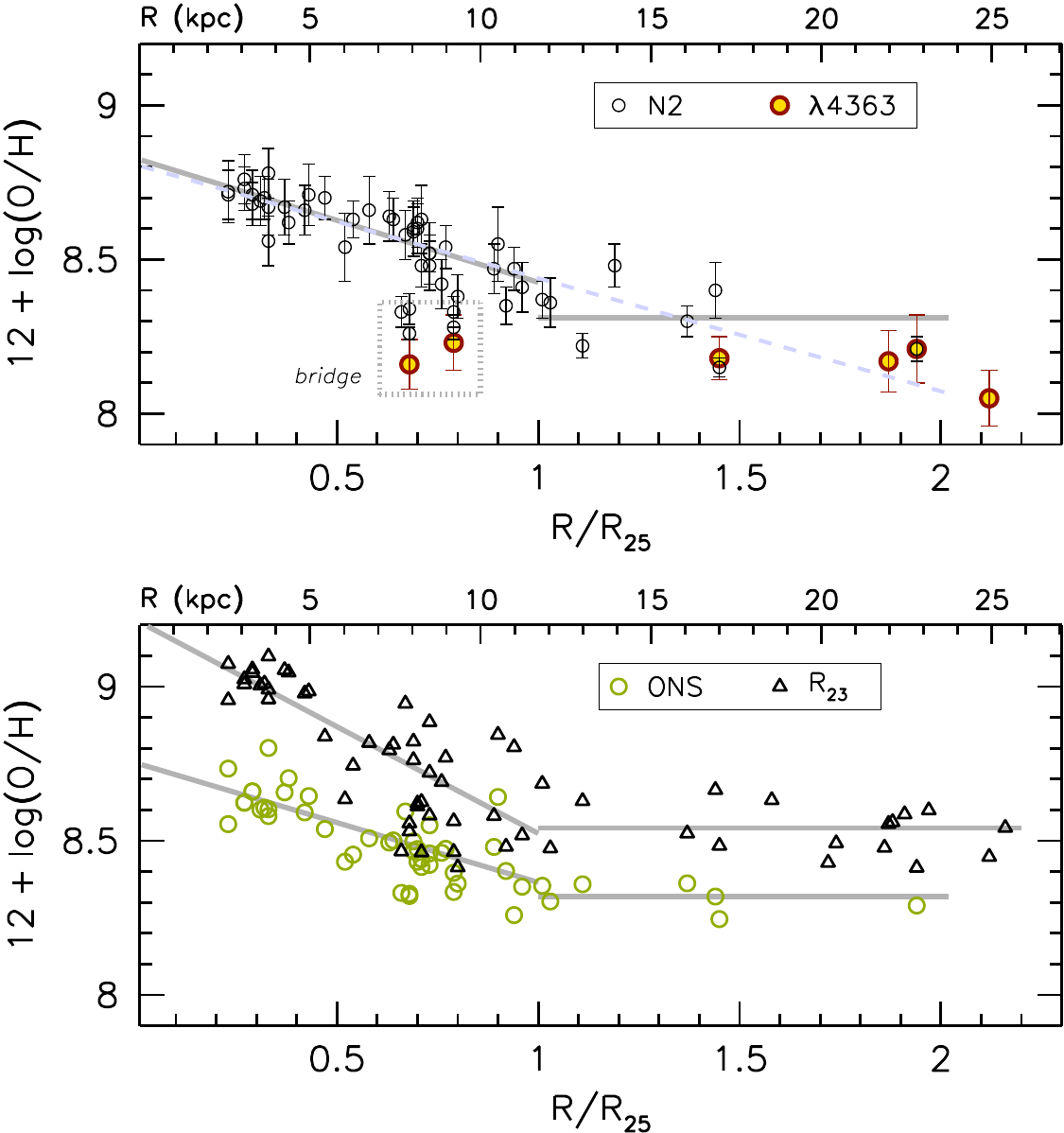}
\caption{{\it (Top)} The galactocentric O/H abundance gradient in NGC~1512, as determined from the N2 diagnostic. Abundances obtained from the \oiii\lin4363 auroral line are shown as yellow dots. The dotted box encloses the 5 \hii\ regions that are located in the `bridge' between NGC~1510 and NGC~1512. The galactocentric distance axis is labeled both in terms of $R$/\rtf\ (bottom scale) and in kpc (top scale).  The dashed line shows a single-slope linear fit for all N2-based data points.
{\it (Bottom)} The galactocentric O/H abundance gradient in NGC~1512, as determined from the \rtwothree\ indicator (open triangles) and from the ONS method (open circles). 
In both panels the full grey lines represent the linear fits to the inner disk \hii\ region radial abundance distribution ($R$\,$<$\,\rtf), and the constant values assigned to the outer disk from the mean O/H ratio for $R$\,$>$\,\rtf. For clarity, error bars are shown only in the top panel.
\label{oh-1512}}
\end{figure}

\begin{figure}
\medskip
\epsscale{1.15}
\plotone{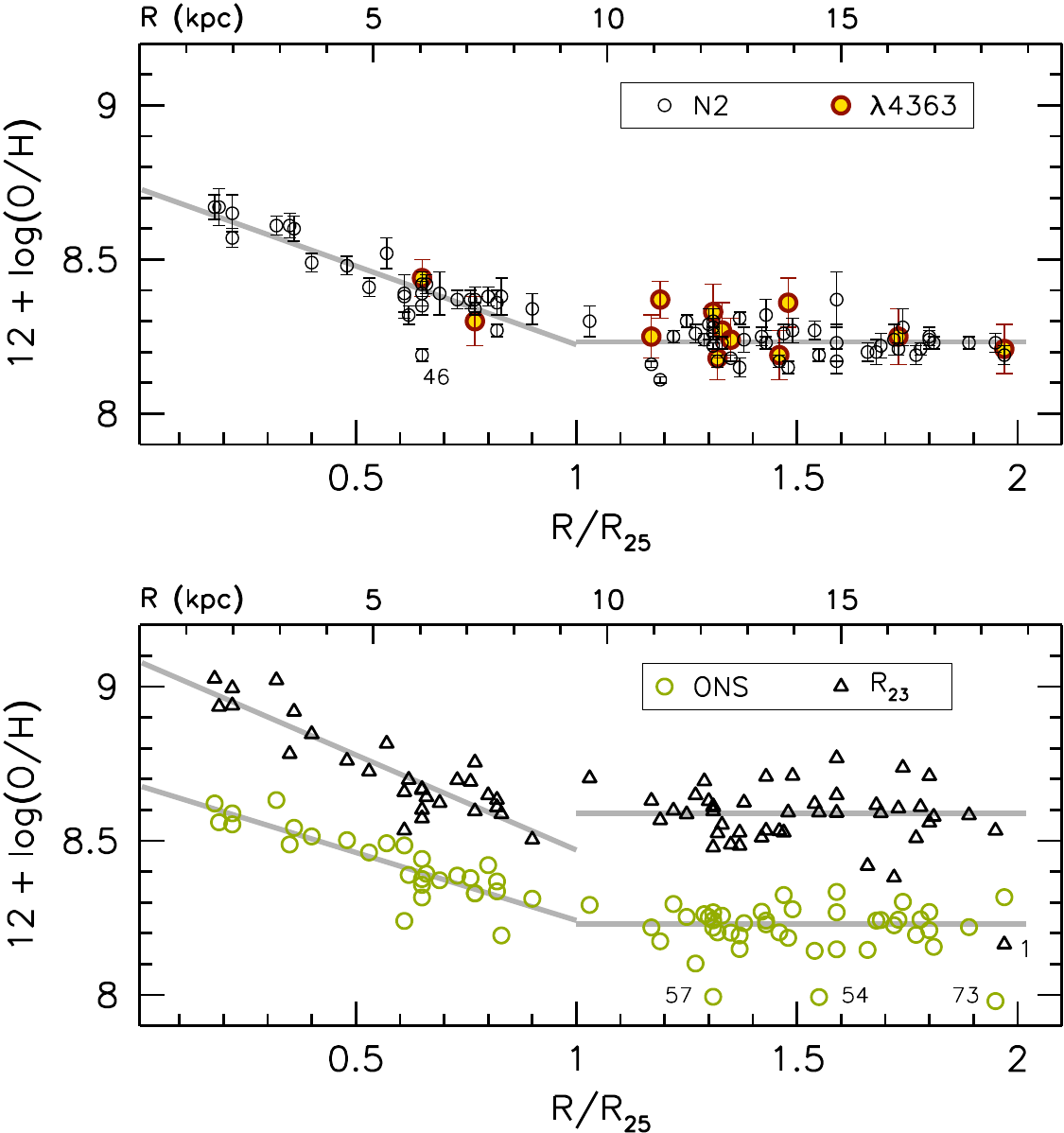}
\caption{Same as Fig.~\ref{oh-1512} for NGC~3621. A few outliers discussed in the text are labeled by their number in Table~\ref{fluxes3621}.
\label{oh-3621}}
\end{figure}

As mentioned in Sect.~\ref{observations}, we do not have red spectra covering the \halpha\ and \nii\ lines for the 10 \hii\ regions located in the outer eastern spiral arm, and therefore the N2 and ONS diagnostics provide an incomplete picture of the abundance distribution in the outer disk of NGC~1512. However, the existing data are in good qualitative agreement with the \rtwothree\ abundances, which clearly show a flat metallicity beyond the isophotal radius. This is also confirmed by the few \oiii\lin4363 detections, even though the O/H value for the outermost data point, corresponding to region 61, is slightly below the rest (although not significantly, considering the error bars; see upper panel of Fig.~\ref{oh-1512}). The N2 and the auroral-line based abundances are in good quantitative agreement for objects in which both methods could be used.

The abundances obtained from the N2 diagnostic reveal that five inner disk \hii\ regions, two of which have consistent \oiii\lin4363-based O/H abundances, lie significantly below the regression line. Interestingly, all of these objects, enclosed in the box drawn in Fig.~\ref{oh-1512} (top) and numbered 53 to 57 in Fig.~\ref{image1},
 are located in the 'bridge' structure identified by \citet{Koribalski:2009} in the zone between the centers of NGC~1512 and NGC~1510, and possibly corresponding to tidal debris (see Fig.~\ref{image1}). These \hii\ regions therefore do not appear to belong to the same population of \hii\ regions found in the main body of NGC~1512, and might instead be related to material that was stripped  during the interaction between these two galaxies. 
The lower O/H abundance ratio measured in the bridge is consistent with the idea that this gas has reached its current location from larger radii, possibly around the isophotal radius (or beyond), where a similar O/H value is found.
 Although this association to tidally stripped material remains speculative, we have removed objects 53-57 from the calculation of the linear regression 
of the metallicity as a function of radius, parameterized as \oh\eq $a$ \,+\, $b$\,($R/R_{25}$). The results obtained for coefficients of this fit from the different abundance diagnostics are summarized in Table~\ref{table:regression}.
It can be seen that the N2 and the ONS methods provide results that are consistent with each other, for both the intercept value, \oh\,$\simeq$\,8.8, and the slope of the abundance gradient in the inner disk, approximately $-0.034$~dex\,kpc$^{-1}$. The \rtwothree\ indicator yields a $\sim$0.4 dex larger intercept value, as expected, but also a steeper slope, although in this case the linear fit displays an rms scatter that is almost double the one measured using the N2 diagnostic.
We finally note that, after discarding the objects in the bridge, the rms abundance scatter obtained for the inner disk of NGC~1512 from the N2 diagnostic, 0.06~dex, is the same value that we measure for  NGC~3621. 

\subsubsection*{NGC~3621}
Thanks to the large number of \hii\ region abundances available in both the inner and the outer disk of this galaxy, the manifestation of a break in the abundance gradient of NGC~3621, and the flat distribution beyond the isophotal radius, are very well established by any of the strong-line metallicity diagnostics we adopted. The N2 abundances 
are in good agreement with those derived from the auroral line method, in those cases where the \oiii\lin4363 line was measured. 
As in the case of NGC~1512, the N2 and ONS diagnostics yield consistent abundance gradients, which are well anchored to the empirical \oiii\lin4363-based metallicities. The overall metallicity is about 0.1 dex lower than in NGC~1512.
While the slope derived from \rtwothree\ is again slightly steeper, it is not significantly different from the one derived from N2. For all the diagnostics, the rms abundance scatter relative to the linear fit is small, $\sim$0.06 dex. The abundance distribution is thus quite tightly defined, and its scatter is essentially determined by the observational errors, rather than true metallicity variations at a given galactocentric radius.

In Fig.~\ref{oh-3621} we identify a few outliers relative to the radial trend defined by the majority of the \hii\ regions. In the top panel, region 46 has an N2-based abundance that is 
about 0.2 dex lower than the \oiii\lin4363-based one. The latter matches the abundance from the linear fit at the corresponding radial distance, so apparently there is something peculiar about its N2 line ratio. In the bottom panel, the metallicity of region~1 estimated from \rtwothree\ is clearly underestimated, a result that is due to its high excitation level. \citet{Bresolin:2011a} showed that this is typical for \hii\ regions displaying nebular \heii\ emission. There are also three objects (54, 57 and 73) for which the ONS method appears to underestimate the O/H abundance ratio (these were excluded from the calculation of the mean 
outer disk abundance given in Table~\ref{table:regression}). \citet{Pilyugin:2010} provide three separate analytical expressions to derive O/H, for different ranges of the \nii\llin6548,\,6583/\hbeta\ and \sii\llin6717,\,6731/\hbeta\ line ratios. Owing to the observed log(\nii/\sii)\,$<$\,$-0.25$, these \hii\ regions are classified as `hot', and are assigned a rather low metallicity. If they were classified as `warm', by a change in the criterion used to separate \hii\ regions in different temperature regimes, the O/H ratio for these three \hii\ regions would agree with the rest of the outer disk sample.
This suggests that some adjustment to the criteria adopted by \citet{Pilyugin:2010} could lead to a better estimation of the O/H ratio at metallicities around \oh\eq8.2.

\input{tab3}

\input{tab4}

\subsection{Nitrogen}
\begin{figure}
\medskip
\epsscale{1.15}
\plotone{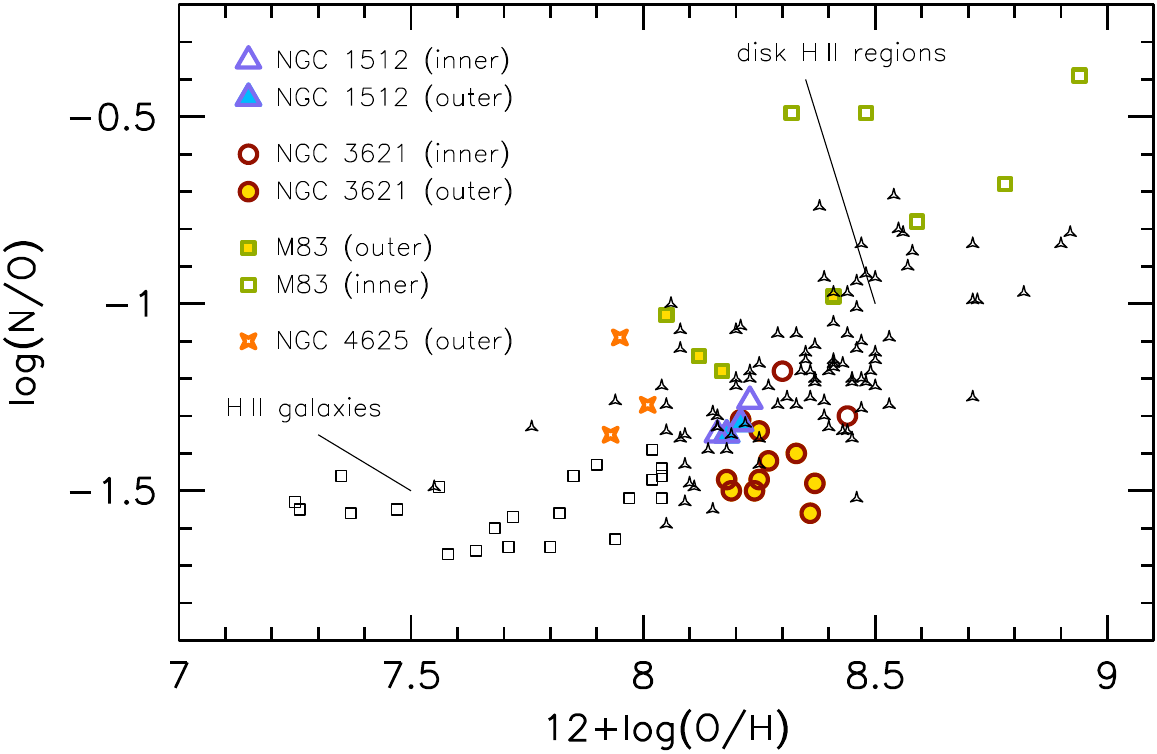}
\caption{O/H \vs\ N/O for extended disk \hii\ regions in NGC~1512, NGC~3621, M83 and NGC~4625, compared with a sample of extragalactic \hii\ regions (triangles) and \hii\ galaxies (open squares) taken from the literature (see text for references). Only objects with abundances obtained from the knowledge of the electron temperatures (derived from auroral lines) are displayed.
\label{no}}
\end{figure}

In Fig.~\ref{no} we show the O/H \vs\ N/O diagram for \hii\ regions in the outer disks of NGC~1512 and NGC~3621, but only for those targets for which we have a firm detection of \oiii\lin4363, and therefore of the electron temperature (those included in Table~\ref{te}). We include targets in the outer disks of M83 (\citealt{Bresolin:2009}) and NGC~4625 (\citealt{Goddard:2011}).
For comparison, we also show a sample of extragalactic \hii\ regions 
(triangles; \citealt{Garnett:1997, Kennicutt:2003, Bresolin:2005, Bresolin:2009a, Esteban:2009, Bresolin:2010})
and \hii\ galaxies (empty squares; \citealt{Izotov:1994, Guseva:2009}). In all cases, we only consider abundances obtained from the knowledge of the electron temperature, as derived from the detection of auroral lines.

The distribution of the points representing the outer disk \hii\ regions is roughly consistent with the well-known general trend in this diagram, in which a flat distribution at low O/H ratios with constant N/O\,$\simeq$\,$-1.5$, attributed to predominant primary nitrogen production at low metallicity,
 is followed by an increase of N/O with O/H, as a signature of an additional, secondary production of nitrogen at higher metallicity. Considering the large spread in N/O at constant O/H, the fact that outer disk \hii\ regions on average do not differ {\it significantly} in a systematic way from disk \hii\ regions in terms of their N/O ratio excludes gross deviations of the initial mass function (IMF) of the embedded star clusters in the extended disks from the canonical one, typical of inner disk clusters. A hypothetical systematic decrease in the upper mass cutoff would lead to an increased N/O ratio, due to the removal of the main producers of oxygen, i.e.~massive hot stars. While we postpone a more detailed discussion to a future work (Bresolin et al., in preparation), we exclude a dramatic reduction of the upper mass cutoff from 100~\msun\ to 25-30~\msun.

However, Fig.~\ref{no} suggests that outer disk \hii\ regions in NGC~3621 lie towards systematically low N/O values, compared to similar objects we studied in M83, NGC~4625 and, in part, NGC~1512. While this might be simply due to small number statistics, it is tantalizing to speculate in terms of the different evolutionary status of the extended disks  of these galaxies. Among the four galaxies, NGC~3621 is the only one which we can consider isolated and non-barred. The others show clear signs of  tidal interactions (as in the case of NGC~1512 and NGC~4625) or might 
have had an interaction in the past (as suggested for M83, \citealt{van-den-Bergh:1980}), and contain stellar bars (NGC~1512, M83). Both of these properties are linked to the presence of radial gas flows, which can presumably mix metals out of the main disk of the galaxy into the outer disk.
We can speculate that the extended disk of NGC~3621, in the absence of such a radial mixing process, is less chemically processed than the remaining extended disks.
Because of the 
longer timescale for nitrogen production compared to that of oxygen, the extended disk of NGC~3621 would therefore display a lower mean N/O ratio.
Of course, we realize that observations of the N/O ratio in a significantly larger sample of extended disk galaxies will be necessary before this hypothesis can be seriously tested.

\section{Discussion}
\begin{figure}
\medskip
\epsscale{1.15}
\plotone{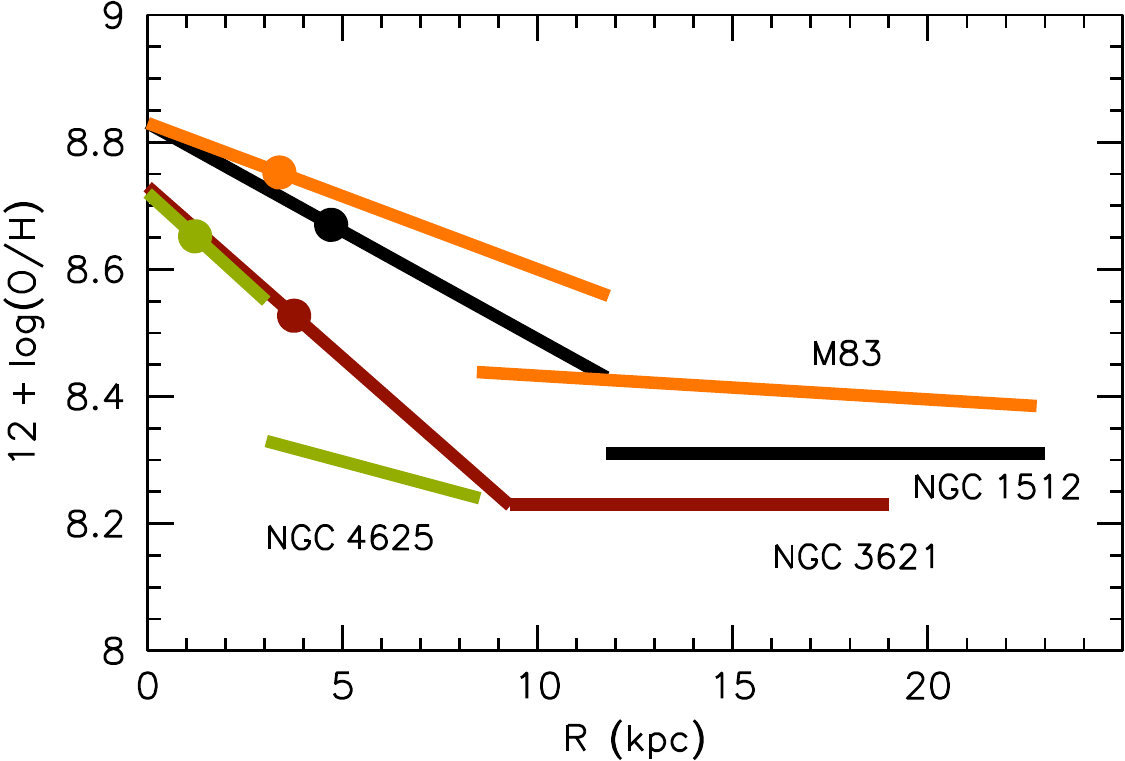}
\caption{Schematic representation of the galactocentric O/H abundance gradients obtained by our group for the extended disk galaxies M83, NGC~4625, NGC~1512 and NGC~3621, using the N2 metallicity indicator. Dots  are shown at $R$\,=\,0.4\,\rtf\ for each galaxy to represent the characteristic abundances of their inner disks.
\label{4gal}}
\end{figure}

In the previous section we illustrated the flattening of the \hii\ region radial abundance gradient occurring in the extended disks of the spiral galaxies NGC~1512 and NGC~3621. The O/H ratio is essentially constant for galactocentric distances larger than $\sim$10~kpc, to the outermost regions we observed at distances of 18~kpc (NGC~3621) and 25~kpc (NGC~1512).  The small abundance scatter that we mesaure, on the order of 0.06 dex, is consistent with what is expected from the observational uncertainties.
Strong evidence for flat outer disk gradients in external spiral galaxies has been recently presented by \citet{Bresolin:2009} for the prototypical XUV disk galaxy M83, and subsequently by \citet{Goddard:2011} for NGC~4625, and \citet{Werk:2011} for a sample of 13, mostly merging or interacting, galaxies, although in the latter case the abundance scatter  is in general considerably larger ($\sim$0.15 dex, compared to 0.06~dex), due to the significantly smaller number of \hii\ regions observed.
A flattening, or possibly an upturn, in the metallicity gradient has also been detected from stellar photometry of the old, resolved stellar population (red giant branch stars) in the outer disks of NGC~300 (\citealt{Vlajic:2009}) and NGC~7793 (\citealt{Vlajic:2011}), even though this result is somewhat affected by the age-metallicity degeneracy. Strong indications for a flat outer abundance gradient have also been found for  the Milky Way
from a variety of studies (e.g.~\citealt{Pedicelli:2009, Lepine:2011}, and references therein). A schematic representation of the galactocentric O/H abundance gradients obtained by our group for the extended disk galaxies M83, NGC~4625, NGC~1512 and NGC~3621, determined from the N2 metallicity indicator, is presented in Fig.~\ref{4gal}. This provides an impression of the similarity in the behavior of the metallicity gradient observed at large galactocentric distances, despite the variety in this (admittedly small) sample of galaxies (for example in terms of physical size, interaction activity and presence of bars). The value of \oh\ in the outer disks ranges between 8.2 and 8.4, in a way that appears correlated with the mean metallicity of the inner disk (as discussed below).

Negative radial abundance gradients are a well-established feature of the inner disks of spiral galaxies (\citealt{Vila-Costas:1992, Zaritsky:1994}). They are understood within the inside-out galaxy formation scenario as a result of the 
increased timescales of the gas infall with radius within a $\Lambda$ cold dark matter context
and the consequent 
radial dependence of  the star formation rate. On the other hand, observational and theoretical investigations of the metal  content in the outer regions of galaxies have only recently begun. With the discovery of UV emission from star-forming regions located in the extended outer disks of a significant fraction of galaxies in the Local Universe (\citealt{Gil-de-Paz:2005, Thilker:2005, Thilker:2007}), the crucial role played by studies of the outer fringes of galaxies in our understanding of galaxy formation and evolution 
has become evident. While recent investigations have focussed especially on the star forming properties of these XUV disks (\citealt{Dong:2008, Bigiel:2010, Goddard:2010, Alberts:2011}), it is essential to include the additional constraints  to galactic evolution studies provided
by the chemical abundance analysis, of \hii\ regions in particular, although this is hindered by observational difficulties, due to the relative scarcity and faintness of the targets.

\subsection{The flatness issue}
For the interpretation of our observational results it is  important to first address the possible reasons for the flat chemical abundance distribution in extended galaxy disks.  
A potential clue is offered by the comparison of the star formation efficiency (SFE) between inner and outer disks of galaxies. \citet{Bigiel:2010} showed that this efficiency, 
defined as the ratio of the surface density of star formation rate to that of the available neutral hydrogen gas, SFE = $\Sigma_{SFR} / \Sigma_{H\,I}$, is considerably flatter in the outer disks, in the radial range between one and two isophotal radii, when compared to the inner ($R<$\,\rtf) disks. Examples where this behavior continues to even larger galactocentric distances can be found in M83 (\citealt{Bigiel:2010a}) and NGC~864 (\citealt{Espada:2011}). It is interesting to note that 
the flattening in M83 occurs around $R$/\rtf\,=\,1.5 (see Fig.~4 of \citealt{Bigiel:2010a}), which corresponds to the approximate radius at which \citet{Bresolin:2009} found the radial abundance gradient to flatten out (for 1\,$<$ $R$/\rtf\ $<$ 1.4 there seems to be a `superposition' of inner and outer disk abundance gradients).

The gaseous oxygen-to-hydrogen abundance ratio per unit disk surface area can be approximated by
\medskip

\begin{equation}
\frac{O}{H}\,=\, \frac{(y_O~ t~ \Sigma_{SFR})} {\mu~\Sigma_{H\,I} } \propto \mbox{SFE}\\
\end{equation}

\noindent \\
where $y_O$ is the net oxygen yield by mass, $\mu=11.81$ is the conversion factor from number ratio to mass fraction,  and $t$ is the timescale for the star formation activity.
This expression  suggests that a slow radial dependence of the star formation efficiency at large galactocentric distances could translate into a flattening of the  metallicity gradient.

A potential complication to this picture is represented by the relatively large O/H value measured in the outer disks.
As pointed out by \citet{Bresolin:2009} in the case of M83, and later confirmed by \citet{Werk:2010a, Werk:2011}, outer disks are overabundant for their (extremely large) gas fractions. Virtually all of the outer disks in which nebular abundances have been derived todate possess very extended neutral hydrogen envelopes, which act as large gas reservoirs for the ongoing low-level star forming activity we witness in the form of UV-emitting knots and \hii\ regions. Due to the preponderance of \hi\ in the calculation of the baryonic mass in the outer disks, the effective yield 
$y_{\mbox{\footnotesize eff}}=Z/(ln\,\mu^{-1})$ (where $\mu$ is the gas fraction) is thus quite high.

Considering the O/H values anchored to the \oiii\lin4363 auroral line-based abundances as lower limits to the nebular oxygen abundance (since the various strong-line methods provide
equal or higher O/H abundances), we see from Table~\ref{table:regression} that \oh\  is $\geq$~8.2 and $\geq$~8.3 in NGC~3621 and NGC~1512, respectively. An estimation of the time necessary to enrich the outer disks to these abundance levels, assuming a constant star formation rate (SFR) equal to the presently observed value, can be 
quickly obtained from the knowledge of the SFR and the surface density of gas (see Eq.~1). While in this way we obviously neglect the (unknown) details of the chemical enrichment history of the galactic disks, the approximation will provide an idea of the timescales involved. The calculation is done first  for the case of the outer disk of NGC~3621, for which the relevant mean quantities are readily available from \citet{Bigiel:2010}: $\Sigma_{SFR} = 10^{-4}$ M$_\odot$\,yr$^{-1}$\,kpc$^{-2}$, $\Sigma_{H\,I} = 5$ M$_\odot$\,pc$^{-2}$. Adopting an oxygen yield of 0.01 (\citealt{Maeder:1992}), the time required to enrich the
 interstellar medium up to \oh\,=\,8.23 is $\sim$10~Gyr. The timescale would more than double if we wished to obtain \oh\,=\,8.6, as measured from the \rtwothree\ abundance indicator. 
 
Even though a timescale that is consistent with the age expected for inner galaxy disks ($>$\,10~Gyr) has been found in this case\footnote{Our calculation for M83 in \citet{Bresolin:2009} contained an error, which led to underestimate the timescale by a factor of $\mu=11.81$. Therefore, for the outer disk of M83 the required timescale for chemical enrichment would be $\sim$$3\times10^{10}$\,yr, 
 much longer than a Hubble time.}, outer disks are unlikely to have formed stars for such long times. 
In the  inside-out scenario for galaxy growth the outer disks of galaxies are considerably younger. 
Cosmological hydrodynamical simulations
indicate that the outer portions of galaxy disks could have mean formation times around 4-6 Gyr ago, while the
inner few kpc are formed $>$\,10 Gyr ago (e.g.~\citealt{Scannapieco:2008, Scannapieco:2009}).
Supporting observational evidence has recently been 
provided by the star formation histories of nearby resolved spiral galaxies, such as M33 (\citealt{Williams:2009,Barker:2011}).
\citet{Gogarten:2010} 
found that while for the inner 3~kpc of NGC~300 50\% of the stars had already formed 11~Gyr ago, 
at a galactocentric distance of 5~kpc (corresponding approximately to the isophotal radius)
the same level of star formation was reached about 6~Gyr ago. Age gradients (younger stars at larger radii) have also been inferred from resolved stellar populations studies  in the outer disks
of spiral galaxies such as M81 (\citealt{Davidge:2009}) and the extended disk galaxy M83 (\citealt{Davidge:2010}).
The absence of a significant population of red giant branch stars in the outer disk of M83 noted by  \citet{Davidge:2010}
points against a significant star formation activity a few Gyr before the present time.  \hi-rich galaxies with star-forming outer disks  have been shown to have color gradients that are consistent with recent gas accretion and an inside-out growth (\citealt{Wang:2011}). The sub-critical density of the gas available in the outer disks for star formation suggests that it is unlikely that significantly higher star formation rates have been experienced in the past.
We therefore argue that the systems we are analyzing
are observed in a phase that  could have started a few Gyr ago (consistent with the age estimates obtained for the oldest UV-bright clusters), during which a large amount of neutral gas has become available for star formation, and during which the
mean star formation rate has remained close to the present-day value (as also suggested by the outer disk simulations by \citealt{Bush:2010}).
The problem of the chemical enrichment of outer disks is exacerbated under these conditions, because of the 
reduced amount of time available for metal production. For example, in NGC~3621, star formation at the currently observed levels would enrich the ISM to \oh\,$\simeq$\,7.8 in 4~Gyr.

For the case of the outer disk of NGC~1512 we have derived the following approximate parameters from \citet{Koribalski:2009}: 
$\Sigma_{SFR} = 2.5\times10^{-4}$ M$_\odot$\,yr$^{-1}$\,kpc$^{-2}$, $\Sigma_{H\,I} = 2$ M$_\odot$\,pc$^{-2}$. The resulting timescale to reach the O/H abundances in Table~\ref{table:regression} is $\sim$2-3~Gyr, depending on the metallicity diagnostic used. The ongoing interaction with NGC~1510 
might explain the shorter value for the estimated timescale compared to NGC~3621 (the star formation rate could have been recently enhanced, thus leading to overestimate the average SFR over time), although this remains rather speculative.

We conclude that, at least for a few galaxies we have analyzed so far (such as M83 and NGC~3621),  it is unlikely that the {\it in situ} star formation could have produced and ejected enough metals to enrich the interstellar medium to the presently observed values. We also note that
O/H self-enrichment of the \hii\ regions from embedded massive (O and Wolf-Rayet) stars  and type II supernovae appears a very unlikely mechanism, especially at metallicities below solar, in light of the theoretical results presented by 
\citet{Kroger:2006} and \citet{Wofford:2009}.

\subsection{Radial metal mixing}
Although we should expect a fraction of the elevated metal amounts found in the outer disks
to be produced {\it in situ}, since stars are known to populate galaxy disks to large galactocentric distances (\citealt{Bland-Hawthorn:2005, Davidge:2010}), another considerable fraction must have reached the outer disks, possibly from the inner disks, via some transport mechanism, efficient out to galactocentric distances of at least 20\,--\,25~kpc, as observed in NGC~1512, NGC~3621 (this work) and M83 (\citealt{Bresolin:2009}).
Several possibilities have already been pointed out by other authors (\citealt{Werk:2010a, Werk:2011, Vlajic:2009, Vlajic:2011}). These include the effects 
of radial gas flows induced by bars (\citealt{Friedli:1994}) and tidal interactions, as already discussed in 
\citet{Bresolin:2009}. Abundance gradient breaks have been reported in  a few barred galaxies (e.g.~NGC~3359, \citealt{Martin:1995, Zahid:2011a}), although it should be noted that such breaks are expected to occur near the bar corotation 
radius, in most cases located within the galaxy optical disk, so that this would not in general explain breaks taking place further out around the isophotal radius.
The flattening of radial abundance gradients due to galaxy tidal interactions and merging  (\citealt{Rupke:2010}) could explain most of the cases investigated by \citet{Werk:2011}, as well as our result for NGC~1512. However, the fact that apparently isolated galaxies (although we cannot exclude past merger events), such as NGC~3621, also display similar uniform metallicity distribution in their outer disks leads us to conclude that alternative mechanisms of metal redistribution must play an important role.

\citet{Werk:2011} argued in favor of the mixing scenario without feedback and outflows proposed by \citet{Tassis:2008}, in which metals  are driven in the hot gas phase to large galactocentric distances by turbulence and large-scale motions resulting from gravitational instabilities, accretion flows and mergers. 
This model was developed to explain known scaling relations of dwarf galaxies.
According to \citet{Tassis:2008}, mergers should be the dominant source of turbulence, therefore it is unclear whether this mechanism can fully explain the chemical enrichment of the  outer disks of nearby, non-dwarf isolated galaxies. The model seems also incompatible with the direct evidence for  gas outflows in dwarf galaxies (e.g.~\citealt{Elson:2011}). 

Additional processes of  angular momentum transport  that can help to explain the radial mixing of metals in galaxy disks include 
radial gas flows (\citealt{Lacey:1985, Goetz:1992, Portinari:2000, Ferguson:2001, Schonrich:2009, Spitoni:2011}, among others), perturbations by satellite galaxy encounters (\citealt{Quillen:2009, Bird:2012}), resonance scattering with transient spiral density waves  (\citealt{Sellwood:2002}), and the overlap of spiral and bar resonances (\citealt{Minchev:2011}).
Stellar radial migrations (\citealt{Roskar:2008}) can lead to a flattening of the metallicity gradients for old stellar populations with increasing radius (\citealt{Roskar:2008b}), as observed for example in NGC~300 (\citealt{Vlajic:2009}).
However, since the effect of radial mixing on age and metallicity as a function of radius is cumulative with time, and since stars and gas are essentially decoupled,
we do not expect the same mechanism to be able to explain  the flat abundance gradients we derive for the gas out to galactocentric distances of $\sim$20~kpc.

\subsection{Enriched infall}
An alternative interpretation of the chemical properties of extended disks involves the predictions from recent cosmological simulations in the hierarchical context concerning the evolution of galaxies that consider cold gas accretion, modulated by feedback, outflows and star formation processes  (\citealt{Keres:2005, Dekel:2006, Ocvirk:2008, Bouche:2010}). 
The direct detection of gas accretion into galaxies is elusive, and has been inferred from gas depletion timescales, which are considerably shorter than the Hubble time  in galaxies at low and high redshift (\citealt{Genzel:2010}), and from considerations concerning misaligned components in disks, such as warps,  polar disks and other ring structures (\citealt{Roskar:2010, Finkelman:2011, Spavone:2011}). Metal-poor Lyman limit systems have been recently proposed as direct observational support  for cold stream accretion (\citealt{Ribaudo:2011, Fumagalli:2011}).

Galactic scale outflows are well established observationally (e.g.~\citealt{Veilleux:2005, Steidel:2010} and references therein). 
The ejection of chemically enriched gas from galaxies appears necessary to explain the enrichment of the  circumgalactic and intergalactic medium via galactic winds (\citealt{Oppenheimer:2008, Tumlinson:2011}), the origin of the mass-metallicity relation (\citealt{Finlator:2008}), and to circumvent discrepancies between predicted and observed baryonic mass fractions locked in galaxies (\citealt{Keres:2009a}). 
In recent outflow models a `wind recycling' process ensues, in which the infalling gas is progressively enriched by metals produced by stars within galactic disks, ejected by outflows over distances of $\sim$80~kpc and reincorporated into the parent star-forming galaxies over timescales on the order of 1~Gyr (\citealt{Oppenheimer:2008}). This process was more efficient at early epochs and high redshift, $z>2$, when accretion and the star formation rate were significantly higher than at $z=0$, giving rise to rapid metal enrichment in the Universe. In the present epoch the chemically enriched material located in the intergalactic medium participates in ongoing, weaker gas inflows. Within this picture, even galaxies that are not massive enough to sustain important outflows can thus accrete enriched gas.

The metallicity expected for infalling gas in the local Universe does not necessarily have to be `primordial' or very low (0.1\,\zsun\ or below). High velocity clouds represent  an important  mode of gas accretion for  the Milky Way, and are also observed in connection with other galaxies (e.g.~\citealt{Sancisi:2008}). Cosmological simulations by \citet{Keres:2009b} show how such clouds can be produced from the filamentary structure in cold gas accretion, providing Milky Way-size galaxies with a continuous source of gas for star formation.
Absorption line measurements  indicate metallicities in the range 0.1\,\zsun\ to supersolar (\citealt{Wakker:2001, Yao:2011}). This range probably reflects the variety in the origin of these systems, from metal-poor intergalactic medium accretion and small mergers to metal-rich gas participating in the galactic fountain (\citealt{Shapiro:1976}). Whatever their origin, enriched gas clouds (0.2 -- 0.5\,\zsun) associated with the Milky Way can be found at galactocentric distances (9 -- 18\,kpc) that place them in the outer Galactic disk (\citealt{Tripp:2011}).

The recent cosmological hydrodynamic simulations by \citet{Dave:2011} highlight the importance of reaccretion of metal-enriched gas in defining the observed galaxy mass-metallicity and mass-gas fraction relations  at redshifts $z<1$. These authors followed the redshift evolution of the gas-phase metallicity (in terms of the star formation-weighted O/H ratio) of both the interstellar medium of model galaxies and the infalling gas. Their momentum-conserving wind models at $z=0$ reach an ISM metallicity approximately between \oh\,=\,8.3 ($2\times10^9$\,\msun\ galaxy mass model) and 
\oh\,=\,8.6 ($10^{10}$\,\msun). The corresponding infall gas metallicities are roughly 0.15 dex lower, with the ratio between the infalling gas metallicity and the inner disk ISM metallicity  $\alpha_Z \simeq 0.7$. This parameter enters directly into the equation for the equilibrium metallicity of the gas under the combined 
effects of star formation and dilution by infall (see \citealt{Dave:2011, Dave:2011a} for details).
If we consider the abundance gradient plots for NGC~1512 (Fig.~\ref{oh-1512}) and NGC~3621 (Fig.~\ref{oh-3621}) we see that the difference between the mean O/H ratio for the inner disks and the outer disk is comparable to this theoretical value. If we use the inner disk metallicity measured at $R=0.4\,$\rtf\, as representative of the `effective' or characteristic metallicity (\citealt[see the dots in Fig.~\ref{4gal}]{Zaritsky:1994}) we obtain, from the results for the N2 metallicity indicator in Table~\ref{table:regression}, $\alpha_Z \simeq 0.5$. Interestingly, if we also consider the extended disk galaxies M83 and NGC~4625 we studied previously, we obtain similar values ($\alpha$\,=\,0.46 and 0.43, respectively). In this limited sample of extended disk galaxies we thus find that the outer disk metallicities are approximately half those representing the mean value for the inner disk.
Therefore, the metallicity we measure in extended disks appear to be quantitatively compatible with the predictions of the metallicity evolution of the Universe made by recent hydrodynamic simulations.

It is interesting to compare the metallicity we measure in outer disks
to the expected metallicity of Damped Lyman-alpha (DLA) systems
evolved to $z=0$ (\citealt{Prochaska:2003, Kulkarni:2005}). The
metallicities of ionized gas in emission and neutral gas in absorption
have been shown to agree in at least one local DLA galaxy
(\citealt{Schulte-Ladbeck:2005}). On face value it appeared that the
metal content of DLAs fell short of that expected from the global star
formation rate of galaxies integrated over time (\citealt{Pettini:1999}). However, further analysis by \citet{Zwaan:2005} and \citet{Chen:2005} found that the metal content of local field galaxies is in
good agreement with low-$z$ DLAs. \citet{Zwaan:2005} applied the
luminosity-metallicity relation, together with known galactic metallicity gradients
(measured only to \rtf, \citealt{Vila-Costas:1992}), to 21\,cm \hi\
maps of local galaxies. They found a median cross section-weighted
metallicity of [O/H]\,=\,$-0.85 \pm 0.2$ (1/7 solar) for \hi\ gas at $z=0$
above the DLA column density limit ($N_{H\,I}=2\times 10^{20}$
cm$^{-2}$). Figure~1 in \citet{Zwaan:2005} shows that typically up to 50\% of the DLA
cross-section lies outside \rtf. If metallicity gradients do not
continue to decline at the same rate beyond \rtf\ - as we have
found in NGC 1512 and 3621 - a higher median would have been
measured. \citet{Chen:2005} used a different approach: they applied average
radial \hi\ profiles of three galaxy types and measured a $N_{HI}$-weighted
mean metallicity of $Z=0.3$~\zsun\ out to the radius where 21\,cm
emission is normally detected. However, not all the gas at these radii meets
the DLA criteria, so that, had the results been corrected for the intrinsic
relationship between $N_{HI}$ and metallicity, a higher mean metallicity
would have been inferred.

Utilizing high-quality 21\,cm maps of NGC 1512 (\citealt{Koribalski:2009}) and NGC 3621 (\citealt{Walter:2008}), we have
established that every \hii\ region in our sample lies in an area
satisfying the DLA column density limit. We also carried out the same
analysis on \hii\ regions in M83 reported in \citet{Bresolin:2009},  overlaying
their positions on the 21\,cm maps of \citet{Walter:2008} and
\citet{Crosthwaite:2002}. Ignoring the three \hii\ regions with
\rtf\,$>$\,2.5, which are beyond the limit of the \hi\ map, the remaining
46 \hii\ regions lie in gas above the DLA cutoff. Although our sample of
three galaxies is small, we are able to compare metallicities and \hi\
measurements in a region of parameter space not covered directly by
\citet{Zwaan:2005} or \citet{Chen:2005}. We find that the [O/H] ratio of all gas
with $N_{HI}=2\times 10^{20}$ cm$^{-2}$ corresponds to  $Z>0.35$ \zsun.

Detailed observations of the metallicity and \hi\ distribution beyond
\rtf\ need to be performed on a larger sample of galaxies to address
this possible mismatch between low redshift metal-poor (0.2 solar)
DLAs and moderate metallicity (0.35 solar) outer disks of galaxies.

\section{Conclusions}
In this paper we have presented the oxygen abundances of a large sample of \hii\ regions in the spiral galaxies NGC~1512 and NGC~3621, extending out to 
approximately 25~kpc and 20~kpc from the galactic centers, respectively. This sample allowed us to obtain for the first time the metallicity of the extended disks of these two 
galaxies, and to compare with the nebular abundances in the inner disk. Similar work has been carried out recently by us and others in a handful of other extended-disk galaxies, where the far-UV and \halpha\ emission from star-forming regions has been found to large galactocentric distances.
We confirm that a flattening of the radial abundances occurs approximately at the isophotal radius, which mimicks the behavior of the star formation efficiency. The inner disks display the familiar negative metallicity gradients, while the outer disks have a virtually constant O/H ratio. In our new work, thanks to the large number of high-quality spectra obtained, we 
find what is arguably the `cleanest' example of such a behavior obtained so far: the abundance scatter in the extended disk of NGC~3621 is a mere 0.06~dex, which is compatible with a perfectly flat metal distribution, within the measurement uncertainties. This result does not depend on the metallicity diagnostic we use.
The O/H ratios we find in the outer disks of NGC~1512 and NGC~3621 is $\sim$0.35$\times$ the solar value, based on measurements of the \oiii\lin4363 auroral lines, or larger if strong-line metallicity diagnostics such as \rtwothree\ are used. This agrees with our earlier findings for M83 and NGC~4625. The statement occasionally found in the literature (e.g.~\citealt{Espada:2011}) that outer disks of spirals have low metallicities on the order of 0.1\,\zsun\ is therefore inaccurate.

We have summarized a few of the mechanisms that can be responsible for  the radial abundance distribution and enrichment of extended disks. They can be broadly 
classified into the following categories: {\it (a)} mixing and turbulence processes; 
{\it (b)} galactic scale outflows and {\it (c)} enriched accretion. The boundary between {\it internally} and {\it externally} driven mechanisms is somewhat blurred, because the accretion material is progressively enriched by outflows that originate in the galaxies themselves. We can envision a situation in which all these processes are at work to produce the chemical signatures we observe in extended disks. Moreover, in some cases, e.g.~for barred or interacting galaxies, the dominant mechanism could be different than for structurally different galaxies, such as isolated, non-barred galaxies. The two galaxies we have studied here are examples of each of these two possibilities, and yet the resulting radial abundance distributions (inner slopes, outer disk metallicities) are very similar. It is possible that a balance between radial mixing and accretion governs the chemical evolution of extended disks.

We have also shown that considering the outer disk metallicities in the context of enriched accretion in cosmological simulations, the empirical values found from \hii\ regions can provide useful constraints to the present-time metallicity of the infalling gas, in particular the proportion of pristine,
metal-poor gas relative to the metal enriched gas that originates from outflows. More in general, future galactic chemical evolution models must be able to account for the different radial trends observed in inner and outer disks. 
In order to distinguish between the different scenarios illustrated here, specific model predictions for the chemical enrichment and the flat gradients in extended disks  would be desirable.

\bigskip

Because of their extreme conditions (e.g.~very low gas densities and long dynamical timescales), outer galaxy disks are sensitive probes of the mechanisms leading to the assembly of galactic disks and their evolution. The study of the metallicity of extended disks complements in a unique way the study of the star formation properties of outer galaxy disks based on UV, optical (H$\alpha$) and radio observations, by providing the insight into the evolutionary status of galactic systems represented by the buildup (and later mixing) of metals in the interstellar medium. Arguably the best way to measure metallicities of extended disks to large galactocentric radii beyond the nearest galaxies in the foreseeable future is by optical spectroscopy of the very faint \hii\ regions found associated with young stellar clusters.

There are several questions that are left unanswered regarding the chemical composition and evolution of extended disks, that will be addressed by ongoing and future work. For example, do all galaxies with an extended disk where metallicities can be derived from \hii\ region emission lines behave like NGC~1512 and NGC~3621, with an abrupt flattening of the abundance gradient roughly at the isophotal radius? Is the ratio between outer and inner metallicities approximately universal? Does it change with galaxy mass? Do the chemical properties of outer disks depend on galaxy properties, such as mass and environment? Are there parallels in the chemical mixing properties between extended disks and low surface brightness galaxies? What is the metal distribution in the outer disks of `rejuvenated' early-type galaxies?

\acknowledgments
We thank Q. Goddard for assistance in the initial stages of this project.
FB gratefully acknowledges the support from the National Science Foundation grants AST-0707911 and AST-1008798.
ERW acknowledges the support of Australian Research Council grant DP1095600.

\medskip
\noindent
{\it Facilities:} \facility{VLT:Antu (FORS2)}\\

\bibliography{n3621}

\appendix

\section{H\,II region data}
In this Appendix we present tables containing the celestial coordinates, the galactocentric distances, line fluxes and extinction coefficients for the \hii\ regions observed in NGC~1512 (Table~\ref{fluxes1512}) and NGC~3621 (Table~\ref{fluxes3621}).

\input{tab5}
\input{tab6a}
\input{tab6b}

\end{document}

%% file: tab1.tex
\begin{deluxetable}{lcc}
\tablecolumns{3}
\tablewidth{0pt}
\tablecaption{Galaxy parameters\label{parameters}}

\tablehead{
\colhead{\phantom{aaaaaaaaaaaaaaaaaaaaaaaa}}	&
\colhead{NGC~1512}	&
\colhead{NGC~3621}	}

\startdata
\\[-2mm]
R.A. (J2000.0)\dotfill			& 	04:03:54.17	 						&	11:18:16.52							\\
Decl. (J2000.0)\dotfill		&	$-$43:20:55.4							&	$-$32:48:50.7							\\
Morphological type\dotfill		&	SB(r)a								&	SA(s)d								\\
Distance\dotfill				&	9.5 Mpc								&	6.6 Mpc								\\
\rtf\dotfill					&	4\farcm26 (11.76~kpc)\tablenotemark{a}		&	4\farcm89 (9.39~kpc)\tablenotemark{b}		\\	
Inclination (deg)\dotfill		&	35\tablenotemark{c}						&	65\tablenotemark{d}						\\			Position angle (deg)\dotfill	&	260\tablenotemark{c}					&	345\tablenotemark{d}					\\		
B$_T^0$\dotfill				&	10.96\tablenotemark{e}					&	9.20\tablenotemark{e}					\\
M$_B^0$\dotfill				&	$-$18.93								&	$-$19.90		
\enddata 
\tablecomments{$^a$ 1 arcsec = 46~pc.  $^b$  1 arcsec = 32~pc. $^c$ \citet{Koribalski:2009}.  $^d$ \citet{de-Blok:2008}. $^e$ \citet{de-Vaucouleurs:1991}.   }
\end{deluxetable}

%% file: tab2.tex
\begin{deluxetable}{lcc}
\tablecolumns{3}
\tablewidth{0pt}
\tablecaption{Positions of SNRs and Emission Line stars\label{snr}}

\tablehead{
\colhead{R.A. (J2000)}	&
\colhead{DEC (J2000)}	&
\colhead{Comment}	}
\startdata
\\[-2mm]
 & Supernova remnant candidates & \\[2mm]
04 03 46.90		&	$-$43 20 32.8	&	in NGC~1512	 \\			
04 03 46.00		&	$-$43 22 44.4	&	''	 \\			
11 18 09.39		&	$-$32 43 20.9	&	in NGC~3621  \\			
11 18 12.02		&	$-$32 46 50.0	& 	''	 \\[1mm]			

 & Emission-line stars (NGC~3621)& \\[1mm]
11 18 24.15		&	$-$32 50 09.8	&		\\		
11 18 19.43		&	$-$32 56 01.4	&		\\		
11 18 12.00		&	$-$32 42 06.5	&				

\enddata 
\tablecomments{Units of right ascension are hours, minutes and seconds, and units of declination
are degrees, arcminutes and arcseconds.}
\end{deluxetable}

%% file: tab3.tex
\begin{deluxetable*}{ccccc}
\tablecolumns{5}
\tablewidth{380pt}
\tablecaption{Direct method electron temperatures and abundances\label{te}}

\tablehead{
\colhead{\phantom{aaaaa}ID\phantom{aaaaa}}	     &
\colhead{\oiii}	 &
\colhead{T$_e$}  &
\colhead{12\,+\,log(O/H)}	 &
\colhead{log(N/O)}	 \\[0.5mm]
\colhead{}       &
\colhead{4363}   &
\colhead{(K)}   &
\colhead{}   &
\colhead{} \\[1mm]
\colhead{(1)}	&
\colhead{(2)}	&
\colhead{(3)}	&
\colhead{(4)}	&
\colhead{(5)}	}
\startdata
\\[-1mm]
\multicolumn{5}{c}{\it NGC~1512}\\[2mm]
1\dotfill	&	5.9 $\pm$ 0.7   & 12090 $\pm$ 750 & 8.21 $\pm$ 0.11 & $-1.32$ $\pm$ 0.18 \\  
12\dotfill	&	5.8 $\pm$ 0.3	& 11980 $\pm$ 460 & 8.18 $\pm$ 0.07 & $-1.35$ $\pm$ 0.13 \\  
45\dotfill	&	3.4 $\pm$ 0.4	& 11390 $\pm$ 610 & 8.17 $\pm$ 0.10 & \nodata \\  
55\dotfill	&	2.3 $\pm$ 0.2	& 11440 $\pm$ 540 & 8.16 $\pm$ 0.08 & $-1.35$ $\pm$ 0.12 \\  
57\dotfill	&	2.4 $\pm$ 0.2	& 10810 $\pm$ 500 & 8.23 $\pm$ 0.09 & $-1.26$ $\pm$ 0.14 \\  
61\dotfill	&	6.9 $\pm$ 0.6	& 13380 $\pm$ 710 & 8.05 $\pm$ 0.09 & \nodata \\[3mm]

\multicolumn{5}{c}{\it NGC~3621}\\[2mm]
1\dotfill	&	15.2 $\pm$ 1.2  & 13980 $\pm$ 740 & 8.21 $\pm$ 0.08 & $-1.31$ $\pm$ 0.14 \\ 
16\dotfill	&	1.7 $\pm$ 0.2   & 10360 $\pm$ 550 & 8.25 $\pm$ 0.09 & $-1.47$ $\pm$ 0.13 \\ 
20\dotfill	&	1.7 $\pm$ 0.2   & 10060 $\pm$ 450 & 8.30 $\pm$ 0.08 & $-1.18$ $\pm$ 0.12 \\ 
46\dotfill	&	2.0 $\pm$ 0.1   & 9250  $\pm$ 270 & 8.44 $\pm$ 0.06 & $-1.30$ $\pm$ 0.09 \\ 
49\dotfill	&	3.3 $\pm$ 0.3   & 11350 $\pm$ 480 & 8.19 $\pm$ 0.08 & $-1.50$ $\pm$ 0.11 \\ 
51\dotfill	&	2.5 $\pm$ 0.2   & 10280 $\pm$ 360 & 8.25 $\pm$ 0.07 & $-1.34$ $\pm$ 0.10 \\
52\dotfill	&	1.8 $\pm$ 0.2   &  9700 $\pm$ 400 & 8.36 $\pm$ 0.08 & $-1.56$ $\pm$ 0.11 \\
53\dotfill	&	2.8 $\pm$ 0.1   &  9890 $\pm$ 270 & 8.37 $\pm$ 0.06 & $-1.48$ $\pm$ 0.08 \\
60\dotfill	&	3.1 $\pm$ 0.3   & 10570 $\pm$ 510 & 8.33 $\pm$ 0.09 & $-1.40$ $\pm$ 0.12 \\
62\dotfill	&	2.1 $\pm$ 0.3   & 10540 $\pm$ 570 & 8.27 $\pm$ 0.09 & $-1.42$ $\pm$ 0.13 \\
63\dotfill	&	3.5 $\pm$ 0.2   & 11250 $\pm$ 440 & 8.24 $\pm$ 0.07 & $-1.50$ $\pm$ 0.10 \\
64\dotfill	&	3.5 $\pm$ 0.2   & 11510 $\pm$ 440 & 8.18 $\pm$ 0.07 & $-1.47$ $\pm$ 0.10 
\enddata
\tablecomments{The reddening-corrected intensity of \oiii\lin4363 in Column (2) is in units of H$\beta$\,=\,100.}
\end{deluxetable*}

%% file: tab4.tex
\begin{deluxetable*}{lccccr}
\tabletypesize{\scriptsize}
\tablewidth{315pt}
\tablecolumns{6}
\tablecaption{Results of linear regression to the abundance gradients\label{table:regression}}
\tablehead{
\colhead{Diagnostic}	 &
\colhead{Intercept ($a$)}  &
\multicolumn{2}{c}{Slope ($b$)} &
\colhead{rms}				&
\colhead{\phantom{aaaaa}Outer disk}\\[0.5mm]
\colhead{}       &
\colhead{}   &
\colhead{dex\,kpc\expone}   &
\colhead{dex\,\rtf\expone}   &
\colhead{}		&
\colhead{\phantom{aaaaa}mean}   \\[1 mm]
\colhead{(1)}	&
\colhead{(2)}	&
\colhead{(3)}	&
\colhead{(4)}	&
\colhead{(5)}	&
\colhead{\phantom{aaaaa}(6)}	}
\startdata
\\[-2mm]
\multicolumn{6}{c}{\it NGC~1512}\\[2mm]
N2					&	$8.83 \pm 0.03$   & $-0.034 \pm 0.004$  		&	$-0.40\pm0.04$	&	0.06				&	8.31$\pm$0.11\\  
ONS					&	$8.75 \pm 0.03$   & $-0.033 \pm 0.005$  		&	$-0.39\pm0.05$	&	0.07				&	8.32$\pm$0.04\\
\rtwothree			&	$9.22 \pm 0.05$   & $-0.059 \pm 0.007$  		&	$-0.69\pm0.08$	&	0.11				&	8.54$\pm$0.08\\[3mm]
\multicolumn{6}{c}{\it NGC~3621}\\[2mm]
N2					&	$8.73 \pm 0.03$   & $-0.054 \pm 0.006$  		&	$-0.51\pm0.05$	&	0.06				&	8.23$\pm$0.05\\  
ONS					&	$8.68 \pm 0.03$   & $-0.047 \pm 0.006$  		&	$-0.44\pm0.05$	&	0.06				&	8.23$\pm$0.05\\
\rtwothree			&	$9.09 \pm 0.04$   & $-0.066 \pm 0.007$  		&	$-0.62\pm0.06$	&	0.07				&	8.59$\pm$0.08
\enddata
\end{deluxetable*}

%% file: tab5.tex
\tabletypesize{\scriptsize}
\begin{deluxetable}{cccccccccc}

\tablecolumns{10}
\tablewidth{0pt}
\tablecaption{NGC~1512: Coordinates and reddening-corrected fluxes\label{fluxes1512}}

\tablehead{
\colhead{ID}	     &
\colhead{R.A.}	 &
\colhead{DEC}	 &
\colhead{$R$/\rtf}	 &
\colhead{\oii}	 &
\colhead{\oiii}  &
\colhead{\nii}	 &
\colhead{\sii}	 &
\colhead{F(H$\alpha$)} &
\colhead{$c$(\hbeta)}   \\[0.5mm]
\colhead{}       &
\colhead{(J2000.0)}       &
\colhead{(J2000.0)}       &
\colhead{}       &
\colhead{3727}   &
\colhead{5007}   &
\colhead{6583}   &
\colhead{6717+6731} &
\colhead{(erg\,s$^{-1}$\,cm$^{-2}$)}     &
\colhead{(mag)}     \\[1mm]
\colhead{(1)}	&
\colhead{(2)}	&
\colhead{(3)}	&
\colhead{(4)}	&
\colhead{(5)}	&
\colhead{(6)}	&
\colhead{(7)}   &
\colhead{(8)}   &
\colhead{(9)}   &
\colhead{(10)} }
\startdata
\\[-2mm]
 1 & 04 03 23.9  & $-43$ 16 10.51  & 1.94 &    260 $\pm$   25 &     515 $\pm$   25 &     20 $\pm$    2 &      28 $\pm$    3 &     1.7 $\times 10^{-15}$ &  0.00 $\pm$ 0.14	\\   
 2 & 04 03 52.4  & $-43$ 17 03.36  & 1.11 &    233 $\pm$   24 &     277 $\pm$   14 &     21 $\pm$    3 &      16 $\pm$    2 &     1.8 $\times 10^{-15}$ &  0.12 $\pm$ 0.15	\\   
 3 & 04 03 25.2  & $-43$ 18 40.37  & 1.44 &    403 $\pm$   56 &      38 $\pm$    3 &     48 $\pm$    8 &      57 $\pm$    7 &     1.7 $\times 10^{-15}$ &  0.06 $\pm$ 0.21	\\   
 4 & 04 04 01.9  & $-43$ 18 59.05  & 0.63 &    289 $\pm$   28 &      47 $\pm$    2 &     93 $\pm$    8 &      61 $\pm$    4 &     1.5 $\times 10^{-15}$ &  0.32 $\pm$ 0.14	\\   
 5 & 04 04 29.6  & $-43$ 19 01.28  & 1.58 &    363 $\pm$   42 &     128 $\pm$    7 &    \nodata        &     \nodata        &     1.1 $\times 10^{-15}$ &  0.00 $\pm$ 0.17	\\   
 6 & 04 04 03.1  & $-43$ 19 02.55  & 0.64 &    275 $\pm$   27 &      45 $\pm$    2 &     91 $\pm$    7 &      62 $\pm$    4 &     1.5 $\times 10^{-15}$ &  0.50 $\pm$ 0.14	\\   
 7 & 04 03 35.0  & $-43$ 19 06.02  & 1.01 &    367 $\pm$   36 &      67 $\pm$    3 &     42 $\pm$    5 &      54 $\pm$    5 &     1.0 $\times 10^{-15}$ &  0.00 $\pm$ 0.14	\\   
 8 & 04 04 01.8  & $-43$ 19 10.30  & 0.58 &    274 $\pm$   30 &      41 $\pm$    2 &     98 $\pm$   12 &      70 $\pm$    6 &     3.1 $\times 10^{-15}$ &  0.11 $\pm$ 0.16	\\   
 9 & 04 04 00.4  & $-43$ 19 11.92  & 0.54 &    297 $\pm$   25 &      88 $\pm$    4 &     91 $\pm$    7 &      55 $\pm$    3 &     6.5 $\times 10^{-15}$ &  0.44 $\pm$ 0.12	\\   
10 & 04 04 05.7  & $-43$ 19 12.46  & 0.67 &    185 $\pm$   17 &      21 $\pm$    1 &     81 $\pm$    8 &      70 $\pm$    5 &     4.0 $\times 10^{-15}$ &  0.07 $\pm$ 0.13	\\   
11 & 04 03 36.2  & $-43$ 19 14.06  & 0.94 &    300 $\pm$   29 &      21 $\pm$    1 &     59 $\pm$    7 &      94 $\pm$    8 &     1.2 $\times 10^{-15}$ &  0.01 $\pm$ 0.14	\\   
12 & 04 03 23.0  & $-43$ 19 16.66  & 1.45 &    191 $\pm$   15 &     515 $\pm$   24 &     14 $\pm$    1 &      18 $\pm$    1 &     9.6 $\times 10^{-15}$ &  0.00 $\pm$ 0.11	\\   
13 & 04 03 56.9  & $-43$ 19 18.21  & 0.47 &    267 $\pm$   23 &      29 $\pm$    1 &    105 $\pm$    8 &      81 $\pm$    5 &     3.3 $\times 10^{-15}$ &  0.38 $\pm$ 0.12	\\   
14 & 04 03 57.9  & $-43$ 19 32.45  & 0.42 &    154 $\pm$   13 &      23 $\pm$    1 &     97 $\pm$    8 &      47 $\pm$    3 &     4.3 $\times 10^{-15}$ &  0.24 $\pm$ 0.12	\\   
15 & 04 03 38.6  & $-43$ 19 34.33  & 0.80 &    377 $\pm$   39 &     364 $\pm$   18 &     44 $\pm$    5 &      82 $\pm$    8 &     1.1 $\times 10^{-15}$ &  0.00 $\pm$ 0.15	\\   
16 & 04 03 53.4  & $-43$ 19 46.93  & 0.33 &    170 $\pm$   15 &      24 $\pm$    1 &    123 $\pm$    9 &      96 $\pm$    6 &     2.6 $\times 10^{-15}$ &  0.37 $\pm$ 0.12	\\   
17 & 04 03 52.5  & $-43$ 19 49.85  & 0.32 &    125 $\pm$   10 &      23 $\pm$    1 &    106 $\pm$    7 &      46 $\pm$    2 &     1.2 $\times 10^{-14}$ &  0.39 $\pm$ 0.11	\\   
18 & 04 04 34.0  & $-43$ 19 56.44  & 1.72 &    524 $\pm$   57 &     168 $\pm$    9 &    \nodata        &     \nodata        &     8.0 $\times 10^{-16}$ &  0.34 $\pm$ 0.16	\\   
19 & 04 03 23.4  & $-43$ 19 58.58  & 1.37 &    378 $\pm$   39 &     231 $\pm$   12 &     31 $\pm$    3 &      38 $\pm$    3 &     1.3 $\times 10^{-15}$ &  0.18 $\pm$ 0.15	\\   
20 & 04 03 37.5  & $-43$ 20 05.45  & 0.77 &    287 $\pm$   25 &      73 $\pm$    3 &     72 $\pm$    7 &      50 $\pm$    4 &     2.6 $\times 10^{-15}$ &  0.04 $\pm$ 0.12	\\   
21 & 04 03 34.4  & $-43$ 20 14.25  & 0.89 &    421 $\pm$   46 &     115 $\pm$    6 &     59 $\pm$    7 &      57 $\pm$    5 &     6.7 $\times 10^{-16}$ &  0.11 $\pm$ 0.16	\\   
22 & 04 03 59.9  & $-43$ 20 17.41  & 0.29 &     87 $\pm$    7 &      16 $\pm$    1 &    106 $\pm$    9 &      53 $\pm$    3 &     1.6 $\times 10^{-14}$ &  0.24 $\pm$ 0.12	\\   
23 & 04 03 37.2  & $-43$ 20 19.89  & 0.76 &    255 $\pm$   30 &     186 $\pm$   10 &     51 $\pm$    7 &      51 $\pm$    5 &     1.5 $\times 10^{-15}$ &  0.01 $\pm$ 0.17	\\   
24 & 04 04 37.9  & $-43$ 20 19.81  & 1.88 &    389 $\pm$   36 &     176 $\pm$    8 &    \nodata        &     \nodata        &     1.7 $\times 10^{-15}$ &  0.44 $\pm$ 0.13	\\   
25 & 04 03 50.6  & $-43$ 20 20.79  & 0.23 &     80 $\pm$    7 &       7 $\pm$    1 &    110 $\pm$   11 &      52 $\pm$    4 &     3.5 $\times 10^{-15}$ &  0.03 $\pm$ 0.12	\\   
26 & 04 03 59.9  & $-43$ 20 26.66  & 0.27 &    116 $\pm$    9 &      19 $\pm$    1 &    119 $\pm$    9 &      64 $\pm$    4 &     9.7 $\times 10^{-15}$ &  0.30 $\pm$ 0.11	\\   
27 & 04 03 49.0  & $-43$ 20 27.06  & 0.27 &    134 $\pm$   13 &      17 $\pm$    1 &    111 $\pm$    8 &      43 $\pm$    2 &     1.7 $\times 10^{-15}$ &  0.56 $\pm$ 0.13	\\   
28 & 04 03 50.0  & $-43$ 20 28.96  & 0.23 &    155 $\pm$   15 &      40 $\pm$    2 &    108 $\pm$    9 &      55 $\pm$    4 &     1.7 $\times 10^{-15}$ &  0.42 $\pm$ 0.14	\\   
29 & 04 04 37.5  & $-43$ 20 29.22  & 1.86 &    393 $\pm$   55 &     268 $\pm$   15 &    \nodata        &     \nodata        &     2.2 $\times 10^{-16}$ &  0.36 $\pm$ 0.21	\\   
30 & 04 03 37.8  & $-43$ 20 53.16  & 0.70 &    313 $\pm$   27 &     206 $\pm$   10 &     89 $\pm$    9 &      80 $\pm$    6 &     4.9 $\times 10^{-15}$ &  0.07 $\pm$ 0.12	\\   
31 & 04 04 34.4  & $-43$ 21 05.13  & 1.74 &    293 $\pm$   27 &     367 $\pm$   17 &    \nodata        &     \nodata        &     1.7 $\times 10^{-15}$ &  0.36 $\pm$ 0.13	\\   
32 & 04 03 37.6  & $-43$ 21 20.08  & 0.71 &    319 $\pm$   29 &     184 $\pm$    9 &     61 $\pm$    7 &      55 $\pm$    4 &     2.9 $\times 10^{-15}$ &  0.00 $\pm$ 0.13	\\   
33 & 04 03 37.3  & $-43$ 21 26.05  & 0.73 &    226 $\pm$   27 &      32 $\pm$    2 &     69 $\pm$    9 &      65 $\pm$    7 &     1.1 $\times 10^{-15}$ &  0.01 $\pm$ 0.17	\\   
34 & 04 03 33.4  & $-43$ 21 39.03  & 0.90 &    272 $\pm$   35 &      13 $\pm$    1 &     75 $\pm$   12 &      49 $\pm$    7 &     3.4 $\times 10^{-16}$ &  0.00 $\pm$ 0.19	\\   
35 & 04 03 37.7  & $-43$ 21 42.09  & 0.73 &    361 $\pm$   35 &     185 $\pm$    9 &     61 $\pm$    7 &      36 $\pm$    3 &     1.0 $\times 10^{-15}$ &  0.00 $\pm$ 0.14	\\   
36 & 04 03 47.4  & $-43$ 21 49.09  & 0.37 &     87 $\pm$    7 &      17 $\pm$    1 &     98 $\pm$    9 &      57 $\pm$    4 &     6.0 $\times 10^{-15}$ &  0.01 $\pm$ 0.11	\\   
37 & 04 03 38.3  & $-43$ 21 49.38  & 0.71 &    558 $\pm$   59 &      74 $\pm$    4 &     90 $\pm$   12 &      97 $\pm$   10 &     1.4 $\times 10^{-15}$ &  0.00 $\pm$ 0.16	\\   
38 & 04 03 50.7  & $-43$ 21 49.60  & 0.29 &    101 $\pm$    9 &      14 $\pm$    1 &    102 $\pm$    8 &      49 $\pm$    3 &     8.8 $\times 10^{-15}$ &  0.36 $\pm$ 0.12	\\   
39 & 04 03 48.0  & $-43$ 21 56.63  & 0.38 &    107 $\pm$    9 &       9 $\pm$    1 &     87 $\pm$    7 &      53 $\pm$    3 &     1.3 $\times 10^{-14}$ &  0.34 $\pm$ 0.12	\\   
40 & 04 03 55.2  & $-43$ 21 59.72  & 0.31 &    129 $\pm$   10 &      24 $\pm$    1 &    103 $\pm$    9 &      67 $\pm$    4 &     2.9 $\times 10^{-14}$ &  0.09 $\pm$ 0.11	\\   
41 & 04 03 51.9  & $-43$ 22 03.63  & 0.33 &     58 $\pm$    6 &       5 $\pm$    1 &     77 $\pm$    8 &      47 $\pm$    4 &     6.1 $\times 10^{-15}$ &  0.00 $\pm$ 0.13	\\   
42 & 04 03 54.2  & $-43$ 22 05.55  & 0.33 &    143 $\pm$   12 &      22 $\pm$    1 &     99 $\pm$    9 &      69 $\pm$    5 &     6.5 $\times 10^{-15}$ &  0.11 $\pm$ 0.12	\\   
43 & 04 03 38.5  & $-43$ 22 06.43  & 0.73 &    326 $\pm$   28 &      78 $\pm$    4 &     69 $\pm$    6 &      50 $\pm$    3 &     4.5 $\times 10^{-15}$ &  0.22 $\pm$ 0.12	\\   
44 & 04 03 48.2  & $-43$ 22 10.89  & 0.43 &    159 $\pm$   17 &      13 $\pm$    1 &    107 $\pm$   11 &      60 $\pm$    5 &     7.3 $\times 10^{-16}$ &  0.28 $\pm$ 0.15	\\   
45 & 04 04 36.3  & $-43$ 22 10.92  & 1.87 &    248 $\pm$   21 &     346 $\pm$   16 &    \nodata        &     \nodata        &     4.8 $\times 10^{-15}$ &  0.23 $\pm$ 0.12	\\   
46 & 04 03 27.5  & $-43$ 22 18.67  & 1.19 &    \nodata        &      86 $\pm$    4 &     61 $\pm$    6 &      73 $\pm$    6 &     2.3 $\times 10^{-15}$ &  0.00 $\pm$ 0.12	\\   
47 & 04 03 40.6  & $-43$ 22 21.98  & 0.69 &    262 $\pm$   26 &      51 $\pm$    3 &     84 $\pm$    9 &      51 $\pm$    4 &     3.4 $\times 10^{-15}$ &  0.11 $\pm$ 0.14	\\   
48 & 04 04 36.9  & $-43$ 22 25.28  & 1.91 &    368 $\pm$   42 &     173 $\pm$    9 &    \nodata        &     \nodata        &     1.1 $\times 10^{-15}$ &  0.32 $\pm$ 0.17	\\   
49 & 04 03 41.3  & $-43$ 22 33.00  & 0.70 &    435 $\pm$   47 &      47 $\pm$    3 &     84 $\pm$    8 &      66 $\pm$    5 &     1.9 $\times 10^{-15}$ &  0.42 $\pm$ 0.16	\\   
50 & 04 03 41.9  & $-43$ 22 35.83  & 0.69 &    261 $\pm$   24 &     110 $\pm$    5 &     82 $\pm$    8 &      53 $\pm$    4 &     3.6 $\times 10^{-15}$ &  0.15 $\pm$ 0.13	\\   
51 & 04 04 37.8  & $-43$ 22 38.24  & 1.97 &    437 $\pm$   50 &      74 $\pm$    4 &    \nodata        &     \nodata        &     6.6 $\times 10^{-16}$ &  0.58 $\pm$ 0.17	\\   
52 & 04 03 52.7  & $-43$ 22 44.27  & 0.52 &    393 $\pm$   50 &      90 $\pm$    5 &     72 $\pm$   11 &      60 $\pm$    7 &     9.6 $\times 10^{-16}$ &  0.00 $\pm$ 0.19	\\   
53 & 04 03 46.1  & $-43$ 22 58.66  & 0.66 &    480 $\pm$   47 &     178 $\pm$    9 &     37 $\pm$    4 &      65 $\pm$    5 &     3.0 $\times 10^{-15}$ &  0.18 $\pm$ 0.14	\\   
54 & 04 03 47.7  & $-43$ 23 11.10  & 0.68 &    469 $\pm$   39 &     113 $\pm$    5 &     38 $\pm$    4 &      63 $\pm$    5 &     9.9 $\times 10^{-15}$ &  0.00 $\pm$ 0.12	\\   
55 & 04 03 48.6  & $-43$ 23 13.75  & 0.68 &    344 $\pm$   20 &     231 $\pm$   11 &     26 $\pm$    2 &      35 $\pm$    2 &     4.6 $\times 10^{-15}$ &  0.00 $\pm$ 0.06	\\   
56 & 04 03 50.1  & $-43$ 23 38.99  & 0.79 &    525 $\pm$   50 &     122 $\pm$    6 &     36 $\pm$    4 &      35 $\pm$    3 &     1.7 $\times 10^{-15}$ &  0.19 $\pm$ 0.14	\\   
57 & 04 03 51.7  & $-43$ 23 42.26  & 0.79 &    286 $\pm$   23 &     291 $\pm$   14 &     28 $\pm$    3 &      38 $\pm$    3 &     1.5 $\times 10^{-14}$ &  0.00 $\pm$ 0.11	\\   
58 & 04 03 53.7  & $-43$ 24 08.78  & 0.92 &    480 $\pm$   67 &     160 $\pm$    9 &     39 $\pm$    5 &      44 $\pm$    4 &     2.9 $\times 10^{-16}$ &  0.40 $\pm$ 0.21	\\   
59 & 04 03 54.8  & $-43$ 24 16.97  & 0.96 &    498 $\pm$   57 &      90 $\pm$    5 &     48 $\pm$    7 &      78 $\pm$    8 &     5.7 $\times 10^{-16}$ &  0.00 $\pm$ 0.17	\\   
60 & 04 03 57.6  & $-43$ 24 28.38  & 1.03 &    553 $\pm$   73 &      62 $\pm$    4 &     41 $\pm$    7 &      58 $\pm$    7 &     3.9 $\times 10^{-16}$ &  0.00 $\pm$ 0.20	\\   
61 & 04 04 34.0  & $-43$ 24 53.60  & 2.12 &    266 $\pm$   23 &     459 $\pm$   21 &    \nodata        &     \nodata        &     3.0 $\times 10^{-15}$ &  0.00 $\pm$ 0.12	\\   
62 & 04 04 32.8  & $-43$ 25 21.42  & 2.16 &    310 $\pm$   27 &     288 $\pm$   13 &    \nodata        &     \nodata        &     1.5 $\times 10^{-15}$ &  0.14 $\pm$ 0.12	  
\enddata
\tablecomments{Units of right ascension are hours, minutes and seconds, and units of declination
are degrees, arcminutes and arcseconds.	Line fluxes are in units of H$\beta$\,=\,100. The \halpha\ flux in column (9) has been measured from narrow-band images, and corrected for extinction.}
\end{deluxetable}
\tabletypesize{\scriptsize}

%% file: tab6a.tex
\tabletypesize{\scriptsize}
\begin{deluxetable}{cccccccccc}

\tablecolumns{10}
\tablewidth{0pt}
\tablecaption{NGC~3621: Coordinates and reddening-corrected fluxes\label{fluxes3621}}

\tablehead{
\colhead{ID}	     &
\colhead{R.A.}	 &
\colhead{DEC}	 &
\colhead{$R$/\rtf}	 &
\colhead{\oii}	 &
\colhead{\oiii}  &
\colhead{\nii}	 &
\colhead{\sii}	 &
\colhead{F(H$\alpha$)} &
\colhead{$c$(\hbeta)}   \\[0.5mm]
\colhead{}       &
\colhead{(J2000.0)}       &
\colhead{(J2000.0)}       &
\colhead{}       &
\colhead{3727}   &
\colhead{5007}   &
\colhead{6583}   &
\colhead{6717+6731} &
\colhead{(erg\,s$^{-1}$\,cm$^{-2}$)}     &
\colhead{(mag)}     \\[1mm]
\colhead{(1)}	&
\colhead{(2)}	&
\colhead{(3)}	&
\colhead{(4)}	&
\colhead{(5)}	&
\colhead{(6)}	&
\colhead{(7)}   &
\colhead{(8)}   &
\colhead{(9)}   &
\colhead{(10)} }
\startdata
\\[-2mm]
 1 & 11 18 13.8  & $-32$ 39 55.10  & 1.97 &    261 $\pm$   18 &     915 $\pm$   43 &     18 $\pm$    1 &      42 $\pm$    3 &     4.3 $\times 10^{-15}$ &  0.00 $\pm$ 0.08	\\   
 2 & 11 18 12.4  & $-32$ 40 06.25  & 1.89 &    353 $\pm$   19 &     192 $\pm$    9 &     21 $\pm$    1 &      45 $\pm$    2 &     6.6 $\times 10^{-15}$ &  0.05 $\pm$ 0.05	\\   
 3 & 11 18 08.0  & $-32$ 40 15.60  & 1.80 &    396 $\pm$   27 &     170 $\pm$    8 &     23 $\pm$    2 &      36 $\pm$    2 &     5.5 $\times 10^{-15}$ &  0.09 $\pm$ 0.08	\\   
 4 & 11 18 05.8  & $-32$ 40 28.99  & 1.77 &    351 $\pm$   28 &     280 $\pm$   14 &     17 $\pm$    2 &      36 $\pm$    2 &     1.5 $\times 10^{-15}$ &  0.23 $\pm$ 0.10	\\   
 5 & 11 18 05.8  & $-32$ 40 44.80  & 1.72 &    308 $\pm$   34 &     495 $\pm$   26 &     24 $\pm$    3 &      63 $\pm$    6 &     3.9 $\times 10^{-15}$ &  0.08 $\pm$ 0.16	\\   
 6 & 11 18 13.2  & $-32$ 40 51.17  & 1.74 &    305 $\pm$   36 &      86 $\pm$    5 &     29 $\pm$    4 &      41 $\pm$    5 &     6.4 $\times 10^{-16}$ &  0.00 $\pm$ 0.17	\\   
 7 & 11 18 05.5  & $-32$ 40 58.76  & 1.68 &    206 $\pm$   20 &     323 $\pm$   16 &     19 $\pm$    2 &      30 $\pm$    3 &     2.0 $\times 10^{-15}$ &  0.19 $\pm$ 0.13	\\   
 8 & 11 18 02.5  & $-32$ 42 03.42  & 1.59 &    321 $\pm$   49 &      34 $\pm$    3 &     42 $\pm$    7 &      41 $\pm$    6 &     1.9 $\times 10^{-16}$ &  0.18 $\pm$ 0.22	\\   
 9 & 11 18 02.2  & $-32$ 42 08.81  & 1.59 &    320 $\pm$   38 &     160 $\pm$    9 &     21 $\pm$    4 &      31 $\pm$   10 &     3.4 $\times 10^{-16}$ &  0.00 $\pm$ 0.17	\\   
10 & 11 17 58.4  & $-32$ 42 19.95  & 1.78 &    334 $\pm$   18 &     184 $\pm$    8 &     19 $\pm$    1 &      29 $\pm$    1 &     1.3 $\times 10^{-14}$ &  0.02 $\pm$ 0.05	\\   
11 & 11 18 04.9  & $-32$ 42 30.21  & 1.43 &    353 $\pm$   33 &      59 $\pm$    3 &     35 $\pm$    4 &      74 $\pm$    7 &     2.0 $\times 10^{-15}$ &  0.00 $\pm$ 0.13	\\   
12 & 11 18 02.5  & $-32$ 42 57.15  & 1.47 &    422 $\pm$   31 &     176 $\pm$    8 &     26 $\pm$    2 &      33 $\pm$    2 &     2.7 $\times 10^{-15}$ &  0.27 $\pm$ 0.09	\\   
13 & 11 18 04.2  & $-32$ 43 34.46  & 1.30 &    398 $\pm$   36 &      91 $\pm$    5 &     30 $\pm$    3 &      60 $\pm$    5 &     2.6 $\times 10^{-15}$ &  0.00 $\pm$ 0.12	\\   
14 & 11 18 02.1  & $-32$ 44 06.32  & 1.37 &    551 $\pm$   42 &      49 $\pm$    2 &     33 $\pm$    2 &      63 $\pm$    2 &     1.2 $\times 10^{-14}$ &  0.75 $\pm$ 0.10	\\   
15 & 11 18 08.9  & $-32$ 44 07.42  & 1.03 &    321 $\pm$   30 &     102 $\pm$    5 &     32 $\pm$    4 &      56 $\pm$    5 &     2.6 $\times 10^{-15}$ &  0.12 $\pm$ 0.13	\\   
16 & 11 17 57.2  & $-32$ 44 17.05  & 1.73 &    297 $\pm$   15 &     231 $\pm$   10 &     19 $\pm$    1 &      30 $\pm$    1 &     6.1 $\times 10^{-15}$ &  0.13 $\pm$ 0.04	\\   
17 & 11 17 56.3  & $-32$ 44 41.94  & 1.80 &    345 $\pm$   29 &      68 $\pm$    3 &     25 $\pm$    2 &      48 $\pm$    3 &     5.4 $\times 10^{-15}$ &  0.26 $\pm$ 0.11	\\   
18 & 11 18 03.4  & $-32$ 44 54.01  & 1.22 &    354 $\pm$   22 &     175 $\pm$    8 &     25 $\pm$    1 &      37 $\pm$    2 &     4.0 $\times 10^{-15}$ &  0.31 $\pm$ 0.07	\\   
19 & 11 18 13.6  & $-32$ 45 10.91  & 0.77 &    315 $\pm$   23 &      57 $\pm$    3 &     43 $\pm$    4 &      62 $\pm$    4 &     3.2 $\times 10^{-15}$ &  0.00 $\pm$ 0.09	\\   
20 & 11 18 14.0  & $-32$ 45 13.00  & 0.77 &    289 $\pm$   14 &     251 $\pm$   11 &     37 $\pm$    2 &      69 $\pm$    3 &     2.3 $\times 10^{-14}$ &  0.00 $\pm$ 0.03	\\   
21 & 11 18 01.9  & $-32$ 45 30.09  & 1.31 &    433 $\pm$   36 &      64 $\pm$    3 &     31 $\pm$    3 &      54 $\pm$    4 &     1.1 $\times 10^{-15}$ &  0.04 $\pm$ 0.11	\\   
22 & 11 18 07.0  & $-32$ 45 49.47  & 0.90 &    516 $\pm$   43 &      81 $\pm$    4 &     37 $\pm$    4 &      54 $\pm$    4 &     2.0 $\times 10^{-15}$ &  0.00 $\pm$ 0.11	\\   
23 & 11 18 14.1  & $-32$ 46 06.85  & 0.57 &    247 $\pm$   15 &      73 $\pm$    3 &     68 $\pm$    4 &      97 $\pm$    5 &     4.6 $\times 10^{-14}$ &  0.00 $\pm$ 0.06	\\   
24 & 11 18 10.5  & $-32$ 46 09.59  & 0.65 &    469 $\pm$   37 &      59 $\pm$    3 &     46 $\pm$    3 &      53 $\pm$    3 &     5.4 $\times 10^{-15}$ &  0.32 $\pm$ 0.10	\\   
25 & 11 18 07.4  & $-32$ 46 24.25  & 0.83 &    470 $\pm$   46 &      23 $\pm$    2 &     44 $\pm$    5 &      62 $\pm$    5 &     1.7 $\times 10^{-15}$ &  0.04 $\pm$ 0.14	\\   
26 & 11 18 07.4  & $-32$ 46 39.53  & 0.82 &    198 $\pm$   12 &     342 $\pm$   16 &     27 $\pm$    1 &      34 $\pm$    1 &     1.8 $\times 10^{-14}$ &  0.26 $\pm$ 0.06	\\   
27 & 11 18 07.4  & $-32$ 46 50.79  & 0.82 &    383 $\pm$   24 &     104 $\pm$    5 &     40 $\pm$    3 &      54 $\pm$    3 &     1.1 $\times 10^{-14}$ &  0.00 $\pm$ 0.07	\\   
28 & 11 18 09.4  & $-32$ 46 54.84  & 0.65 &    348 $\pm$   26 &     112 $\pm$    5 &     50 $\pm$    4 &      53 $\pm$    3 &     4.8 $\times 10^{-15}$ &  0.09 $\pm$ 0.09	\\   
29 & 11 18 14.2  & $-32$ 46 55.88  & 0.40 &    242 $\pm$   12 &      50 $\pm$    2 &     63 $\pm$    3 &      59 $\pm$    2 &     3.1 $\times 10^{-14}$ &  0.00 $\pm$ 0.03	\\   
30 & 11 18 14.4  & $-32$ 47 11.69  & 0.35 &    296 $\pm$   15 &      51 $\pm$    2 &     86 $\pm$    4 &      60 $\pm$    2 &     5.1 $\times 10^{-14}$ &  0.23 $\pm$ 0.04	\\   
31 & 11 18 09.1  & $-32$ 47 21.04  & 0.66 &    312 $\pm$   16 &     175 $\pm$    8 &     51 $\pm$    3 &      84 $\pm$    3 &     2.8 $\times 10^{-14}$ &  0.12 $\pm$ 0.04	\\   
32 & 11 18 13.5  & $-32$ 47 31.20  & 0.32 &    119 $\pm$    6 &      19 $\pm$    1 &     87 $\pm$    3 &      68 $\pm$    2 &     2.9 $\times 10^{-14}$ &  0.41 $\pm$ 0.03	\\   
33 & 11 18 09.6  & $-32$ 47 39.30  & 0.61 &    450 $\pm$   48 &     134 $\pm$    7 &     44 $\pm$    5 &      32 $\pm$    3 &     5.0 $\times 10^{-15}$ &  0.00 $\pm$ 0.15	\\   
34 & 11 18 15.6  & $-32$ 47 57.61  & 0.19 &    185 $\pm$   11 &      30 $\pm$    1 &     99 $\pm$    6 &      46 $\pm$    2 &     2.9 $\times 10^{-14}$ &  0.00 $\pm$ 0.06	\\   
35 & 11 18 10.6  & $-32$ 48 02.40  & 0.53 &    291 $\pm$   14 &     113 $\pm$    5 &     49 $\pm$    2 &      43 $\pm$    1 &     7.1 $\times 10^{-14}$ &  0.18 $\pm$ 0.03	\\   
36 & 11 18 12.5  & $-32$ 48 09.63  & 0.36 &    188 $\pm$    9 &      40 $\pm$    2 &     83 $\pm$    4 &      46 $\pm$    2 &     5.5 $\times 10^{-14}$ &  0.00 $\pm$ 0.03	\\   
37 & 11 18 14.0  & $-32$ 48 33.44  & 0.22 &    129 $\pm$    9 &      31 $\pm$    2 &     95 $\pm$    6 &      55 $\pm$    3 &     2.8 $\times 10^{-14}$ &  0.11 $\pm$ 0.07	\\   
38 & 11 18 18.2  & $-32$ 48 46.25  & 0.18 &    110 $\pm$    8 &      22 $\pm$    1 &    100 $\pm$    5 &      55 $\pm$    2 &     6.9 $\times 10^{-14}$ &  0.68 $\pm$ 0.08	\\   
39 & 11 18 19.0  & $-32$ 49 10.93  & 0.22 &    149 $\pm$    8 &      59 $\pm$    3 &     79 $\pm$    3 &      68 $\pm$    2 &     1.4 $\times 10^{-13}$ &  0.50 $\pm$ 0.04	\\   
40 & 11 18 24.4  & $-32$ 49 17.86  & 0.73 &    341 $\pm$   19 &      86 $\pm$    4 &     43 $\pm$    3 &      51 $\pm$    2 &     3.5 $\times 10^{-14}$ &  0.00 $\pm$ 0.05	\\   
41 & 11 18 21.9  & $-32$ 49 53.13  & 0.48 &    320 $\pm$   15 &      44 $\pm$    2 &     61 $\pm$    3 &      45 $\pm$    1 &     2.2 $\times 10^{-13}$ &  0.18 $\pm$ 0.02	\\   
42 & 11 18 23.9  & $-32$ 50 04.77  & 0.65 &    235 $\pm$   13 &     232 $\pm$   11 &     39 $\pm$    2 &      53 $\pm$    2 &     6.4 $\times 10^{-14}$ &  0.18 $\pm$ 0.05	\\   
43 & 11 18 12.3  & $-32$ 50 12.93  & 0.62 &    328 $\pm$   22 &      99 $\pm$    5 &     35 $\pm$    3 &      33 $\pm$    2 &     4.3 $\times 10^{-14}$ &  0.00 $\pm$ 0.08	\\   
44 & 11 18 25.5  & $-32$ 50 26.18  & 0.80 &    386 $\pm$   23 &      83 $\pm$    4 &     44 $\pm$    3 &      41 $\pm$    2 &     3.9 $\times 10^{-14}$ &  0.00 $\pm$ 0.06	\\   
45 & 11 18 24.0  & $-32$ 51 02.15  & 0.69 &    425 $\pm$   45 &      62 $\pm$    4 &     45 $\pm$    5 &      48 $\pm$    4 &     1.2 $\times 10^{-14}$ &  0.00 $\pm$ 0.15	\\   
46 & 11 18 22.2  & $-32$ 51 39.81  & 0.65 &    162 $\pm$    8 &     399 $\pm$   18 &     18 $\pm$    1 &      17 $\pm$    1 &     5.0 $\times 10^{-14}$ &  0.00 $\pm$ 0.02	\\   
47 & 11 18 19.7  & $-32$ 51 45.83  & 0.61 &    413 $\pm$   39 &      28 $\pm$    2 &     46 $\pm$    5 &      65 $\pm$    5 &     1.1 $\times 10^{-14}$ &  0.00 $\pm$ 0.13	\\   
48 & 11 18 20.4  & $-32$ 52 28.62  & 0.76 &    366 $\pm$   23 &      61 $\pm$    3 &     43 $\pm$    3 &      44 $\pm$    2 &     6.3 $\times 10^{-15}$ &  0.20 $\pm$ 0.07	\\   
49 & 11 18 32.8  & $-32$ 52 31.07  & 1.46 &    276 $\pm$   14 &     340 $\pm$   15 &     15 $\pm$    1 &      25 $\pm$    1 &     1.3 $\times 10^{-14}$ &  0.13 $\pm$ 0.04	\\   
50 & 11 18 32.2  & $-32$ 52 44.55  & 1.42 &    350 $\pm$   24 &     278 $\pm$   13 &     24 $\pm$    2 &      44 $\pm$    2 &     9.4 $\times 10^{-15}$ &  0.07 $\pm$ 0.08	\\   
51 & 11 18 28.9  & $-32$ 52 51.41  & 1.17 &    166 $\pm$    8 &     357 $\pm$   16 &     15 $\pm$    1 &      22 $\pm$    1 &     2.0 $\times 10^{-14}$ &  0.15 $\pm$ 0.03	\\   
52 & 11 18 32.8  & $-32$ 53 01.62  & 1.48 &    241 $\pm$   12 &     310 $\pm$   14 &     14 $\pm$    1 &      22 $\pm$    1 &     4.2 $\times 10^{-14}$ &  0.02 $\pm$ 0.03	\\   
53 & 11 18 28.8  & $-32$ 53 07.35  & 1.19 &    160 $\pm$    7 &     445 $\pm$   20 &     11 $\pm$    1 &      17 $\pm$    1 &     2.4 $\times 10^{-13}$ &  0.14 $\pm$ 0.01	\\   
54 & 11 18 33.2  & $-32$ 53 49.27  & 1.55 &    255 $\pm$   19 &     293 $\pm$   14 &     18 $\pm$    1 &      47 $\pm$    3 &     4.9 $\times 10^{-15}$ &  0.24 $\pm$ 0.09	\\   
55 & 11 18 28.9  & $-32$ 53 49.29  & 1.27 &    419 $\pm$   38 &      33 $\pm$    2 &     27 $\pm$    2 &      62 $\pm$    4 &     2.3 $\times 10^{-15}$ &  0.25 $\pm$ 0.12	\\   
56 & 11 18 28.6  & $-32$ 54 04.33  & 1.29 &    284 $\pm$   17 &     152 $\pm$    7 &     23 $\pm$    1 &      37 $\pm$    2 &     1.3 $\times 10^{-14}$ &  0.09 $\pm$ 0.06	\\   
57 & 11 18 28.7  & $-32$ 54 15.30  & 1.31 &    406 $\pm$   29 &     103 $\pm$    5 &     29 $\pm$    2 &      95 $\pm$    5 &     7.5 $\times 10^{-15}$ &  0.17 $\pm$ 0.08	\\   
58 & 11 18 33.1  & $-32$ 54 21.71  & 1.59 &    370 $\pm$   38 &     164 $\pm$    8 &     16 $\pm$    2 &      37 $\pm$    3 &     8.8 $\times 10^{-15}$ &  0.21 $\pm$ 0.15	\\   
59 & 11 18 28.3  & $-32$ 54 22.05  & 1.31 &    365 $\pm$   27 &     166 $\pm$    8 &     25 $\pm$    1 &      45 $\pm$    2 &     5.4 $\times 10^{-15}$ &  0.68 $\pm$ 0.09	\\   
60 & 11 18 27.8  & $-32$ 54 30.96  & 1.31 &    282 $\pm$   16 &     398 $\pm$   18 &     21 $\pm$    1 &      46 $\pm$    1 &     1.5 $\times 10^{-14}$ &  0.53 $\pm$ 0.05	   
\enddata
\tablecomments{\em (Continues)}
\end{deluxetable}
\tabletypesize{\scriptsize}

%% file: tab6b.tex
\tabletypesize{\scriptsize}
\begin{deluxetable}{cccccccccc}

\tablecolumns{10}
\tablewidth{0pt}
\tablecaption{NGC~3621: Coordinates and reddening-corrected fluxes}
\tablenum{6}

\tablehead{
\colhead{ID}	     &
\colhead{R.A.}	 &
\colhead{DEC}	 &
\colhead{$R$/\rtf}	 &
\colhead{\oii}	 &
\colhead{\oiii}  &
\colhead{\nii}	 &
\colhead{\sii}	 &
\colhead{F(H$\alpha$)} &
\colhead{$c$(\hbeta)}   \\[0.5mm]
\colhead{}       &
\colhead{(J2000.0)}       &
\colhead{(J2000.0)}       &
\colhead{}       &
\colhead{3727}   &
\colhead{5007}   &
\colhead{6583}   &
\colhead{6717+6731} &
\colhead{(erg\,s$^{-1}$\,cm$^{-2}$)}     &
\colhead{(mag)}     \\[1mm]
\colhead{(1)}	&
\colhead{(2)}	&
\colhead{(3)}	&
\colhead{(4)}	&
\colhead{(5)}	&
\colhead{(6)}	&
\colhead{(7)}   &
\colhead{(8)}   &
\colhead{(9)}   &
\colhead{(10)} }
\startdata
\\[-2mm]
61 & 11 18 18.5  & $-32$ 54 31.71  & 1.25 &    428 $\pm$   22 &     102 $\pm$    5 &     31 $\pm$    1 &      67 $\pm$    2 &     2.1 $\times 10^{-14}$ &  0.38 $\pm$ 0.04	\\   
62 & 11 18 27.8  & $-32$ 54 37.12  & 1.33 &    317 $\pm$   17 &     268 $\pm$   12 &     23 $\pm$    1 &      40 $\pm$    1 &     2.5 $\times 10^{-14}$ &  0.36 $\pm$ 0.04	\\   
63 & 11 18 27.8  & $-32$ 54 46.24  & 1.35 &    294 $\pm$   15 &     372 $\pm$   17 &     16 $\pm$    1 &      30 $\pm$    1 &     1.9 $\times 10^{-14}$ &  0.48 $\pm$ 0.04	\\   
64 & 11 18 23.4  & $-32$ 55 06.61  & 1.32 &    273 $\pm$   13 &     352 $\pm$   16 &     16 $\pm$    1 &      27 $\pm$    1 &     3.0 $\times 10^{-14}$ &  0.05 $\pm$ 0.03	\\   
65 & 11 18 26.6  & $-32$ 55 11.70  & 1.38 &    403 $\pm$   52 &      89 $\pm$    4 &     23 $\pm$    2 &      40 $\pm$    3 &     8.4 $\times 10^{-15}$ &  0.60 $\pm$ 0.20	\\   
66 & 11 18 23.5  & $-32$ 55 21.06  & 1.37 &    197 $\pm$   15 &     446 $\pm$   21 &     14 $\pm$    1 &      32 $\pm$    2 &     4.3 $\times 10^{-15}$ &  0.17 $\pm$ 0.09	\\   
67 & 11 18 24.8  & $-32$ 55 35.23  & 1.43 &    383 $\pm$   21 &     214 $\pm$   10 &     22 $\pm$    1 &      47 $\pm$    2 &     6.5 $\times 10^{-15}$ &  0.37 $\pm$ 0.05	\\   
68 & 11 18 15.1  & $-32$ 55 50.76  & 1.69 &    411 $\pm$   37 &     118 $\pm$    6 &     21 $\pm$    2 &      34 $\pm$    3 &     3.8 $\times 10^{-15}$ &  0.14 $\pm$ 0.12	\\   
69 & 11 18 24.6  & $-32$ 55 54.09  & 1.49 &    287 $\pm$   30 &     131 $\pm$    7 &     27 $\pm$    3 &      44 $\pm$    3 &     1.0 $\times 10^{-15}$ &  0.33 $\pm$ 0.15	\\   
70 & 11 18 23.1  & $-32$ 56 10.95  & 1.54 &    441 $\pm$   39 &      37 $\pm$    2 &     27 $\pm$    2 &      51 $\pm$    3 &     3.9 $\times 10^{-15}$ &  0.34 $\pm$ 0.12	\\   
71 & 11 18 20.6  & $-32$ 56 34.97  & 1.66 &    581 $\pm$   60 &     103 $\pm$    5 &     19 $\pm$    2 &      43 $\pm$    3 &     2.2 $\times 10^{-15}$ &  0.45 $\pm$ 0.15	\\   
72 & 11 18 20.0  & $-32$ 57 08.01  & 1.81 &    455 $\pm$   27 &      75 $\pm$    3 &     22 $\pm$    1 &      50 $\pm$    2 &     8.2 $\times 10^{-15}$ &  0.11 $\pm$ 0.06	\\   
73 & 11 18 17.5  & $-32$ 57 24.66  & 1.95 &    452 $\pm$   33 &     133 $\pm$    6 &     22 $\pm$    2 &      60 $\pm$    4 &     4.5 $\times 10^{-15}$ &  0.00 $\pm$ 0.09	  
\enddata
\tablecomments{Units of right ascension are hours, minutes and seconds, and units of declination
are degrees, arcminutes and arcseconds.  Line fluxes are in units of H$\beta$\,=\,100. The \halpha\ flux in column (9) has been measured from narrow-band images, and corrected for extinction.}
\end{deluxetable}
\tabletypesize{\scriptsize}

%% file: ms.bbl
\begin{thebibliography}{151}
\expandafter\ifx\csname natexlab\endcsname\relax\def\natexlab#1{#1}\fi


\bibitem[{{Alberts} {et~al.}(2011)}]{Alberts:2011}
{Alberts}, S., {et~al.} 2011, \apj, 731, 28

\bibitem[{{Asplund} {et~al.}(2009){Asplund}, {Grevesse}, {Sauval}, \&
  {Scott}}]{Asplund:2009}
{Asplund}, M., {Grevesse}, N., {Sauval}, A.~J., \& {Scott}, P. 2009, \araa, 47,
  481

\bibitem[{{Barker} {et~al.}(2011){Barker}, {Ferguson}, {Cole}, {Ibata},
  {Irwin}, {Lewis}, {Smecker-Hane}, \& {Tanvir}}]{Barker:2011}
{Barker}, M.~K., {Ferguson}, A.~M.~N., {Cole}, A.~A., {Ibata}, R., {Irwin}, M.,
  {Lewis}, G.~F., {Smecker-Hane}, T.~A., \& {Tanvir}, N.~R. 2011, \mnras, 410,
  504

\bibitem[{{Bertin} \& {Arnouts}(1996)}]{Bertin:1996}
{Bertin}, E., \& {Arnouts}, S. 1996, \aaps, 117, 393

\bibitem[{{Bertin} \& {Amorisco}(2010)}]{Bertin:2010}
{Bertin}, G., \& {Amorisco}, N.~C. 2010, \aap, 512, A17

\bibitem[{{Bigiel} {et~al.}(2010{\natexlab{a}}){Bigiel}, {Leroy}, {Seibert},
  {Walter}, {Blitz}, {Thilker}, \& {Madore}}]{Bigiel:2010a}
{Bigiel}, F., {Leroy}, A., {Seibert}, M., {Walter}, F., {Blitz}, L., {Thilker},
  D., \& {Madore}, B. 2010{\natexlab{a}}, \apjl, 720, L31

\bibitem[{{Bigiel} {et~al.}(2010{\natexlab{b}}){Bigiel}, {Leroy}, {Walter},
  {Blitz}, {Brinks}, {de Blok}, \& {Madore}}]{Bigiel:2010}
{Bigiel}, F., {Leroy}, A., {Walter}, F., {Blitz}, L., {Brinks}, E., {de Blok},
  W.~J.~G., \& {Madore}, B. 2010{\natexlab{b}}, \aj, 140, 1194

\bibitem[{{Bird} {et~al.}(2012){Bird}, {Kazantzidis}, \&
  {Weinberg}}]{Bird:2012}
{Bird}, J.~C., {Kazantzidis}, S., \& {Weinberg}, D.~H. 2012, \mnras, 420, 913

\bibitem[{{Bland-Hawthorn} {et~al.}(2005){Bland-Hawthorn}, {Vlaji{\'c}},
  {Freeman}, \& {Draine}}]{Bland-Hawthorn:2005}
{Bland-Hawthorn}, J., {Vlaji{\'c}}, M., {Freeman}, K.~C., \& {Draine}, B.~T.
  2005, \apj, 629, 239

\bibitem[{{Boissier} \& {Prantzos}(1999)}]{Boissier:1999}
{Boissier}, S., \& {Prantzos}, N. 1999, \mnras, 307, 857

\bibitem[{{Boissier} {et~al.}(2007)}]{Boissier:2007}
{Boissier}, S., {et~al.} 2007, \apjs, 173, 524

\bibitem[{{Bouch{\'e}} {et~al.}(2010){Bouch{\'e}}, {Dekel}, {Genzel}, {Genel},
  {Cresci}, {F{\"o}rster Schreiber}, {Shapiro}, {Davies}, \&
  {Tacconi}}]{Bouche:2010}
{Bouch{\'e}}, N., {Dekel}, A., {Genzel}, R., {Genel}, S., {Cresci}, G.,
  {F{\"o}rster Schreiber}, N.~M., {Shapiro}, K.~L., {Davies}, R.~I., \&
  {Tacconi}, L. 2010, \apj, 718, 1001

\bibitem[{{Bresolin}(2008)}]{Bresolin:2008}
{Bresolin}, F. 2008, in The Metal-Rich Universe, ed. {G.~Israelian \&
  G.~Meynet}, 155

\bibitem[{{Bresolin}(2011{\natexlab{a}})}]{Bresolin:2011}
{Bresolin}, F. 2011{\natexlab{a}}, \apj, 729, 56

\bibitem[{{Bresolin}(2011{\natexlab{b}})}]{Bresolin:2011a}
---. 2011{\natexlab{b}}, \apj, 730, 129

\bibitem[{{Bresolin} {et~al.}(2009{\natexlab{a}}){Bresolin}, {Gieren},
  {Kudritzki}, {Pietrzy{\'n}ski}, {Urbaneja}, \& {Carraro}}]{Bresolin:2009a}
{Bresolin}, F., {Gieren}, W., {Kudritzki}, R., {Pietrzy{\'n}ski}, G.,
  {Urbaneja}, M.~A., \& {Carraro}, G. 2009{\natexlab{a}}, \apj, 700, 309

\bibitem[{{Bresolin} {et~al.}(1999){Bresolin}, {Kennicutt}, \&
  {Garnett}}]{Bresolin:1999}
{Bresolin}, F., {Kennicutt}, Jr., R.~C., \& {Garnett}, D.~R. 1999, \apj, 510,
  104

\bibitem[{{Bresolin} {et~al.}(2001){Bresolin}, {Kudritzki}, {Mendez}, \&
  {Przybilla}}]{Bresolin:2001}
{Bresolin}, F., {Kudritzki}, R.-P., {Mendez}, R.~H., \& {Przybilla}, N. 2001,
  \apjl, 548, L159

\bibitem[{{Bresolin} {et~al.}(2009{\natexlab{b}}){Bresolin}, {Ryan-Weber},
  {Kennicutt}, \& {Goddard}}]{Bresolin:2009}
{Bresolin}, F., {Ryan-Weber}, E., {Kennicutt}, R.~C., \& {Goddard}, Q.
  2009{\natexlab{b}}, \apj, 695, 580

\bibitem[{{Bresolin} {et~al.}(2005){Bresolin}, {Schaerer}, {Gonz{\'a}lez
  Delgado}, \& {Stasi{\'n}ska}}]{Bresolin:2005}
{Bresolin}, F., {Schaerer}, D., {Gonz{\'a}lez Delgado}, R.~M., \&
  {Stasi{\'n}ska}, G. 2005, \aap, 441, 981

\bibitem[{{Bresolin} {et~al.}(2010){Bresolin}, {Stasi{\'n}ska},
  {V{\'{\i}}lchez}, {Simon}, \& {Rosolowsky}}]{Bresolin:2010}
{Bresolin}, F., {Stasi{\'n}ska}, G., {V{\'{\i}}lchez}, J.~M., {Simon}, J.~D.,
  \& {Rosolowsky}, E. 2010, \mnras, 404, 1679

\bibitem[{{Bush} {et~al.}(2010){Bush}, {Cox}, {Hayward}, {Thilker},
  {Hernquist}, \& {Besla}}]{Bush:2010}
{Bush}, S.~J., {Cox}, T.~J., {Hayward}, C.~C., {Thilker}, D., {Hernquist}, L.,
  \& {Besla}, G. 2010, \apj, 713, 780

\bibitem[{{Bush} {et~al.}(2008){Bush}, {Cox}, {Hernquist}, {Thilker}, \&
  {Younger}}]{Bush:2008}
{Bush}, S.~J., {Cox}, T.~J., {Hernquist}, L., {Thilker}, D., \& {Younger},
  J.~D. 2008, \apjl, 683, L13

\bibitem[{{Buta}(1988)}]{Buta:1988}
{Buta}, R. 1988, \apjs, 66, 233

\bibitem[{{Chakrabarti} {et~al.}(2011){Chakrabarti}, {Bigiel}, {Chang}, \&
  {Blitz}}]{Chakrabarti:2011}
{Chakrabarti}, S., {Bigiel}, F., {Chang}, P., \& {Blitz}, L. 2011, \apj, 743,
  35

\bibitem[{{Chen} {et~al.}(2005){Chen}, {Kennicutt}, \& {Rauch}}]{Chen:2005}
{Chen}, H.-W., {Kennicutt}, Jr., R.~C., \& {Rauch}, M. 2005, \apj, 620, 703

\bibitem[{{Crosthwaite} {et~al.}(2002){Crosthwaite}, {Turner}, {Buchholz},
  {Ho}, \& {Martin}}]{Crosthwaite:2002}
{Crosthwaite}, L.~P., {Turner}, J.~L., {Buchholz}, L., {Ho}, P.~T.~P., \&
  {Martin}, R.~N. 2002, \aj, 123, 1892

\bibitem[{{Dav{\'e}} {et~al.}(2011{\natexlab{a}}){Dav{\'e}}, {Finlator}, \&
  {Oppenheimer}}]{Dave:2011a}
{Dav{\'e}}, R., {Finlator}, K., \& {Oppenheimer}, B.~D. 2011{\natexlab{a}},
  ArXiv e-prints, 1108.0426

\bibitem[{{Dav{\'e}} {et~al.}(2011{\natexlab{b}}){Dav{\'e}}, {Finlator}, \&
  {Oppenheimer}}]{Dave:2011}
---. 2011{\natexlab{b}}, \mnras, 1158

\bibitem[{{Davidge}(2009)}]{Davidge:2009}
{Davidge}, T.~J. 2009, \apj, 697, 1439

\bibitem[{{Davidge}(2010)}]{Davidge:2010}
---. 2010, \apj, 718, 1428

\bibitem[{{de Blok} {et~al.}(2008){de Blok}, {Walter}, {Brinks},
  {Trachternach}, {Oh}, \& {Kennicutt}}]{de-Blok:2008}
{de Blok}, W.~J.~G., {Walter}, F., {Brinks}, E., {Trachternach}, C., {Oh},
  S.-H., \& {Kennicutt}, R.~C. 2008, \aj, 136, 2648

\bibitem[{{de Vaucouleurs} {et~al.}(1991){de Vaucouleurs}, {de Vaucouleurs},
  {Corwin}, {Buta}, {Paturel}, \& {Fouque}}]{de-Vaucouleurs:1991}
{de Vaucouleurs}, G., {de Vaucouleurs}, A., {Corwin}, Jr., H.~G., {Buta},
  R.~J., {Paturel}, G., \& {Fouque}, P. 1991, {Third reference catalogue of
  bright galaxies} (Springer-Verlag Berlin Heidelberg New York)

\bibitem[{{Dekel} \& {Birnboim}(2006)}]{Dekel:2006}
{Dekel}, A., \& {Birnboim}, Y. 2006, \mnras, 368, 2

\bibitem[{{Dicaire} {et~al.}(2008)}]{Dicaire:2008}
{Dicaire}, I., {et~al.} 2008, \mnras, 385, 553

\bibitem[{{Dong} {et~al.}(2008){Dong}, {Calzetti}, {Regan}, {Thilker},
  {Bianchi}, {Meurer}, \& {Walter}}]{Dong:2008}
{Dong}, H., {Calzetti}, D., {Regan}, M., {Thilker}, D., {Bianchi}, L.,
  {Meurer}, G.~R., \& {Walter}, F. 2008, \aj, 136, 479

\bibitem[{{Dopita} {et~al.}(2006)}]{Dopita:2006a}
{Dopita}, M.~A., {et~al.} 2006, \apjs, 167, 177

\bibitem[{{Elson} {et~al.}(2011){Elson}, {de Blok}, \&
  {Kraan-Korteweg}}]{Elson:2011}
{Elson}, E.~C., {de Blok}, W.~J.~G., \& {Kraan-Korteweg}, R.~C. 2011, \mnras,
  411, 200

\bibitem[{{Espada} {et~al.}(2011)}]{Espada:2011}
{Espada}, D., {et~al.} 2011, \apj, 736, 20

\bibitem[{{Esteban} {et~al.}(2009){Esteban}, {Bresolin}, {Peimbert},
  {Garc{\'{\i}}a-Rojas}, {Peimbert}, \& {Mesa-Delgado}}]{Esteban:2009}
{Esteban}, C., {Bresolin}, F., {Peimbert}, M., {Garc{\'{\i}}a-Rojas}, J.,
  {Peimbert}, A., \& {Mesa-Delgado}, A. 2009, \apj, 700, 654

\bibitem[{{Ferguson} \& {Clarke}(2001)}]{Ferguson:2001}
{Ferguson}, A.~M.~N., \& {Clarke}, C.~J. 2001, \mnras, 325, 781

\bibitem[{{Ferguson} {et~al.}(1998){Ferguson}, {Gallagher}, \&
  {Wyse}}]{Ferguson:1998a}
{Ferguson}, A.~M.~N., {Gallagher}, J.~S., \& {Wyse}, R.~F.~G. 1998, \aj, 116,
  673

\bibitem[{{Finkelman} {et~al.}(2011){Finkelman}, {Moiseev}, {Brosch}, \&
  {Katkov}}]{Finkelman:2011}
{Finkelman}, I., {Moiseev}, A., {Brosch}, N., \& {Katkov}, I. 2011, \mnras,
  418, 1834

\bibitem[{{Finlator} \& {Dav{\'e}}(2008)}]{Finlator:2008}
{Finlator}, K., \& {Dav{\'e}}, R. 2008, \mnras, 385, 2181

\bibitem[{{Freedman} {et~al.}(2001)}]{Freedman:2001}
{Freedman}, W.~L., {et~al.} 2001, \apj, 553, 47

\bibitem[{{Friedli} {et~al.}(1994){Friedli}, {Benz}, \&
  {Kennicutt}}]{Friedli:1994}
{Friedli}, D., {Benz}, W., \& {Kennicutt}, R. 1994, \apjl, 430, L105

\bibitem[{{Fumagalli} {et~al.}(2011){Fumagalli}, {Prochaska}, {Kasen}, {Dekel},
  {Ceverino}, \& {Primack}}]{Fumagalli:2011}
{Fumagalli}, M., {Prochaska}, J.~X., {Kasen}, D., {Dekel}, A., {Ceverino}, D.,
  \& {Primack}, J.~R. 2011, \mnras, 418, 1796

\bibitem[{{Garnett} {et~al.}(1997){Garnett}, {Shields}, {Skillman}, {Sagan}, \&
  {Dufour}}]{Garnett:1997}
{Garnett}, D.~R., {Shields}, G.~A., {Skillman}, E.~D., {Sagan}, S.~P., \&
  {Dufour}, R.~J. 1997, \apj, 489, 63

\bibitem[{{Genzel} {et~al.}(2010)}]{Genzel:2010}
{Genzel}, R., {et~al.} 2010, \mnras, 407, 2091

\bibitem[{{Gil de Paz} {et~al.}(2005)}]{Gil-de-Paz:2005}
{Gil de Paz}, A., {et~al.} 2005, \apjl, 627, L29

\bibitem[{{Goddard} {et~al.}(2011){Goddard}, {Bresolin}, {Kennicutt},
  {Ryan-Weber}, \& {Rosales-Ortega}}]{Goddard:2011}
{Goddard}, Q.~E., {Bresolin}, F., {Kennicutt}, R.~C., {Ryan-Weber}, E.~V., \&
  {Rosales-Ortega}, F.~F. 2011, \mnras, 412, 1246

\bibitem[{{Goddard} {et~al.}(2010){Goddard}, {Kennicutt}, \&
  {Ryan-Weber}}]{Goddard:2010}
{Goddard}, Q.~E., {Kennicutt}, R.~C., \& {Ryan-Weber}, E.~V. 2010, \mnras, 405,
  2791

\bibitem[{{Goetz} \& {Koeppen}(1992)}]{Goetz:1992}
{Goetz}, M., \& {Koeppen}, J. 1992, \aap, 262, 455

\bibitem[{{Gogarten} {et~al.}(2010)}]{Gogarten:2010}
{Gogarten}, S.~M., {et~al.} 2010, \apj, 712, 858

\bibitem[{{Guseva} {et~al.}(2011){Guseva}, {Izotov}, {Stasi{\'n}ska}, {Fricke},
  {Henkel}, \& {Papaderos}}]{Guseva:2011}
{Guseva}, N.~G., {Izotov}, Y.~I., {Stasi{\'n}ska}, G., {Fricke}, K.~J.,
  {Henkel}, C., \& {Papaderos}, P. 2011, \aap, 529, A149

\bibitem[{{Guseva} {et~al.}(2009){Guseva}, {Papaderos}, {Meyer}, {Izotov}, \&
  {Fricke}}]{Guseva:2009}
{Guseva}, N.~G., {Papaderos}, P., {Meyer}, H.~T., {Izotov}, Y.~I., \& {Fricke},
  K.~J. 2009, \aap, 505, 63

\bibitem[{{Hawarden} {et~al.}(1979){Hawarden}, {van Woerden}, {Goss}, {Mebold},
  \& {Peterson}}]{Hawarden:1979}
{Hawarden}, T.~G., {van Woerden}, H., {Goss}, W.~M., {Mebold}, U., \&
  {Peterson}, B.~A. 1979, \aap, 76, 230

\bibitem[{{Herbert-Fort} {et~al.}(2009)}]{Herbert-Fort:2009}
{Herbert-Fort}, S., {et~al.} 2009, \apj, 700, 1977

\bibitem[{{Hillier} \& {Miller}(1998)}]{Hillier:1998}
{Hillier}, D.~J., \& {Miller}, D.~L. 1998, \apj, 496, 407

\bibitem[{{Hunter} {et~al.}(2011){Hunter}, {Elmegreen}, {Oh}, {Anderson},
  {Nordgren}, {Massey}, {Wilsey}, \& {Riabokin}}]{Hunter:2011}
{Hunter}, D.~A., {Elmegreen}, B.~G., {Oh}, S.-H., {Anderson}, E., {Nordgren},
  T.~E., {Massey}, P., {Wilsey}, N., \& {Riabokin}, M. 2011, \aj, 142, 121

\bibitem[{{Izotov} {et~al.}(1994){Izotov}, {Thuan}, \&
  {Lipovetsky}}]{Izotov:1994}
{Izotov}, Y.~I., {Thuan}, T.~X., \& {Lipovetsky}, V.~A. 1994, \apj, 435, 647

\bibitem[{{Kehrig} {et~al.}(2011)}]{Kehrig:2011}
{Kehrig}, C., {et~al.} 2011, \aap, 526, A128+

\bibitem[{{Kennicutt} {et~al.}(2003{\natexlab{a}}){Kennicutt}, {Bresolin}, \&
  {Garnett}}]{Kennicutt:2003}
{Kennicutt}, R.~C., {Bresolin}, F., \& {Garnett}, D.~R. 2003{\natexlab{a}},
  \apj, 591, 801

\bibitem[{{Kennicutt}(1989)}]{Kennicutt:1989}
{Kennicutt}, Jr., R.~C. 1989, \apj, 344, 685

\bibitem[{{Kennicutt} {et~al.}(2000){Kennicutt}, {Bresolin}, {French}, \&
  {Martin}}]{Kennicutt:2000}
{Kennicutt}, Jr., R.~C., {Bresolin}, F., {French}, H., \& {Martin}, P. 2000,
  \apj, 537, 589

\bibitem[{{Kennicutt} {et~al.}(2003{\natexlab{b}})}]{Kennicutt:2003a}
{Kennicutt}, Jr., R.~C., {et~al.} 2003{\natexlab{b}}, \pasp, 115, 928

\bibitem[{{Kere{\v s}} \& {Hernquist}(2009)}]{Keres:2009b}
{Kere{\v s}}, D., \& {Hernquist}, L. 2009, \apjl, 700, L1

\bibitem[{{Kere{\v s}} {et~al.}(2009){Kere{\v s}}, {Katz}, {Dav{\'e}},
  {Fardal}, \& {Weinberg}}]{Keres:2009a}
{Kere{\v s}}, D., {Katz}, N., {Dav{\'e}}, R., {Fardal}, M., \& {Weinberg},
  D.~H. 2009, \mnras, 396, 2332

\bibitem[{{Kere{\v s}} {et~al.}(2005){Kere{\v s}}, {Katz}, {Weinberg}, \&
  {Dav{\'e}}}]{Keres:2005}
{Kere{\v s}}, D., {Katz}, N., {Weinberg}, D.~H., \& {Dav{\'e}}, R. 2005,
  \mnras, 363, 2

\bibitem[{{Kewley} \& {Ellison}(2008)}]{Kewley:2008}
{Kewley}, L.~J., \& {Ellison}, S.~L. 2008, \apj, 681, 1183

\bibitem[{{Kinman}(1978)}]{Kinman:1978}
{Kinman}, T.~D. 1978, \aj, 83, 764

\bibitem[{{Koribalski} \& {L{\'o}pez-S{\'a}nchez}(2009)}]{Koribalski:2009}
{Koribalski}, B.~S., \& {L{\'o}pez-S{\'a}nchez}, {\'A}.~R. 2009, \mnras, 400,
  1749

\bibitem[{{Koribalski} {et~al.}(2004)}]{Koribalski:2004}
{Koribalski}, B.~S., {et~al.} 2004, \aj, 128, 16

\bibitem[{{Kraan-Korteweg} \& {Tammann}(1979)}]{Kraan-Korteweg:1979}
{Kraan-Korteweg}, R.~C., \& {Tammann}, G.~A. 1979, Astronomische Nachrichten,
  300, 181

\bibitem[{{Kr{\"o}ger} {et~al.}(2006){Kr{\"o}ger}, {Hensler}, \&
  {Freyer}}]{Kroger:2006}
{Kr{\"o}ger}, D., {Hensler}, G., \& {Freyer}, T. 2006, \aap, 450, L5

\bibitem[{{Kulkarni} {et~al.}(2005){Kulkarni}, {Fall}, {Lauroesch}, {York},
  {Welty}, {Khare}, \& {Truran}}]{Kulkarni:2005}
{Kulkarni}, V.~P., {Fall}, S.~M., {Lauroesch}, J.~T., {York}, D.~G., {Welty},
  D.~E., {Khare}, P., \& {Truran}, J.~W. 2005, \apj, 618, 68

\bibitem[{{Kuzio de Naray} {et~al.}(2004){Kuzio de Naray}, {McGaugh}, \& {de
  Blok}}]{Kuzio-de-Naray:2004}
{Kuzio de Naray}, R., {McGaugh}, S.~S., \& {de Blok}, W.~J.~G. 2004, \mnras,
  355, 887

\bibitem[{{Lacey} \& {Fall}(1985)}]{Lacey:1985}
{Lacey}, C.~G., \& {Fall}, S.~M. 1985, \apj, 290, 154

\bibitem[{{Leitherer} {et~al.}(1999)}]{Leitherer:1999}
{Leitherer}, C., {et~al.} 1999, \apjs, 123, 3

\bibitem[{{Lemonias} {et~al.}(2011)}]{Lemonias:2011}
{Lemonias}, J.~J., {et~al.} 2011, \apj, 733, 74

\bibitem[{{L{\'e}pine} {et~al.}(2011)}]{Lepine:2011}
{L{\'e}pine}, J.~R.~D., {et~al.} 2011, \mnras, 417, 698

\bibitem[{{Maeder}(1992)}]{Maeder:1992}
{Maeder}, A. 1992, \aap, 264, 105

\bibitem[{{Martin} \& {Roy}(1995)}]{Martin:1995}
{Martin}, P., \& {Roy}, J.-R. 1995, \apj, 445, 161

\bibitem[{{Matteucci} \& {Francois}(1989)}]{Matteucci:1989}
{Matteucci}, F., \& {Francois}, P. 1989, \mnras, 239, 885

\bibitem[{{McCall} {et~al.}(1985){McCall}, {Rybski}, \&
  {Shields}}]{McCall:1985}
{McCall}, M.~L., {Rybski}, P.~M., \& {Shields}, G.~A. 1985, \apjs, 57, 1

\bibitem[{{McGaugh}(1991)}]{McGaugh:1991}
{McGaugh}, S.~S. 1991, \apj, 380, 140

\bibitem[{{Menzel} {et~al.}(1941){Menzel}, {Aller}, \& {Hebb}}]{Menzel:1941}
{Menzel}, D.~H., {Aller}, L.~H., \& {Hebb}, M.~H. 1941, \apj, 93, 230

\bibitem[{{Meurer} {et~al.}(2006)}]{Meurer:2006}
{Meurer}, G.~R., {et~al.} 2006, \apjs, 165, 307

\bibitem[{{Meynet} {et~al.}(1994){Meynet}, {Maeder}, {Schaller}, {Schaerer}, \&
  {Charbonnel}}]{Meynet:1994}
{Meynet}, G., {Maeder}, A., {Schaller}, G., {Schaerer}, D., \& {Charbonnel}, C.
  1994, \aaps, 103, 97

\bibitem[{{Minchev} \& {Famaey}(2010)}]{Minchev:2010}
{Minchev}, I., \& {Famaey}, B. 2010, \apj, 722, 112

\bibitem[{{Minchev} {et~al.}(2011){Minchev}, {Famaey}, {Combes}, {Di Matteo},
  {Mouhcine}, \& {Wozniak}}]{Minchev:2011}
{Minchev}, I., {Famaey}, B., {Combes}, F., {Di Matteo}, P., {Mouhcine}, M., \&
  {Wozniak}, H. 2011, \aap, 527, A147

\bibitem[{{Moffett} {et~al.}(2011){Moffett}, {Kannappan}, {Baker}, \&
  {Laine}}]{Moffett:2011}
{Moffett}, A.~J., {Kannappan}, S.~J., {Baker}, A.~J., \& {Laine}, S. 2011, in
  EAS Publications Series, Vol.~48, EAS Publications Series, ed. {M.~Koleva,
  P.~Prugniel, \& I.~Vauglin}, 419--421

\bibitem[{{Naab} \& {Ostriker}(2006)}]{Naab:2006}
{Naab}, T., \& {Ostriker}, J.~P. 2006, \mnras, 366, 899

\bibitem[{{Ocvirk} {et~al.}(2008){Ocvirk}, {Pichon}, \&
  {Teyssier}}]{Ocvirk:2008}
{Ocvirk}, P., {Pichon}, C., \& {Teyssier}, R. 2008, \mnras, 390, 1326

\bibitem[{{Oppenheimer} \& {Dav{\'e}}(2008)}]{Oppenheimer:2008}
{Oppenheimer}, B.~D., \& {Dav{\'e}}, R. 2008, \mnras, 387, 577

\bibitem[{{Pagel} {et~al.}(1979){Pagel}, {Edmunds}, {Blackwell}, {Chun}, \&
  {Smith}}]{Pagel:1979}
{Pagel}, B.~E.~J., {Edmunds}, M.~G., {Blackwell}, D.~E., {Chun}, M.~S., \&
  {Smith}, G. 1979, \mnras, 189, 95

\bibitem[{{Paturel} {et~al.}(2003){Paturel}, {Petit}, {Prugniel}, {Theureau},
  {Rousseau}, {Brouty}, {Dubois}, \& {Cambr{\'e}sy}}]{Paturel:2003}
{Paturel}, G., {Petit}, C., {Prugniel}, P., {Theureau}, G., {Rousseau}, J.,
  {Brouty}, M., {Dubois}, P., \& {Cambr{\'e}sy}, L. 2003, \aap, 412, 45

\bibitem[{{Pauldrach} {et~al.}(2001){Pauldrach}, {Hoffmann}, \&
  {Lennon}}]{Pauldrach:2001}
{Pauldrach}, A.~W.~A., {Hoffmann}, T.~L., \& {Lennon}, M. 2001, \aap, 375, 161

\bibitem[{{Pedicelli} {et~al.}(2009)}]{Pedicelli:2009}
{Pedicelli}, S., {et~al.} 2009, \aap, 504, 81

\bibitem[{{Pettini}(1999)}]{Pettini:1999}
{Pettini}, M. 1999, in Chemical Evolution from Zero to High Redshift, ed.
  {J.~R.~Walsh \& M.~R.~Rosa}, 233

\bibitem[{{Pettini} \& {Pagel}(2004)}]{Pettini:2004}
{Pettini}, M., \& {Pagel}, B.~E.~J. 2004, \mnras, 348, L59

\bibitem[{{Pilyugin} {et~al.}(2010){Pilyugin}, {V{\'{\i}}lchez}, \&
  {Thuan}}]{Pilyugin:2010}
{Pilyugin}, L.~S., {V{\'{\i}}lchez}, J.~M., \& {Thuan}, T.~X. 2010, \apj, 720,
  1738

\bibitem[{{Portinari} \& {Chiosi}(2000)}]{Portinari:2000}
{Portinari}, L., \& {Chiosi}, C. 2000, \aap, 355, 929

\bibitem[{{Prochaska} {et~al.}(2003){Prochaska}, {Gawiser}, {Wolfe}, {Castro},
  \& {Djorgovski}}]{Prochaska:2003}
{Prochaska}, J.~X., {Gawiser}, E., {Wolfe}, A.~M., {Castro}, S., \&
  {Djorgovski}, S.~G. 2003, \apjl, 595, L9

\bibitem[{{Quillen} {et~al.}(2009){Quillen}, {Minchev}, {Bland-Hawthorn}, \&
  {Haywood}}]{Quillen:2009}
{Quillen}, A.~C., {Minchev}, I., {Bland-Hawthorn}, J., \& {Haywood}, M. 2009,
  \mnras, 397, 1599

\bibitem[{{Ribaudo} {et~al.}(2011){Ribaudo}, {Lehner}, {Howk}, {Werk}, {Tripp},
  {Prochaska}, {Meiring}, \& {Tumlinson}}]{Ribaudo:2011}
{Ribaudo}, J., {Lehner}, N., {Howk}, J.~C., {Werk}, J.~K., {Tripp}, T.~M.,
  {Prochaska}, J.~X., {Meiring}, J.~D., \& {Tumlinson}, J. 2011, \apj, 743, 207

\bibitem[{{Ro{\v s}kar} {et~al.}(2010){Ro{\v s}kar}, {Debattista}, {Brooks},
  {Quinn}, {Brook}, {Governato}, {Dalcanton}, \& {Wadsley}}]{Roskar:2010}
{Ro{\v s}kar}, R., {Debattista}, V.~P., {Brooks}, A.~M., {Quinn}, T.~R.,
  {Brook}, C.~B., {Governato}, F., {Dalcanton}, J.~J., \& {Wadsley}, J. 2010,
  \mnras, 408, 783

\bibitem[{{Ro{\v s}kar} {et~al.}(2008{\natexlab{a}}){Ro{\v s}kar},
  {Debattista}, {Quinn}, {Stinson}, \& {Wadsley}}]{Roskar:2008b}
{Ro{\v s}kar}, R., {Debattista}, V.~P., {Quinn}, T.~R., {Stinson}, G.~S., \&
  {Wadsley}, J. 2008{\natexlab{a}}, \apjl, 684, L79

\bibitem[{{Ro{\v s}kar} {et~al.}(2008{\natexlab{b}}){Ro{\v s}kar},
  {Debattista}, {Stinson}, {Quinn}, {Kaufmann}, \& {Wadsley}}]{Roskar:2008}
{Ro{\v s}kar}, R., {Debattista}, V.~P., {Stinson}, G.~S., {Quinn}, T.~R.,
  {Kaufmann}, T., \& {Wadsley}, J. 2008{\natexlab{b}}, \apjl, 675, L65

\bibitem[{{Rupke} {et~al.}(2010){Rupke}, {Kewley}, \& {Barnes}}]{Rupke:2010}
{Rupke}, D.~S.~N., {Kewley}, L.~J., \& {Barnes}, J.~E. 2010, \apjl, 710, L156

\bibitem[{{Salim} \& {Rich}(2010)}]{Salim:2010}
{Salim}, S., \& {Rich}, R.~M. 2010, \apjl, 714, L290

\bibitem[{{Sancisi} {et~al.}(2008){Sancisi}, {Fraternali}, {Oosterloo}, \& {van
  der Hulst}}]{Sancisi:2008}
{Sancisi}, R., {Fraternali}, F., {Oosterloo}, T., \& {van der Hulst}, T. 2008,
  \aapr, 15, 189

\bibitem[{{Scannapieco} {et~al.}(2008){Scannapieco}, {Tissera}, {White}, \&
  {Springel}}]{Scannapieco:2008}
{Scannapieco}, C., {Tissera}, P.~B., {White}, S.~D.~M., \& {Springel}, V. 2008,
  \mnras, 389, 1137

\bibitem[{{Scannapieco} {et~al.}(2009){Scannapieco}, {White}, {Springel}, \&
  {Tissera}}]{Scannapieco:2009}
{Scannapieco}, C., {White}, S.~D.~M., {Springel}, V., \& {Tissera}, P.~B. 2009,
  \mnras, 396, 696

\bibitem[{{Schlegel} {et~al.}(1998){Schlegel}, {Finkbeiner}, \&
  {Davis}}]{Schlegel:1998}
{Schlegel}, D.~J., {Finkbeiner}, D.~P., \& {Davis}, M. 1998, \apj, 500, 525

\bibitem[{{Sch{\"o}nrich} \& {Binney}(2009)}]{Schonrich:2009}
{Sch{\"o}nrich}, R., \& {Binney}, J. 2009, \mnras, 396, 203

\bibitem[{{Schulte-Ladbeck} {et~al.}(2005){Schulte-Ladbeck}, {K{\"o}nig},
  {Miller}, {Hopkins}, {Drozdovsky}, {Turnshek}, \&
  {Hopp}}]{Schulte-Ladbeck:2005}
{Schulte-Ladbeck}, R.~E., {K{\"o}nig}, B., {Miller}, C.~J., {Hopkins}, A.~M.,
  {Drozdovsky}, I.~O., {Turnshek}, D.~A., \& {Hopp}, U. 2005, \apjl, 625, L79

\bibitem[{{Seaton}(1979)}]{Seaton:1979}
{Seaton}, M.~J. 1979, \mnras, 187, 73P

\bibitem[{{Sellwood} \& {Binney}(2002)}]{Sellwood:2002}
{Sellwood}, J.~A., \& {Binney}, J.~J. 2002, \mnras, 336, 785

\bibitem[{{Shapiro} \& {Field}(1976)}]{Shapiro:1976}
{Shapiro}, P.~R., \& {Field}, G.~B. 1976, \apj, 205, 762

\bibitem[{{Spavone} {et~al.}(2011){Spavone}, {Iodice}, {Arnaboldi}, {Longo}, \&
  {Gerhard}}]{Spavone:2011}
{Spavone}, M., {Iodice}, E., {Arnaboldi}, M., {Longo}, G., \& {Gerhard}, O.
  2011, \aap, 531, A21

\bibitem[{{Spitoni} \& {Matteucci}(2011)}]{Spitoni:2011}
{Spitoni}, E., \& {Matteucci}, F. 2011, \aap, 531, A72

\bibitem[{{Steidel} {et~al.}(2010){Steidel}, {Erb}, {Shapley}, {Pettini},
  {Reddy}, {Bogosavljevi{\'c}}, {Rudie}, \& {Rakic}}]{Steidel:2010}
{Steidel}, C.~C., {Erb}, D.~K., {Shapley}, A.~E., {Pettini}, M., {Reddy}, N.,
  {Bogosavljevi{\'c}}, M., {Rudie}, G.~C., \& {Rakic}, O. 2010, \apj, 717, 289

\bibitem[{{Sutherland} \& {Dopita}(1993)}]{Sutherland:1993}
{Sutherland}, R.~S., \& {Dopita}, M.~A. 1993, \apjs, 88, 253

\bibitem[{{Tassis} {et~al.}(2008){Tassis}, {Kravtsov}, \&
  {Gnedin}}]{Tassis:2008}
{Tassis}, K., {Kravtsov}, A.~V., \& {Gnedin}, N.~Y. 2008, \apj, 672, 888

\bibitem[{{Terlevich} {et~al.}(1991){Terlevich}, {Melnick}, {Masegosa},
  {Moles}, \& {Copetti}}]{Terlevich:1991}
{Terlevich}, R., {Melnick}, J., {Masegosa}, J., {Moles}, M., \& {Copetti},
  M.~V.~F. 1991, \aaps, 91, 285

\bibitem[{{Thilker} {et~al.}(2005)}]{Thilker:2005}
{Thilker}, D.~A., {et~al.} 2005, \apjl, 619, L79

\bibitem[{{Thilker} {et~al.}(2007)}]{Thilker:2007}
---. 2007, \apjs, 173, 538

\bibitem[{{Thilker} {et~al.}(2010)}]{Thilker:2010}
---. 2010, \apjl, 714, L171

\bibitem[{{Tripp} \& {Song}(2011)}]{Tripp:2011}
{Tripp}, T.~M., \& {Song}, L. 2011, ArXiv e-prints, 1101.1107

\bibitem[{{Tumlinson} {et~al.}(2011)}]{Tumlinson:2011}
{Tumlinson}, J., {et~al.} 2011, Science, 334, 948

\bibitem[{{van den Bergh}(1980)}]{van-den-Bergh:1980}
{van den Bergh}, S. 1980, \pasp, 92, 122

\bibitem[{{van Zee} {et~al.}(1998){van Zee}, {Salzer}, {Haynes}, {O'Donoghue},
  \& {Balonek}}]{van-Zee:1998a}
{van Zee}, L., {Salzer}, J.~J., {Haynes}, M.~P., {O'Donoghue}, A.~A., \&
  {Balonek}, T.~J. 1998, \aj, 116, 2805

\bibitem[{{Veilleux} {et~al.}(2005){Veilleux}, {Cecil}, \&
  {Bland-Hawthorn}}]{Veilleux:2005}
{Veilleux}, S., {Cecil}, G., \& {Bland-Hawthorn}, J. 2005, \araa, 43, 769

\bibitem[{{Vila-Costas} \& {Edmunds}(1992)}]{Vila-Costas:1992}
{Vila-Costas}, M.~B., \& {Edmunds}, M.~G. 1992, \mnras, 259, 121

\bibitem[{{Vlaji{\'c}} {et~al.}(2009){Vlaji{\'c}}, {Bland-Hawthorn}, \&
  {Freeman}}]{Vlajic:2009}
{Vlaji{\'c}}, M., {Bland-Hawthorn}, J., \& {Freeman}, K.~C. 2009, \apj, 697,
  361

\bibitem[{{Vlaji{\'c}} {et~al.}(2011){Vlaji{\'c}}, {Bland-Hawthorn}, \&
  {Freeman}}]{Vlajic:2011}
---. 2011, \apj, 732, 7

\bibitem[{{Wakker}(2001)}]{Wakker:2001}
{Wakker}, B.~P. 2001, \apjs, 136, 463

\bibitem[{{Walter} {et~al.}(2008){Walter}, {Brinks}, {de Blok}, {Bigiel},
  {Kennicutt}, {Thornley}, \& {Leroy}}]{Walter:2008}
{Walter}, F., {Brinks}, E., {de Blok}, W.~J.~G., {Bigiel}, F., {Kennicutt},
  R.~C., {Thornley}, M.~D., \& {Leroy}, A. 2008, \aj, 136, 2563

\bibitem[{{Wang} {et~al.}(2011){Wang}, {Kauffmann}, {Overzier}, {Catinella},
  {Schiminovich}, {Heckman}, {Moran}, {Haynes}, {Giovanelli}, \&
  {Kong}}]{Wang:2011}
{Wang}, J., {Kauffmann}, G., {Overzier}, R., {Catinella}, B., {Schiminovich},
  D., {Heckman}, T.~M., {Moran}, S.~M., {Haynes}, M.~P., {Giovanelli}, R., \&
  {Kong}, X. 2011, \mnras, 412, 1081

\bibitem[{{Werk} {et~al.}(2011){Werk}, {Putman}, {Meurer}, \&
  {Santiago-Figueroa}}]{Werk:2011}
{Werk}, J.~K., {Putman}, M.~E., {Meurer}, G.~R., \& {Santiago-Figueroa}, N.
  2011, \apj, 735, 71

\bibitem[{{Werk} {et~al.}(2010){Werk}, {Putman}, {Meurer}, {Thilker}, {Allen},
  {Bland-Hawthorn}, {Kravtsov}, \& {Freeman}}]{Werk:2010a}
{Werk}, J.~K., {Putman}, M.~E., {Meurer}, G.~R., {Thilker}, D.~A., {Allen},
  R.~J., {Bland-Hawthorn}, J., {Kravtsov}, A., \& {Freeman}, K. 2010, \apj,
  715, 656

\bibitem[{{Williams} {et~al.}(2009){Williams}, {Dalcanton}, {Dolphin},
  {Holtzman}, \& {Sarajedini}}]{Williams:2009}
{Williams}, B.~F., {Dalcanton}, J.~J., {Dolphin}, A.~E., {Holtzman}, J., \&
  {Sarajedini}, A. 2009, \apjl, 695, L15

\bibitem[{{Wofford}(2009)}]{Wofford:2009}
{Wofford}, A. 2009, \mnras, 395, 1043

\bibitem[{{Yao} {et~al.}(2011){Yao}, {Shull}, \& {Danforth}}]{Yao:2011}
{Yao}, Y., {Shull}, J.~M., \& {Danforth}, C.~W. 2011, \apjl, 728, L16

\bibitem[{{Zahid} \& {Bresolin}(2011)}]{Zahid:2011a}
{Zahid}, H.~J., \& {Bresolin}, F. 2011, \aj, 141, 192

\bibitem[{{Zaritsky} \& {Christlein}(2007)}]{Zaritsky:2007}
{Zaritsky}, D., \& {Christlein}, D. 2007, \aj, 134, 135

\bibitem[{{Zaritsky} {et~al.}(1994){Zaritsky}, {Kennicutt}, \&
  {Huchra}}]{Zaritsky:1994}
{Zaritsky}, D., {Kennicutt}, Jr., R.~C., \& {Huchra}, J.~P. 1994, \apj, 420, 87

\bibitem[{{Zwaan} {et~al.}(2005){Zwaan}, {van der Hulst}, {Briggs},
  {Verheijen}, \& {Ryan-Weber}}]{Zwaan:2005}
{Zwaan}, M.~A., {van der Hulst}, J.~M., {Briggs}, F.~H., {Verheijen}, M.~A.~W.,
  \& {Ryan-Weber}, E.~V. 2005, \mnras, 364, 1467

\end{thebibliography}
